\documentclass[aps,prd,twocolumn]{revtex4-2}

\usepackage{amsmath}
\usepackage{amssymb}
\usepackage{CJK}
\usepackage{graphicx}
\usepackage[colorlinks,citecolor=blue,linkcolor=blue,urlcolor=blue,anchorcolor=blue]{hyperref}
\usepackage{multirow}
\usepackage{tikz}
\usepackage{xcolor}

\makeatletter
\long\def\@hidebookmark#1#2#3#4#5{}
\renewcommand\theequation{\arabic{section}.\arabic{equation}}
\@addtoreset{equation}{section}
\makeatother
\allowdisplaybreaks

\newtheorem{theorem}{Theorem}

\newcommand\dx{{\rm d}}
\newcommand\p{\partial}
\newcommand\HubbleUnit{{\mathrm{km}/\mathrm{s}/\mathrm{Mpc}}}

\newcommand\stablenode{\tikz \foreach \i in {1,...,8}{\pgfmathsetmacro{\angle}{360/8*(\i-1)}\draw[line width=0.45pt] (0,0) -- (\angle:1.05mm);};\,\raisebox{0.8pt}{:}\;}
\newcommand\stablespiral{\tikz \draw [line width=0.45pt,domain=0:25.1327,variable=\t,smooth,samples=100] plot ({\t r}: {0.0002*\t*\t});\,\raisebox{0.8pt}{:}\;}
\newcommand\sgn{{\rm sgn}}
\newcommand\subcm{{\rm c.m.}}
\newcommand\subb{{\rm b}}
\newcommand\subEDE{{\rm EDE}}
\newcommand\subeq{{\rm eq}}
\newcommand\subF{\textsc{f}}
\newcommand\subH{\textsc{h}}
\newcommand\subini{{\rm ini}}
\newcommand\substd{{\rm std}}
\newcommand\subtot{{\rm tot}}

\newcommand\subRmax{{\rm \textsc{r}.max}}
\newcommand\subRmin{{\rm \textsc{r}.min}}
\newcommand\subSsta{{\rm \textsc{s}.sta}}
\newcommand\subSmax{{\rm \textsc{s}.max}}
\newcommand\subSend{{\rm \textsc{s}.end}}
\newcommand\subSinf{{\rm \textsc{s}.inf}}
\newcommand\submax{{\rm max}}

\newcommand\AaA{Astron. Astrophys.}
\newcommand\ApJ{Astrophys. J.}
\newcommand\ApJL{Astrophys. J. Lett.}
\newcommand\ApJS{Astrophys. J. Suppl.}

\newcommand\CQG{Classical Quantum Gravity}
\newcommand\JCAP{J. Cosmol. Astropart. Phys.}
\newcommand\LRR{Living Rev. Relativity}
\newcommand\MNRAS{Mon. Not. R. Astron. Soc.}
\newcommand\PLB{Phys. Lett. B}

\newcommand\PRD{Phys. Rev. D}

\newcommand\PRL{Phys. Rev. Lett.}

\newcommand\RMP{Rev. Mod. Phys.}

\newcommand\AmJPhys{Am. J. Phys.}

\newcommand\AnnPhysNY{Ann. Phys. (N.Y.)}
\newcommand\AnnuRevNuclPartSci{Annu. Rev. Nucl. Part. Sci.}

\newcommand\AstronJ{Astron. J.}
\newcommand\AstropartPhys{Astropart. Phys.}

\newcommand\CommunPhys{Commun. Phys.}

\newcommand\ComputPhysCommun{Comput. Phys. Commun.}

\newcommand\CRPhys{C. R. Phys.}

\newcommand\EurPhysJC{Eur. Phys. J. C}

\newcommand\GenRelativGravit{Gen. Relativ. Gravit.}

\newcommand\IntJTheorPhys{Int. J. Theor. Phys.}
\newcommand\JETPLett{JETP Lett.}
\newcommand\JHighEnergyPhys{J. High Energy Phys.}

\newcommand\JMathPhysNY{J. Math. Phys. (N.Y.)}

\newcommand\NatAstron{Nat. Astron.}

\newcommand\Nature{Nature (London)}

\newcommand\NewJPhys{New J. Phys.}

\newcommand\NuclPhysB{Nucl. Phys. B}

\newcommand\OpenJAstrophys{Open J. Astrophys.}
\newcommand\PhilosMag{Philos. Mag.}
\newcommand\PhilosTransRSocLondon{Philos. Trans. R. Soc. London}
\newcommand\PhilTransRSocA{Phil. Trans. R. Soc. A}
\newcommand\PhysDarkUniverse{Phys. Dark Universe}

\newcommand\PhysRep{Phys. Rep.}
\newcommand\PhysRev{Phys. Rev.}

\newcommand\ProcNatlAcadSciUSA{Proc. Natl. Acad. Sci. U.S.A.}
\newcommand\ProcPhysSocLondon{Proc. Phys. Soc. London}

\newcommand\ProcRSocLondA{Proc. R. Soc. Lond. A.}

\newcommand\Science{Science}

\newcommand\SovPhysJETP{Sov. Phys. JETP}
\newcommand\SovPhysUsp{Sov. Phys. Usp.}

\begin{document}

\begin{CJK*}{UTF8}{gbsn}
\title{Gravitational caloric theory: From early dark energy to a wide variety of \\ gravitational phenomena}
\author{S. X. Tian (田树旬)}
\affiliation{Department of Astronomy, School of Physics and Astronomy, Beijing Normal University, 100875 Beijing, China}
\date{\today}
\begin{abstract}
  We propose gravitational caloric theory (GCT) --- an extension of general relativity that features a vector field $S_\mu$ sourced by and non-minimally coupled to the fluid sector while preserving covariant conservation of the standard fluid energy-momentum tensor. Our initial motivation is to trigger early dark energy (EDE) using the total fluid equation of state that encodes the cosmic radiation-matter transition. This mechanism provides a natural resolution of the EDE coincidence problem. The cosmological background dynamics are analyzed in detail by casting the evolution equations into dynamical-system form and, in relevant reduced cases, using Poincar\'e compactification to uncover the corresponding global phase-space structure. Beyond EDE, GCT admits two novel cosmological applications associated with critical points at infinity. Both arise from a class of energy-cancelling solutions in which conventional energy components preferentially excite $S_\mu$ rather than source spacetime curvature. One is a $\Lambda$-cancelling solution that realizes the self-tuning mechanism for the old cosmological constant problem. However, the present realization does not reenter the standard hot Big Bang phase and is therefore incomplete. The other is a fluid-cancelling solution that serves as the basis for our proposed \textit{early static hot Universe}. In this scenario, $S_\mu$ offsets the gravitational effect of ordinary hot gas, yielding quasi-static expansion with a decreasing comoving Hubble radius that can address the horizon problem. This offers an alternative to inflation. Furthermore, its \textit{hot} ingredient distinguishes this scenario from other quasi-static early Universe models and may leave observable signatures in primordial fluctuations. To probe the viability of GCT beyond cosmology, we further analyze linear perturbations about Minkowski spacetime and investigate static spherically symmetric (strong-field) systems. The perturbative analysis singles out a special parameter point at which the theory is free from instabilities and, under standard asymptotically flat boundary conditions, recovers Newtonian gravity in the Solar System regime and admits six independent luminal gravitational-wave polarizations in the radiative sector. For static spherically symmetric configurations, the solution space exhibits rich structures, including wormholes, naked singularities, and effective $\Lambda$-like behavior, whereas no black hole solution other than Schwarzschild is found in the cases analyzed. Our findings reveal a wide variety of nontrivial gravitational phenomena in GCT, underscoring its promise for future investigations.
\end{abstract}
\makeatletter
\begingroup
  \let\Hy@writebookmark\@hidebookmark
  \maketitle
\endgroup
\makeatother
\end{CJK*}
\tableofcontents

\section{Introduction}\label{sec:I}
General relativity \cite{Einstein1916.NotEngJ.354.769} is the standard theory of gravity. Over the past century, a wide variety of alternative theories have been proposed. Several factors underlie this diversity. First, most existing observational and experimental constraints on gravity take the form of upper-limits, and current data still leave considerable freedom in the description of gravity \cite{Fischbach1999.book,Will2014.LRR.17.4}. Second, the mathematical foundation (notably the variational principle) for generalizing general relativity is well-established, providing systematic routes such as introducing additional fields \cite{Brans1961.PhysRev.124.925} or higher-order curvatures \cite{Buchdahl1970.MNRAS.150.1}. Third, motivations for modifying gravity arise from multiple branches, including astrophysics \cite{Starobinsky1980.PLB.91.99,Milgrom1983.ApJ.270.365,Carroll2004.PRD.70.043528}, particle physics \cite{Lee1955.PhysRev.98.1501,ArkaniHamed1998.PLB.429.263}, and the unified field theory \cite{Green1984.PLB.149.117}. The broad landscape of models can make it challenging to identify which ingredients are essential and which are artefacts of subjective modeling choices. Modern astronomical observations continue to invigorate the study of modified gravity. To confront the diversity challenge, one may reasonably expect from any new theory a clear physical motivation, a broad range of astrophysical applications, and, ideally, testable predictions. In this paper, we propose a new theory named the \textit{gravitational caloric theory} (GCT). Our aim here is not to provide definitive observational tests, but to set up the basic picture of the theory and to outline directions where testable predictions may emerge.

This paper is organized as follows. Section~\ref{sec:I.A} describes our initial motivation for GCT --- from the Hubble tension to early dark energy (EDE) and its coincidence problem. Section~\ref{sec:I.B} discusses the broader aims and scope beyond EDE --- pursuing a more comprehensive study and identifying novel physical insights. For pedagogical purposes, Sec.~\ref{sec:I.C} summarizes the technical background needed for this work. GCT modifies gravity from the fluid side. The motivation for adopting this approach is presented in Secs.~\ref{sec:I.A} and \ref{sec:I.B}. Section~\ref{sec:II} constructs GCT by introducing a new vector field $S_\mu$ and specifying its dynamics such that the conventional fluid energy-momentum tensor remains covariantly conserved even in the presence of direct vector-fluid interactions. To address our initial motivation, Sec.~\ref{sec:III.A} analyzes the sourceless evolution of $S_\mu$ in the radiation and pressureless-matter eras and the triggering of EDE during the cosmic radiation-matter transition. Section~\ref{sec:III.B} then includes the observed cosmological constant $\Lambda$ to analyze the late-time and future evolution of $S_\mu$ within an otherwise standard $\Lambda$CDM framework. In the far future, $S_\mu$ can counteract $\Lambda$, resulting in a small residual Hubble expansion rate. This $\Lambda$-cancelling behavior motivates a separate discussion of the self-tuning mechanism for the old cosmological constant problem. Inspired by a special critical point presented in the EDE analysis, Sec.~\ref{sec:III.C} proposes an early static hot Universe (ESHU) scenario, offering a new solution to the horizon problem and a potential alternative to inflation for describing the real early Universe. Mathematically, centre manifold theory is employed to derive the attractor solutions associated with the self-tuning dynamics and ESHU. Beyond the cosmological background analysis, Sec.~\ref{sec:IV} follows the standard perturbative approach around Minkowski spacetime to discuss the stability issue, Newtonian approximation, and gravitational waves. Section~\ref{sec:V} investigates static spherically symmetric systems as a first step toward compact-object phenomenology. However, due to the complexity of the structure equations, we have only obtained solutions for specific cases, leaving a complete analysis to future work. Conclusions and prospects are summarized in Sec.~\ref{sec:VI}. The Appendix provides basic mathematics for dynamical systems.

Terminology should be mentioned here. \textit{Caloric} originally referred to an invisible field used to describe heat, emerging from 19th-century thermodynamics research \cite{Callendar1910.ProcPhysSocLondon.23.153}. This concept laid the foundation for Carnot's principle, though it was eventually replaced by molecular dynamics. In this paper, we employ the term \textit{gravitational caloric} to signify that the vector field $S_\mu$ is sourced by fluid thermodynamic quantities and that it is involved in the gravitational interaction. The presence of thermodynamic quantities in the gravitational interaction is not weird --- internal energy can act as a source for gravity in general relativity. GCT permits thermodynamic quantities to act beyond the energy-momentum tensor of the fluid. We refer to the EDE model proposed in this paper as \textit{caloric early dark energy} to distinguish it from the other models outlined in Sec.~\ref{sec:I.A}.

\subsection{Initial motivation}\label{sec:I.A}
Since Hubble's discovery of the cosmic expansion \cite{Hubble1929.ProcNatlAcadSciUSA.15.168}, precisely measuring the Hubble constant $H_0$ has been a central pursuit in cosmology \cite{Sandage1958.ApJ.127.513,Tully1977.AaA.54.661,Freedman2001.ApJ.553.47}. Several probes now enable percent-level measurements. One employs low-redshift Type Ia supernovae (SN\,Ia) as a direct distance-redshift probe, which is local and cosmological-model-independent. In contrast, another probe is global and model-dependent: it involves fitting the standard cosmological model~\footnote{The $\Lambda$-cold-dark-matter ($\Lambda$CDM) model together with the flat Friedmann-Lema\^itre-Robertson-Walker (FLRW) background.} to cosmic microwave background (CMB) data and extracts information from the early Universe. These two approaches have dominated the development over the past fifteen years, culminating in the discovery of the Hubble tension --- a significant discrepancy between the local measured value and global fitted value of $H_0$. Here is a brief summary. In 2011, on the eve of the storm, the seven-year \textit{Wilkinson Microwave Anisotropy Probe} result $H_0=70.4\pm2.5\,\HubbleUnit$ \cite{Komatsu2011.ApJS.192.18} and the \textit{Supernova and $H_0$ for the Equation of State} (SH0ES) result $73.8\pm2.4\,\HubbleUnit$ \cite{Riess2011.ApJ.730.119} were in remarkable agreement. Data accumulated over the next five years made the tension clear, mainly solidified by the \textit{Planck} 2013 result \cite{Ade2014.AaA.571.A16}, the \textit{Planck} 2015 result \cite{Ade2016.AaA.594.A13}, and the SH0ES 2016 result \cite{Riess2016.ApJ.826.56}. The tension reached a statistical significance of $5\sigma$ with refined measurements from the \textit{Planck} 2018 result $67.4\pm0.5\,\HubbleUnit$ \cite{Aghanim2020.AaA.641.A6} and the SH0ES 2022 result $73.0\pm1.0\,\HubbleUnit$ \cite{Riess2022.ApJL.934.L7}. Currently, no observationally confirmed systematic errors have been identified that can account for the inconsistency \cite{Efstathiou2021.OpenJAstrophys.4.8,Brout2022.ApJ.938.110,Anderson2024,DiValentino2025.PhysDarkUniverse.49.101965}. Furthermore, new independent and precise measurements from the TRGB-SBF project \cite{Jensen2025.ApJ.987.87} and the Atacama Cosmology Telescope collaboration \cite{Louis2025.JCAP.11.062} confirm the existence of Hubble tension, making a dominant systematic error in previous results (notably the distance ladder calibration) increasingly less tenable. The Hubble tension has become a real crisis and is one of the most significant problems in modern cosmology.

The Hubble tension has motivated extensive research into possible departures from the standard model (encompassing both the $\Lambda$CDM and FLRW sides). Soon after the release of \textit{Planck} 2013 result \cite{Ade2014.AaA.571.A16}, it was recognized that the tension with the SH0ES 2011 result \cite{Riess2011.ApJ.730.119} could potentially be resolved by late-time generalized dark energy models \cite{Salvatelli2013.PRD.88.023531,Xia2013.PRD.88.063501}, local void \cite{Marra2013.PRL.110.241305,Keenan2013.ApJ.775.62}, early dark energy (EDE, \cite{Hojjati2013.PRL.111.041301}), and many other possible mechanisms. This was followed by a substantial body of in-depth research (see Refs.~\cite{Schoneberg2022.PhysRep.984.1,DiValentino2025.PhysDarkUniverse.49.101965} for reviews), yet the conclusions have become highly contentious. The controversy may stem from~\footnote{For clarity, the references cited under each keyword are limited to those addressing a specific type of solution to the Hubble tension. Our aim here is not to provide a complete literature review on the origins of the controversy, but rather to offer illustrative examples. Such a comprehensive review is urgently needed for the community, but appears to be a formidable challenge to accomplish.} divergent modeling strategies for the same physical effect \cite{BenDayan2014.PRL.112.221301,Odderskov2016.JCAP.02.001}, updates to observational data and the emergence of new tensions \cite{Poulin2019.PRL.122.221301,Hill2020.PRD.102.043507,Hill2022.PRD.105.123536,Smith2022.PRD.106.043526,Efstathiou2024.PRL.132.221002}, distinct statistical methodologies \cite{Zhao2017.NatAstron.1.627,Wang2018.ApJL.869.L8}, potential misuse of data \cite{Camarena2021.MNRAS.504.5164,Efstathiou2021.MNRAS.505.3866}, along with other ambiguous factors. No proposal has yet emerged as a widely accepted solution that simultaneously resolves the Hubble tension and satisfies all major cosmological constraints. Furthermore, recently, \citet{Calabrese2025.JCAP.11.063} concluded that no extended model is statistically favored over $\Lambda$CDM. Nevertheless, systematic comparative studies \cite{Knox2020.PRD.101.043533,Schoneberg2022.PhysRep.984.1,Khalife2024.JCAP.04.059} still identify weak relative advantages for some models by fairly comparing multiple candidates under consistent datasets, methodologies, and technical settings. At current stage, EDE stands as a prominent example among the promising candidates, with a clear mechanism that operates by modulating the sound horizon formed prior to cosmic recombination to adjust the CMB-inferred $H_0$ value. The shortcomings of EDE are also clear. Aside from its limitations in data fitting \cite{Hill2020.PRD.102.043507,Jedamzik2021.CommunPhys.4.123}, the EDE scenario is further plagued by the theoretical coincidence problem. This raises the question of whether the coincidence problem can guide a physically motivated refinement of the EDE model, with possible implications for its observational viability.

The coincidence problem concerns why EDE emerges precisely around the time of matter-radiation equality (redshift $\sim3400$) \cite{Lin2019.PRD.100.063542,Sakstein2020.PRL.124.161301}. This poses a severe puzzle in the single canonical scalar field models \cite{Kamionkowski2014.PRL.113.251302,Poulin2018.PRD.98.083525,Lin2019.PRD.100.063542}, since the scalar field dynamics therein are almost~\footnote{While ordinary matter can exert an indirect influence on scalar field dynamics through Hubble friction, this does not provide a mechanism to control the emergence time of EDE.} decoupled from both radiation and pressureless matter. The emergence time in such canonical scalar EDE models hinges on an energy scale in its potential that is essentially arbitrary. This issue also afflicts some other EDE proposals \cite{Niedermann2021.PRD.103.L041303,Oikonomou2021.PRD.103.044036}. The key to resolving the coincidence problem lies in identifying a suitable trigger~\footnote{\citet{Zumalacarregui2020.PRD.102.023523} constructed a trigger-free solution, in which the key ingredient is an energy density that dilutes faster than matter but slower than radiation. This naturally forces the peak of its relative energy density to coincide with the epoch of matter-radiation equality. However, the slower dilution rate compared to radiation (after the peak) appears disfavored by cosmological constraints. Furthermore, this scenario requires a fine-tuning of the initial energy density of the EDE component in the earlier universe to achieve $\Omega_\subEDE\approx10\%$ near equality.} around the epoch of matter-radiation equality to replace the artificially imposed energy scale. Proposed triggers include the relativistic-to-non-relativistic transition of massive neutrinos \cite{Sakstein2020.PRL.124.161301,CarrilloGonzalez2021.JCAP.04.063}, the onset of dark matter dominance \cite{Karwal2022.PRD.105.063535,Lin2023.PRD.107.103523}, and the total equation of state (EoS) of the fluid \cite{Tian2021.PRD.103.043518,Tian2023.PRD.107.103507} or spacetime dynamics \cite{Jing2024.PRD.109.044016} that encodes cosmic radiation-matter transition. However, from the perspectives of theoretical consistency and practical utility, none of the existing models appears to be fully satisfactory. For the neutrino-assisted EDE scenario, \citet{Maziashvili2023.AstropartPhys.145.102792} and \citet{deSouza2023.PRD.108.083512} argued that the neutrino sector does not suffice to trigger the observationally required amount of EDE~\footnote{\citet{CarrilloGonzalez2023} dispute these arguments. However, they do not provide a concrete and viable parameter setting (in their Sec.~II\,C) to directly address the issue, particularly the $\Delta$ criterion proposed in Ref.~\cite{deSouza2023.PRD.108.083512}. Separately, \citet{Kamionkowski2025.PRD.111.063551} proposed a new neutrino-triggered EDE model, in which the dependence on initial conditions leaves the coincidence problem fundamentally open.}. Technically, such a model requires modifications to the Boltzmann hierarchy in cosmological perturbation analysis \cite{Ichiki2008.JCAP.06.005,Oldengott2015.JCAP.04.016}, and the difficulty has likely slowed progress in this direction. In the dark matter-triggered EDE~\footnote{As stated by \citet{Lin2023.PRD.107.103523}, the model proposed by \citet{Karwal2022.PRD.105.063535} does not correctly resolve the EDE coincidence problem. Here we refer to the model proposed by \citet{Lin2023.PRD.107.103523}.}, the coincidence problem appears to be resolved, as the timing of EDE emergence is primarily set by when the square of Hubble expansion rate becomes comparable to the dark matter energy density (in suitable units). However, this mechanism requires a bare potential with a flat $\mathrm{eV}$-scale plateau, with the EDE amplitude being set by this energy scale. Such a low-energy-scale plateau may introduce a cosmological constant-like fine-tuning problem at the beginning of the Big Bang \cite{Carroll2001.LRR.4.1}. The EoS-triggered EDE requires modification of gravitational theory from the fluid side \cite{Tian2021.PRD.103.043518,Tian2023.PRD.107.103507}. Inheriting a common drawback of $f(R,T)$-like theory \cite{Harko2010.EurPhysJC.70.373,Harko2011.PRD.84.024020}, this framework generally exhibits non-conservation of the fluid energy-momentum tensor. This inevitably affects the Boltzmann hierarchy for conventional matter, particularly the collision terms. However, no self-consistent theoretical framework currently exists to calculate these modifications quantitatively. The spacetime-triggered EDE can be embedded into the scalar-tensor theory of gravitation. The model proposed by \citet{Jing2024.PRD.109.044016} may introduce a singularity problem~\footnote{See discussions above Eq.~(8) in Ref.~\cite{Jing2024.PRD.109.044016}. While introducing a cosmological constant can provide a temporary fix, it makes the model theoretically unappealing.} due to the geometric curvatures appearing in the denominator. In addition, as a modified gravity, this model requires further scrutiny to ensure its compatibility with a broader set of tests.

Given the shortcomings in existing models, it is necessary to construct a better EDE model capable of resolving the coincidence problem. Section~\ref{sec:II} builds upon our previous approach \cite{Tian2021.PRD.103.043518,Tian2023.PRD.107.103507} to refine the model with modifications to the fluid sector. To avoid potential fine-tuning problems, we impose a strong rule~\footnote{Cosmic evolution spans a wide range of energy scales. A theory that solves every observational anomaly by inserting new physics at a specific energy scale becomes increasingly cumbersome. Furthermore, a collection of phenomenological patches may conceal a more fundamental, unified principle. One should not introduce too many anthropic energy scales into a gravitational system. It could be argued that this statement is challenged by the case of dark energy. Nevertheless, despite its limited acceptance, there are dark energy models that avoid any ad-hoc energy scale \cite{Dodelson2000.PRL.85.5276,Tian2020.PRD.101.063531}.}: no additional parameters (including model parameters or initial conditions of the new field) are introduced to set the energy scale of EDE, and all parameters should be dimensionless and of order unity. Compared with our previous modified-fluid theory, the new model ensures the conservation of the energy-momentum tensor of conventional fluid, thereby facilitating its future integration into Einstein-Boltzmann solvers \cite{Lewis2000.ApJ.538.473,Blas2011.JCAP.07.034}. The analysis of EDE dynamics is presented in Sec.~\ref{sec:III.A}.

Besides the utility as an EDE trigger, a further motivation for adopting the fluid approach is provided by the $S_8$ tension, which characterizes the amplitude of matter clustering on cosmological scales \cite{Peebles1980.book,Martinez2002.book}. Observationally, a tension between high- and low-redshift probes has been reported at a confidence level of $2\sim3\sigma$ \cite{Ade2016.AaA.594.A24,Aghanim2020.AaA.641.A6,Asgari2021.AaA.645.A104,Abbott2022.PRD.105.023520,NoS8Tension}. There are numerous attempts in the literature to address this tension \cite{DiValentino2025.PhysDarkUniverse.49.101965}. One proposed solution is simply to modify the evolution of (dark) matter density fluctuations in the late-time Universe \cite{Poulin2023.PRD.107.123538,Lin2024.PRD.109.063523}. This approach can be incorporated into our theoretical framework --- since (dark) matter is treated as fluid in cosmology, modifying gravity from the fluid side should directly impact the evolution of matter fluctuations. It should be noted that this statement does not conflict with the energy-momentum conservation of conventional fluids in our theory --- the modification is applied to the Poisson equation on cosmological scales, rather than to the geodesic equation. The analysis of cosmological perturbations for our theory, including its application to the $S_8$ tension, will be presented in a separate publication.

\subsection{Broader aims and scope}\label{sec:I.B}
A gravity theory can admit applications in a wide variety of gravitational systems beyond cosmology \cite{Zeldovich1971.book,Misner1973.book}. In particular, our theory incorporates the rule that only dimensionless model parameters of order unity are introduced. This makes it possible for the applications to broadly cover gravitational regimes ranging from weak- to strong-field systems. Owing to both astronomical observations and laboratory experiments, precise measurements are now available for some cases \cite{Will2014.LRR.17.4,deRham2017.RMP.89.025004,Baker2021.RMP.93.015003,Abbott2019.PRL.123.011102}. If a theory is to be considered a real physical theory rather than a toy model, it must be able to pass these tests. To this end, Secs.~\ref{sec:IV} and \ref{sec:V} present preliminary studies of weak-field and compact-object phenomenology in GCT. Section~\ref{sec:IV} develops the gauge-invariant linear perturbation theory around Minkowski spacetime as a first weak-field consistency check. The resulting framework is used to diagnose perturbative stability, recover the Newtonian limit relevant to Solar-System tests, and characterize gravitational-wave physics. The primary aim of Sec.~\ref{sec:V} is to take a first step toward compact-object phenomenology by studying static spherically symmetric systems in the strong-field regime. Because the full structure equations are highly complex, the present analysis is restricted to two fluid-free vacuum subsectors. Within the explored cases, the only black-hole branch identified so far is the Schwarzschild solution \cite{NoHairCaveat}, while the remaining solutions include naked singularities, wormholes, and effective-$\Lambda$-like behavior. These preliminary results leave several concrete paths for future analysis.

Apart from agreement with established tests, the predictive power of a theory for new physics represents a key criterion of its value. For this pursuit, mathematical equations and solutions can guide the direction of human thought. Historically, a convincing example supporting this claim comes from Dirac's prediction of the positron, based on the negative-energy solutions of his equation \cite{Dirac1931.ProcRSocLondA.133.60}. In cosmology, parallel instances exist, albeit with less influence. For example, before \citet{Guth1981.PRD.23.347} proposed the inflationary theory, the possibility that the early Universe could undergo exponential expansion had been independently identified by \citet{Starobinsky1980.PLB.91.99} through a solution of quantum-corrected Einstein equations and by \citet{Sato1981.MNRAS.195.467} in the context of studying vacuum phase transitions~\footnote{Unfortunately, both \citet{Starobinsky1980.PLB.91.99} and \citet{Sato1981.MNRAS.195.467} failed to recognize that the exponential expansion could resolve the horizon and flatness problems.}. A lesson from history is that some novel or even unnatural mathematical solutions may harbor important new physics. This guiding philosophy underpins our work in Secs.~\ref{sec:III.B} and \ref{sec:III.C}, where physical scenarios were constructed specifically to interpret the mathematical results. This methodology differs fundamentally from that employed in other sections of this paper. Section \ref{sec:III.B} numerically computes the cosmological evolution of $S_\mu$ within the case incorporating $\Lambda$ (see Fig.~\ref{fig:cosmos:05} Right). We then identify this solution as the self-tuning mechanism --- a process where field dynamics cancel the gravitational effect of a large bare $\Lambda$ \cite{Dolgov1985.JETPLett.41.345,Ford1987.PRD.35.2339,Charmousis2012.PRL.108.051101}. Section \ref{sec:III.C} is devoted to exploring the possible cosmological applications of critical points other than $F_1$ in Fig.~\ref{fig:cosmos:02}, which describe the dynamics of $S_\mu$ with a perfect fluid. The critical point $I_3$ at infinity caught our attention and led us to propose a new physical scenario --- ESHU. The remarkable aspect is that the ESHU can resolve the horizon problem and possesses the potential to replace inflation in describing the real early Universe.

As discussed in Sec.~\ref{sec:I.A}, both Hubble tension and $S_8$ tension may be regarded as signals to modify gravity theory from the fluid side. However, this approach is not unique and may encounter greater challenges (particularly concerning energy-momentum conservation). Given this, why do we still persist with the fluid approach? There are three additional reasons. First, the fluid non-conservation issue can be resolved through the introduction of a new vector field. Second, the reasoning is motivated by a historical perspective. Newton's law of universal gravitation states that mass generates gravity \cite{Newton1687.book}. Einstein's general relativity formulates gravity as being sourced by all components of the energy-momentum tensor, encompassing both internal energy and pressure \cite{Einstein1916.NotEngJ.354.769,Misner1973.book}. Our objective is to study other possible gravitational couplings of fluid thermodynamic quantities and explore their implications for modern astrophysics. The third aim is to establish a connection between the Newtonian gravitational constant $G$ and other fundamental physical constants, as Maxwell did for the speed of light \cite{Maxwell1861.PhilosMag.90.11,Maxwell1865.PhilosTransRSocLondon.155.459}. Many notable efforts have been made in this direction. For example, \citet{Dirac1937.Nature.139.323} and \citet{Dicke1959.Science.129.621} argued that the value of $G$ depends on cosmological variables, while \citet{Sakharov1967.NotEngJ.177.70} attempted to derive it from quantum field theory. Our idea is that if the source of gravity were to be fundamentally replaced by other thermodynamic quantities of the fluid, it could yield a novel gravitational coupling constant, which might resolve certain challenges in this field. This goal requires a self-consistent gravitational theory incorporating modified fluid, though its complete resolution lies far beyond the scope of this work.

\subsection{Technical background}\label{sec:I.C}
The cosmological analysis in this paper concentrates on the background evolution, which is mainly investigated through dynamical analysis. We recommend three models to master the basic techniques. One is the canonical scalar field with an exponential potential, serving as a model for inflation \cite{Lucchin1985.PRD.32.1316} and dark energy \cite{Ratra1988.PRD.37.3406}. \citet{Copeland1998.PRD.57.4686} introduced two key variables and established a foundational framework for dynamical analysis, with the corresponding mathematics being addressed using linear stability theory. Another model that describes phantom dark energy also employs an exponential potential but is characterized by a negative kinetic term \cite{Hao2003.PRD.67.107303}. \citet{Bahamonde2018.PhysRep.775.1} projected the two-dimensional phase portrait \cite{UrenaLopez2005.JCAP.09.013} onto the Poincar\'e sphere to analyze the possible past attractors of this system. The mathematics involved here concerns the critical points at infinity and their stability analysis. The third is the chaotic inflation model with a quadratic potential \cite{Linde1983.PLB.129.177}. \citet{Belinsky1985.PLB.155.232} discussed the properties of this dynamical system at infinity, while \citet{Rendall2002.GenRelativGravit.34.1277} demonstrated how to employ centre manifold theory to analyze the saddle-type attractor solutions~\footnote{The \textit{saddle-type} indicates a critical point that attracts trajectories in one direction and repels in the other. The \textit{attractor} in this context specifically refers to the stable manifold --- a special trajectory along the repulsive direction that attracts nearby orbits. Figure 1b in Ref.~\cite{Belinsky1985.PLB.155.232} provides an illustration.}. The preceding discussion provides the foundation for the three tools used in Sec.~\ref{sec:III}: linear stability theory, centre manifold theory, and the analysis of critical points at infinity. A brief summary can be found in Ref.~\cite{Bahamonde2018.PhysRep.775.1}, while for a more rigorous mathematical treatment, we refer the reader to Refs.~\cite{Lefschetz1957.book,Carr1981.book,Perko2001.book,Meiss2007.book}. We highlight that centre manifold theory proves to be a powerful tool for analyzing attractor solutions in cosmology. Note that the stability theorem at infinity presented in Refs.~\cite{Bahamonde2018.PhysRep.775.1,Perko2001.book} involves a sign dependency whose direct application would render the analysis in Table \ref{tab:03} of this paper unduly cumbersome. To circumvent this, Appendix~\ref{sec:App.A.3} provides the results of the coordinate transformations for our dynamical system, which resolves the sign ambiguity inherent in the approach of Refs.~\cite{Bahamonde2018.PhysRep.775.1,Perko2001.book}. This derivation also serves to highlight the importance of \textit{time rescaling} and the explicit $y_i\rightarrow z_i$ projection (see Fig. \ref{fig:cosmos:06}).

In a metric theory of gravity, linear perturbations are important and possess well-defined analytical techniques. Einstein employed this tool extensively in the early stages of general relativity \cite{Einstein1916.NotEngJ.354.769,Einstein1916.NotEngJ.NotEngJ.688}. Our analysis adopts the modern standard approach. Mathematically, linear perturbations of the Minkowski metric mainly involve scalar-vector-tensor decomposition, constructing gauge invariants, and decoupling evolution equations. \citet{Flanagan2005.NewJPhys.7.204} summarized the details for general relativity. We can draw from the metric decomposition and relevant gauge invariants. The spatial components of the vector field $S_\mu$ can be treated via the Helmholtz decomposition. The constraint equations in the tensor decomposition can guide the decoupling of the equations. The physical applications do not rely on specific mathematical tools. For the relevant theoretical background, we recommend Ref.~\cite{Delhom.arXiv.2207.13431} for instability analysis, Ref.~\cite{Bertschinger2011.PhilTransRSocA.369.4947} for the Newtonian approximation, and Refs.~\cite{Eardley1973.PRD.8.3308,Eardley1973.PRL.30.884} for gravitational waves. 

Static spherically symmetric systems are intended to analyze compact objects such as black holes and neutron stars. Following the Tolman-Oppenheimer-Volkoff approach \cite{Tolman1939.PhysRev.55.364,Oppenheimer1939.PhysRev.55.374}, it is straightforward to write down the general structure equations for GCT. However, the equations are too complex to be solved exactly, and moreover, we lack effective tools for a systematic analysis. We thus focus on the simplest possible cases. The cores of simplification are equation decoupling and reducing the order of differential equations through appropriate variable transformations. Fortunately, in the simplest case, the structure equations can be reduced to an Abel equation of the first kind \cite{Polyanin2018.book}, which is a first-order nonlinear ordinary differential equation in one variable. A slight extension can be described by a system of first-order differential equations in two variables. Mathematical analysis involves numerical computation, series expansions, and the construction of special exact solutions. To determine the nature of the metric (especially the singularities), we compute curvature scalars and employ series expansions to establish connections with known metrics. The latter is essential to our discussion of wormholes, for which we recommend Refs.~\cite{Ellis1973.JMathPhysNY.14.104,*Ellis1974,Morris1988.AmJPhys.56.395} for the physics.

The software and programs employed in this work are listed below. Analytical derivations about Riemannian geometry were carried out using a custom \texttt{Maple} program \cite{Tian2020} as well as the \texttt{xAct}~\footnote{\url{https://www.xact.es}} package (including \texttt{xTensor}, \texttt{xPert} \cite{Brizuela2008.GenRelativGravit.41.2415}, \texttt{xCoba} and \texttt{xTras} \cite{Nutma2013.ComputPhysCommun.185.1719}). Other mathematical simplifications and numerical calculations, including solving differential equations and matrix analysis, were performed with \texttt{Mathematica} and \texttt{Maple}. For stiff differential equations, we ensure the correctness of the results by increasing the numerical precision. Figures were plotted using \texttt{Mathematica} and \texttt{Ti\textit{k}Z}.

\section{The theory}\label{sec:II}
As a classical field theory, the first step is to formulate the gravitational field equations. This section presents our theory. Traditional methods for obtaining the field equations encompass direct construction and the variational principle. The former is the approach employed by Einstein in his formulation of general relativity \cite{Einstein1916.NotEngJ.354.769}. The latter has been widely adopted in modified gravities \cite{Brans1961.PhysRev.124.925,Buchdahl1970.MNRAS.150.1,Horndeski1974.IntJTheorPhys.10.363}, as it provides a well-defined strategy that yields field equations consistent with conservation laws \cite{Bargmann1957.RMP.29.161,Bekenstein2004.PRD.70.083509}. However, for the gravitational theories that modify the fluid sector \cite{Sotiriou2008.CQG.25.205002,Harko2010.EurPhysJC.70.373,Harko2011.PRD.84.024020,Pinto2022.PRD.106.044043,Tian2023.PRD.107.103507}, the variational principle generally fails to ensure conventional energy-momentum conservation~\footnote{Conformal matter coupling can preserve the usual matter conservation law in the corresponding physical metric \cite{Damour1992.CQG.9.2093}. However, it operates through a metric rescaling rather than direct couplings between a new field and fluid thermodynamic variables, which limits its applicability to our EDE purpose. A recent Herglotz-type extension of $f(R,\mathrm{Matter})$ gravity may offer another route by introducing an additional closed one-form whose contribution cancels the non-conservation current \cite{TamasGeza2026}. This is distinct from our direct-construction approach.}. We therefore adopt the method of direct construction.

Throughout this paper, we adopt SI units and retain all physical constants, unless otherwise specified. For the conventions of Riemannian geometry, we refer the reader to footnote [32] in Ref.~\cite{Tian2023.PRD.107.103507}. For clarity, we use the metric signature $(-,+,+,+)$ and write $\Box\equiv\nabla_\alpha\nabla^\alpha$. Table~\ref{tab:01} summarizes the main symbols used in Secs.~\ref{sec:II} and \ref{sec:III} and the corresponding reference equations. The symbols used in Secs.~\ref{sec:IV} and \ref{sec:V} are not fully consistent with the preceding sections, as they adhere to the established conventions of the specific fields discussed therein.

\begin{table*}[!t]
  \centering
  \begin{minipage}{0.88\textwidth}
  \caption{Main symbols used in Secs.~\ref{sec:II} and \ref{sec:III}.}
  \label{tab:01}
  \begin{tabular*}{\hsize}{@{\ \ }@{\extracolsep{\fill}}lll@{\ \ }}
    \hline\hline
    Symbols                     &  Physical Interpretations                                &  References  \\
    \hline
    $G_{\mu\nu}$                &  Einstein tensor                                         &  (\ref{eq:2.1})  \\
    $H$                         &  Hubble parameter                                        &  (\ref{eq:3.1})  \\
    $\kappa$                    &  $\equiv8\pi G/c^4$                                      &  (\ref{eq:2.1}), Sec.~\ref{sec:IV.C.1}  \\
    \hline
    $T_{\mu\nu}$                &  Energy-momentum tensor of perfect fluid plus $\Lambda$  &  (\ref{eq:2.2})  \\
    $w_\subF$                   &  Equation of state of perfect fluid                      &  (\ref{eq:2.2})  \\
    $J_\mu$                     &  $\equiv\sqrt{\kappa\rho_\subF}\,u_\mu$                  &  Sec.~\ref{sec:II.A}  \\[0.3mm]
    $\varrho_\subF$             &  $\equiv\displaystyle\sqrt{8\pi G\rho_\subF}$            &  (\ref{eq:3.1})  \\[0.1mm]
    \hline
    $S_\mu$                     &  Vector caloric field                                    &  (\ref{eq:2.1})  \\
    $S_{\mu\nu}$                &  Effective gravitational tensor of $S_\mu$               &  (\ref{eq:2.2})  \\
    $c_i,\,\tilde{c}_i$         &  Model parameters                                        &  (\ref{eq:2.1}), (\ref{eq:3.1})  \\
    $\Omega_i$                  &  Time-dependent density parameter                        &  (\ref{eq:3.3})  \\
    $x_i,\,y_i,\,z_i$           &  Variables for the dynamical systems                     &  (\ref{eq:3.2}), App.~\ref{sec:App.A.3}  \\
    $t,\,N,\,\sigma$            &  Cosmic time, \textit{e}-folding number, the rescaled $N$  &  (\ref{eq:3.1}), (\ref{eq:3.5}), App.~\ref{sec:App.A.3}  \\
    $F_i,\,I_i,\,\mathcal{F}_i,\,\mathcal{I}_i$  &  Critical points in the finite region and at infinity  &  Tables \ref{tab:02} \& \ref{tab:03}  \\
    $J$                         &  Jacobian matrix                                         &  (\ref{eq:3.7})  \\
    $\alpha_i$                  &  the coefficients in $u_h=\phi(u_c)=\sum_{i=2}^{\infty}\alpha_i u_c^i$  &  (\ref{eq:3.26}), App.~\ref{sec:App.A.2}  \\
    $\alpha,\,\beta_\subcm$     &  Parameters used in Sec.~\ref{sec:III.C.3}               &  (\ref{eq:3.39}), (\ref{eq:3.44})  \\[1mm]
    \hline
    Subscripts                                    \\
    $\subF$                     &  Perfect fluid                                           &  (\ref{eq:2.2})  \\
    $\Lambda$                   &  Cosmological constant                                   &  (\ref{eq:2.2})  \\
    $\substd$                   &  Sources in the standard $\Lambda$CDM model ($\subF+\Lambda$)  &  Sec.~\ref{sec:III.A.1}  \\
    $\subEDE$                   &  Early dark energy                                       &  (\ref{eq:3.3})  \\
    $\subtot$                   &  All sources in our model ($\substd+\subEDE$)            &  Sec.~\ref{sec:III.B.1}  \\
    $\subini,\,\subeq$          &  Initial time, cosmic matter-radiation equality          &  (\ref{eq:3.9}), (\ref{eq:3.13})  \\
    $\subcm$                    &  Centre manifold  &  (\ref{eq:3.20})  \\
    \hline\hline
  \end{tabular*}
  \end{minipage}
\end{table*}

\subsection{Field equations}\label{sec:II.A}
The key to constructing the field equations lies in finding a covariantly conserved rank-2 symmetric tensor that involves the fluid thermodynamic quantities but is distinct from the energy-momentum tensor. It seems impossible for the tensor to be comprised solely of the fluid. If there is an unknown field, we can enforce a vanishing covariant divergence of the tensor by appropriately specifying the equation of motion of the field. The vanishing divergence gives rise to four constraint equations. Therefore, the most straightforward choice is a vector field $S_\mu$ with four components, rather than a scalar field.

What should be the mathematical structure of the symmetric tensor? The electromagnetic field may not serve as a useful guide, as it is incompatible with a homogeneous and isotropic Universe \cite{Jacobs1968.ApJ.153.661,Heisenberg2019.PhysRep.796.1}. A mathematical reason for this fact is that the strength tensor $F_{\mu\nu}\equiv\p_\mu A_\nu-\p_\nu A_\mu$ is antisymmetric, which hides the $A_0$ component in the cosmic background. In contrast, nonlocal RT gravity \cite{Maggiore2014.PRD.89.043008,Dirian2014.JCAP.06.033,Tian2019.PRD.100.124059} provides a symmetric tensor $\nabla_\mu S_\nu+\nabla_\nu S_\mu$, which gives a nonzero cosmological contribution of $S_0$ and is compatible with a homogeneous and isotropic Universe. Such a tensor is the starting point of our construction.

As we discussed in Sec.~\ref{sec:I}, model parameters should be dimensionless. Therefore, the Einstein tensor $G_{\mu\nu}$ and $\nabla_\mu S_\nu$ must have the same dimension, which in turn gives $[S_\mu]=[\nabla_\mu]=\mathrm{length}^{-1}$ \footnote{For simplicity, in Sec.~\ref{sec:II}, we can assume that the metric is dimensionless and $[\dx x^\mu]=\mathrm{length}$. This assumption breaks down for the FLRW metric, which would assign a different dimension to $S_0$ in Sec.~\ref{sec:III}. Nevertheless, this inconsistency does not affect the discussion in Sec.~\ref{sec:II}.}, where $[\cdots]$ denotes the dimension. The philosophy of assigning dimension to the field instead of model parameters is the same as that adopted by \citet{Brans1961.PhysRev.124.925}. We believe this is a sound principle in constructing a theory of gravity. At this stage, we can write the preliminary forms of the gravitational field equation and vector field equation as
\begin{gather}
  G_{\mu\nu} + \nabla_\mu S_\nu + \nabla_\nu S_\mu + \cdots = \kappa T_{\mu\nu}, \nonumber\\
  \Box S_\mu + \nabla^\nu\nabla_\mu S_\nu + \cdots = 0. \nonumber
\end{gather}
For the pure vector field, dimensional analysis allows us to further add the $S_\mu S_\nu$, $g_{\mu\nu}\nabla_\alpha S^\alpha$, and $g_{\mu\nu}S_\alpha S^\alpha$ terms to the gravitational field equation. Consequently, these terms should also be included in the vector equation.

We now turn to the source term of the EDE. This must correctly represent the cosmic radiation-matter transition. Considering the dimensions, a suitable choice is a term like $(\p_\mu w_\subF)^2$, where $w_\subF$ is the EoS for all matter that can be modeled as a perfect fluid. This term acts as a source in the vector field equation, which is the mathematical counterpart of the \textit{gravitational caloric} interpretation discussed in Sec.~\ref{sec:I}. The source term vanishes in both the standard radiation- and (pressureless) matter-dominated eras due to the constancy of $w_\subF$, while being nonzero during the transition. In the EDE scenario, we should also expect the energy of $S_\mu$ to approach zero in both the radiation- and matter-dominated eras. Mathematically, we require a specific critical point to be stable (see Sec.~\ref{sec:III.A} for details).

Unfortunately, our calculations show that the present terms are unable to ensure the required stability~\footnote{More specifically, the $\nabla_\mu S_\nu$, $c_2$, $c_4$ and $c_6$ terms in Eq.~(\ref{eq:2.1}) are unable to ensure the stability of the critical point $F_1$ in Fig.~\ref{fig:cosmos:02}.}. This stability analysis depends on the linear terms in $S_\mu$ that appear in the gravitational field equations, such as the $\nabla_\mu S_\nu$ and $g_{\mu\nu}\nabla_\alpha S^\alpha$ terms. Nonlinear terms such as $S_\mu S_\nu$ do not contribute to EDE but serve other applications and have therefore been retained. Now, for the EDE purpose, we need more linear terms. Dimensional analysis indicates that terms similar to $G_{\mu\nu}^{1/2}$ and $(\kappa T_{\mu\nu})^{1/2}$ can be used to replace $\nabla_\mu$. The former is not an optimal choice, as it complicates the analysis of the vacuum case. Furthermore, it appears that we cannot define a rank-1 tensor analogous to $G_{\mu\nu}^{1/2}$. The latter does not suffer from these issues. Notably, it enables us to investigate more diverse forms of fluid-gravity coupling. Here we introduce $J_\mu\equiv\sqrt{\kappa\rho_\subF}\,u_\mu$ with dimension $\mathrm{length}^{-1}$, where $\rho_\subF$ is the fluid density and $u_\mu$ is the four-velocity. Finally, based on dimensional analysis, we can add the $J_\mu S_\nu$, $g_{\mu\nu}J_\alpha S^\alpha$, $J_\mu J_\nu$, and $g_{\mu\nu}J_\alpha J^\alpha$ terms to the gravitational field equation, in which the first two are linear in $S_\mu$.

In summary, the field equations could be written as
\begin{subequations}\label{eq:2.1}
\begin{align}
  & G_{\mu\nu} + S_{\mu\nu} = \kappa T_{\mu\nu}, \label{eq:2.1a}\\
  & \Box S_\mu + \nabla^\nu\nabla_\mu S_\nu + c_5\nabla^\nu(J_\mu S_\nu + J_\nu S_\mu) \nonumber\\ 
  & \quad + \nabla^\nu(c_6 S_\mu S_\nu + c_7 J_\mu J_\nu) = \nabla_\mu X, \label{eq:2.1b}
\end{align}
\end{subequations}
where
\begin{subequations}\label{eq:2.2}
\begin{align}
  S_{\mu\nu} &= \nabla_\mu S_\nu + \nabla_\nu S_\mu + c_5(J_\mu S_\nu+J_\nu S_\mu) \nonumber\\
             &  \quad + c_6 S_\mu S_\nu + c_7 J_\mu J_\nu - g_{\mu\nu}X, \\
  T_{\mu\nu} &= (\rho_\subF+p_\subF/c^2)u_\mu u_\nu +p_\subF g_{\mu\nu} - \Lambda g_{\mu\nu}/\kappa, \label{eq:2.2b}
\end{align}
\end{subequations}
and $X \equiv c_1\p^\mu w_\subF\p_\mu w_\subF + (c_2 \nabla_\mu + c_3 J_\mu + c_4 S_\mu) S^\mu + c_8 J_\mu J^\mu$, the constant $\kappa\equiv8\pi G/c^4$ \footnote{Generally, a rescaled $G$ is required in modified gravity \cite{Tian2019.PRD.99.064044}. However, Sec. \ref{sec:IV} shows that it is not needed for GCT.}, and $p_\subF$ is the fluid pressure, $\Lambda$ is the cosmological constant, $c$ is the speed of light, $c_i$ are dimensionless model parameters~\footnote{It is unnecessary to introduce a free coefficient for the $\nabla_\mu S_\nu + \nabla_\nu S_\mu$ term in $S_{\mu\nu}$ due to a scaling property --- if $S_{\mu\nu}=c_0(\nabla_\mu S_\nu + \nabla_\nu S_\mu)+\cdots$, then the transformation $\{\tilde{S}_\mu=c_0S_\mu,\, \tilde{c}_1=c_1,\, \tilde{c}_2=c_2/c_0,\, \tilde{c}_3=c_3/c_0,\, \tilde{c}_4=c_4/c_0^2,\, \tilde{c}_5=c_5/c_0,\, \tilde{c}_6=c_6/c_0^2,\, \tilde{c}_7=c_7,\, \tilde{c}_8=c_8\}$ absorbs the coefficient and leaves the effective gravitational tensor of the vector field unchanged.}. The symmetric tensor $S_{\mu\nu}$ can be regarded as the effective gravitational tensor for $S_\mu$. Together with the Bianchi identity, Eq.~(\ref{eq:2.1b}) serves to maintain consistency with $\nabla^\nu T_{\mu\nu}=0$. This conservation law implies that the kinematic equations of particles in curved spacetime remain unchanged. Consequently, the thermodynamic properties (including $w_\subF$) of conventional matter can be calculated in the same way as in general relativity. Note that $T_{\mu\nu}$ encompasses all energy components in the standard $\Lambda$CDM model, e.g., photon gas, neutrino gas, baryons, dark matter, and dark energy ($\Lambda$). In contrast, $J_\mu$ comprises only the fluid components, i.e., the first four, but excludes dark energy. This is an assumption of our theory without any physical motivation.

Besides the EDE application, the linear terms in $S_{\mu\nu}$ (especially minimally coupled ones like $\nabla_\mu S_\nu$) grant the theory intrinsic merit. Comparison can highlight its importance. In general relativity, $G_{\mu\nu}$ contains the linear terms of the metric~\footnote{One may recall the linearized Einstein field equations.}. In contrast, no linear term appears in the effective energy-momentum tensor of existing vector field theories, such as Maxwell theory \cite{Maxwell1861.PhilosMag.90.11,Maxwell1865.PhilosTransRSocLondon.155.459}, Proca theory \cite{Proca1936.NotEngJ.7.347}, and many generalizations \cite{Tasinato2014.JHighEnergyPhys.04.067,Heisenberg2014.JCAP.05.015}. Our construction, as well as nonlocal RT gravity \cite{Maggiore2014.PRD.89.043008}, may suggest a new mathematical structure or symmetry for the vector field.

One thing should be declared here. The $\{c_7,c_8\}$ terms in Eq.~(\ref{eq:2.1}) were introduced after we had completed all calculations in this paper. The calculations were not updated accordingly. We retain these two terms to facilitate future investigation of their role in the $S_8$ tension (see Sec.~\ref{sec:I.A}). Henceforth, we will set $c_7=c_8=0$ by default and will not repeatedly state the settings of these two parameters.

\subsection{Comment on the Lagrangian}
There are significant advantages to finding the Lagrangian corresponding to the field equations. At the classical level, it facilitates the analysis of symmetries; at the quantum level, it permits the quantization process. However, we failed to find a Lagrangian for GCT. Especially, it is too difficult to construct a Lagrangian for the pure kinetic term $\nabla_\mu S_\nu + \nabla_\nu S_\mu$. An example of failed attempt is $\mathcal{L}\propto \nabla_\mu S^\mu$, which yields null results in the variation. Similar difficulties also arise in nonlocal RT gravity \cite{Maggiore2014.PRD.89.043008}.

Constructing the Lagrangian for a rank-2 tensor can be viewed as an inverse problem of Noether's theorem \cite{Deser2018.PLB.790.533}. \citet{Deser2023.AnnPhysNY.448.169163} argued that \textit{all identically conserved gravitational (geometric) tensors are metric variations of an action}. It is crucial to note that this statement is restricted to geometric tensors. For our case, the introduction of additional vector fields and fluids further complicates this problem, rendering it currently intractable.

The absence of a Lagrangian, or equivalently an action principle, is a limitation of the present formulation, but not by itself an immediate inconsistency. This restricts what can currently be claimed about the fundamental status of GCT. In the usual effective-field-theory sense, the low-energy dynamics is encoded in a local effective action organized as an expansion in energy and valid below a cutoff scale \cite{Donoghue1994.PRD.50.3874,Burgess2007.AnnuRevNuclPartSci.57.329}. Without such an action, GCT in its current form should be regarded as a phenomenological classical field-equation framework rather than an effective field theory. Consequently, quantum stability and radiative stability cannot be assessed in the standard way. A possible way forward is provided by non-Lagrangian quantization, where the equations of motion are supplemented by a compatible Lagrange anchor \cite{Kazinski2005.JHighEnergyPhys.07.076,Lyakhovich2007.JHighEnergyPhys.01.047}. Investigating whether Eq.~(\ref{eq:2.1}) admits such a structure, and whether the resulting quantization is stable, is left for future work.

\section{Cosmology}\label{sec:III}
This section is dedicated to the analysis of cosmological applications, with a focus on the background level. Beyond our initial motivation, a roadmap for exploring broader applications was outlined at the beginning of Sec.~\ref{sec:I}. For the busy reader, the \textit{time} evolution plots provide an intuitive summary of the key outcomes, such as the EDE and self-tuning dynamics in Fig.~\ref{fig:cosmos:05}, and the ESHU scenario in Fig.~\ref{fig:cosmos:08}. Mapping the evolution into phase space can reveal the global properties of the system. Technically, the stability of critical points plays a key role in shaping the phase portrait and provides essential guidance for constructing models and setting parameters. In our specific realization, EDE is primarily characterized by the stability of point $F_1$ in Fig.~\ref{fig:cosmos:02}, leading to the introduction of the $J_\mu S_\nu$-type terms in Sec.~\ref{sec:II}. The self-tuning mechanism is analyzed through $\mathcal{I}_4$ in Fig.~\ref{fig:cosmos:06}, which serves as a simple application of the centre manifold theory. The ESHU scenario is defined by $I_3$ in Fig.~\ref{fig:cosmos:02}, while its successful exit also relies on the stability of $F_1$. As a powerful application of the centre manifold theory, we show that the ESHU gives rise to rich dynamics in the evolution of the comoving Hubble radius.

The background analysis in this section suggests one possible cosmological history in GCT:
\begin{quote}
  radiation-like expanding stage ($\Omega_\subF=\mathcal{O}(1)$) 
  $\rightarrow$ \textit{ESHU} (Sec.~\ref{sec:III.C}) 
  $\rightarrow$ standard radiation era 
  $\rightarrow$ \textit{EDE near equality} (Sec.~\ref{sec:III.A}) 
  $\rightarrow$ matter era 
  $\rightarrow$ current dark energy era ($\Lambda$) 
  $\rightarrow$ \textit{linear expansion in the far future} (Sec.~\ref{sec:III.B}).
\end{quote}
The \textit{italicized stages} highlight possible new ingredients introduced by GCT relative to the current standard cosmological picture. Note that, in this suggested cosmological history, no huge bare vacuum energy is assumed, while $\Lambda$ represents the observed dark energy in the late-time Universe.

\subsection{Early dark energy}\label{sec:III.A}
In this subsection, we begin by deriving the complete cosmological evolution equations, followed by the presentation of an equivalent dynamical system. The EDE requires a stable critical point $F_1$ (see Fig.~\ref{fig:cosmos:02}) to ensure its dissipation in the standard radiation- and matter-dominated eras. The stability analysis allows us to place the first constraints on the viable parameter space. During the cosmic radiation-matter transition, the source term becomes significant and kicks the vector field $S_\mu$ away from $F_1$. We investigate the EDE dynamics during the transition with two approaches --- numerical calculations and analytical approximations. 

Here is a summary of the roles of model parameters in the EDE dynamics. The amplitude during the cosmic transition depends mainly on $c_1$. The dissipation rate in the radiation- and matter-dominated eras depends on $c_2$, $c_3$, $c_5$ (corresponding to the linear terms in $S_\mu$) and the source term. Parameters $c_4$ and $c_6$ (corresponding to the nonlinear terms in $S_\mu$) play a negligible role.

\subsubsection{Background evolution equations}\label{sec:III.A.1}
We adopt the flat FLRW metric $\dx s^2=-c^2\dx t^2+a^2\dx\mathbf{x}^2$, where $a=a(t)$, and assume the vector field of the form $S_\mu=[S_0(t),0,0,0]$ with the dimension $[S_0]={\rm time}^{-1}$. We use the subscript $\subF$ to denote the perfect fluid (comprising photons, neutrinos, baryons and dark matter for the EDE purpose), with its EoS given by $w_\subF(t)\equiv p_\subF/(\rho_\subF c^2)$. Note that $w_\subF$ is set by the thermodynamics of the fluid, rather than a free scalar. The fluid proper time is defined through $\dx s^2=-c^2\dx\tau^2$, and the four-velocity is defined as $u^\mu\equiv\dx x^\mu/\dx\tau$. In the cosmological background, we have $u^\mu=(1,0,0,0)$. Substituting the above configurations into Eq.~(\ref{eq:2.2b}) yields $T^\mu_{\ \nu}={\rm diag}\{-\rho_\substd c^2,p_\substd,p_\substd,p_\substd\}$, where $\rho_\substd=\rho_\subF+\Lambda/(\kappa c^2)$ and $p_\substd=p_\subF-\Lambda/\kappa$. Within the standard $\Lambda$CDM model, the total EoS $w_\substd\equiv p_\substd/(\rho_\substd c^2)$. Equation (\ref{eq:2.1}) gives
\begin{subequations}\label{eq:3.1}
\begin{align}
  & H^2-\frac{1}{3}\big(c_1\dot{w}_\subF^2 - \tilde{c}_2\dot{S}_0 + 3c_2HS_0 - \tilde{c}_{35}\varrho_\subF S_0 + \tilde{c}_{46}S_0^2\big) \nonumber\\
  & \quad = \frac{8\pi G}{3}\rho_\substd, \label{eq:3.1a}\\
  & \dot{H}+\frac{3}{2}H^2-\frac{1}{2}\big[c_1\dot{w}_\subF^2 + c_2\dot{S}_0 + (3c_2-2)HS_0 \nonumber\\
  & \quad - c_3\varrho_\subF S_0 + c_4S_0^2\big] = -\frac{4\pi G}{c^2}p_\substd, \label{eq:3.1b}\\
  & \tilde{c}_2\ddot{S}_0 + 3\tilde{c}_2H\dot{S}_0 - 3(c_2h+2)H^2S_0 - 6c_5\varrho_\subF HS_0 \nonumber\\
  & \quad - 2\tilde{c}_{46}S_0\dot{S}_0 + 3c_6HS_0^2 + \tilde{c}_{35}\varrho_\subF\big(\dot{S}_0+\frac{\dot{\rho}_\subF S_0}{2\rho_\subF}\big) \nonumber\\
  & \quad = 2c_1\dot{w}_\subF\ddot{w}_\subF, \label{eq:3.1c}
\end{align}
\end{subequations}
where $\dot{}\equiv\dx/\dx t$, $H=\dot{a}/a$, $\dot{H} = \ddot{a}/a-H^2$, $h=\dot{H}/H^2$, $\tilde{c}_2=2-c_2$, $\tilde{c}_{35}=c_3-2c_5$, $\tilde{c}_{46}=c_4-c_6$, $\varrho_\subF=\sqrt{8\pi G\rho_\subF}$. Energy conservation equation reads $\dot{\rho}_i+3H(\rho_i+p_i/c^2)=0$, where $i=\subF,\,\substd$. The above equations are not independent of each other, and a prudent choice can simplify the subsequent derivations.

To construct the dynamical system, inspired by the Friedmann equation (\ref{eq:3.1a}), we introduce four dimensionless variables
\begin{equation}\label{eq:3.2}
  x_1 = \frac{S_0}{H},\
  x_2 = \frac{\dot{S}_0}{H^2},\
  x_3 = \frac{\varrho_\subF}{H}, \
  x_4 = \frac{\Lambda c^2}{3H^2},
\end{equation}
and the \textit{e}-folding number $N\equiv\log(a/a_0)$, where $a_0$ is the present-day scale factor \cite{Copeland1998.PRD.57.4686}. The differential relation $\dx N=H\dx t$ allows us to transform the time derivative $\dx/\dx t$ into $\dx/\dx N$. For instance, we have $\dot{w}_\subF = H w_\subF^\prime$ and $\ddot{w}_\subF/H^2 = w_\subF^{\prime\prime} + hw_\subF^\prime$, where ${}^{\prime}\equiv\dx/\dx N$. At this stage, we can write down the (effective) relative energy density for each component: $\Omega_\subF\equiv8\pi G\rho_\subF/(3H^2)=x_3^2/3$, $\Omega_\Lambda=x_4$, and
\begin{equation}\label{eq:3.3}
  \Omega_\subEDE = \frac{c_1(w_\subF^\prime)^2 - \tilde{c}_2x_2 + 3c_2x_1 - \tilde{c}_{35}x_1x_3 + \tilde{c}_{46}x_1^2}{3},
\end{equation}
in which $\Omega_\subEDE$ is defined via Eq.~(\ref{eq:3.1a}). One constraint equation is $\Omega_\subF+\Omega_\Lambda+\Omega_\subEDE=1$, which admits a unique solution $x_2=[c_1(w_\subF^\prime)^2 + 3c_2x_1 - \tilde{c}_{35}x_1x_3 + \tilde{c}_{46}x_1^2 + x_3^2 + 3x_4 - 3]/\tilde{c}_2$. Since $H$ appears in the denominator of $x_i$, the expression of $\dot{H}/H^2$ is essential for simplifying $x_i^\prime$. Substituting Eq.~(\ref{eq:3.1a}) and the definition of $w_\substd$ into Eq.~(\ref{eq:3.1b}), we obtain
\begin{align}\label{eq:3.4}
  h &= - \frac{3(1+w_\substd)}{2} + \frac{1}{2}\Big[c_1(w_\subF^\prime)^2 + (3c_2-2)x_1 \nonumber\\
    &\quad + c_2x_2 - c_3x_1x_3 + c_4x_1^2 + 3w_\substd\Omega_\subEDE\Big].
\end{align}
With these preparations, we can directly derive the dynamical equations, yielding
\begin{subequations}\label{eq:3.5}
\begin{align}
  x_1^\prime &= x_2-hx_1, \\
  x_2^\prime &= - (3+2h)x_2 + \frac{1}{2\tilde{c}_2}\big\{6(2+c_2h)x_1 - 6c_6x_1^2 \nonumber\\
    &\quad + 4\tilde{c}_{46}x_1x_2 + [12c_5+3\tilde{c}_{35}(1+w_\subF)]x_1x_3 \nonumber\\
    &\quad - 2\tilde{c}_{35}x_2x_3 + 4c_1w_\subF^\prime(w_\subF^{\prime\prime}+hw_\subF^\prime)\big\}, \\
  x_3^\prime &= -\frac{x_3}{2}\big[3(1+w_\subF)+2h\big], \\
  x_4^\prime &= -2hx_4.
\end{align}
\end{subequations}
The above equations are valid for the whole parameter space with the exception of $c_2=2$, which is not considered in this paper. The self-consistency of the formalism can be verified by noting that Eq.~(\ref{eq:3.5}) identically satisfies the constraint equation discussed below Eq.~(\ref{eq:3.3}). In specific cosmological applications, we can reduce the dimensionality of the system from 4 to 2 by applying the constraint equation and considering the limiting cases such as $\Lambda\rightarrow0$. Note that $w_\substd$ appears in the expression of $h$, while $w_\subF$ appears elsewhere. 

\subsubsection{Dynamics for constant EoS}\label{sec:III.A.2}
\begin{figure*}[!t]
  \centering
  \includegraphics[width=0.99\textwidth]{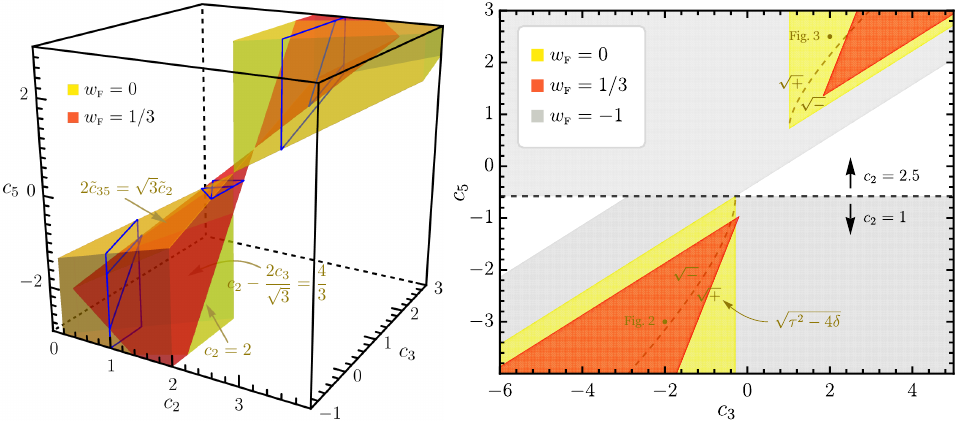}
  \caption{Left: Stability region of the critical point $F_1$ in the 3D parameter space $(c_2,c_3,c_5)$, as governed by Eq.~(\ref{eq:3.6}), for $w_\subF=0$ (yellow) and $1/3$ (red). The boundaries are planes given by Eq.~(\ref{eq:3.8}), and the case of $w_\subF=0$ is labeled in the figure. The blue curves, from left to right, represent the slices at $c_2=1$, $c_5=0$, and $c_2=2.5$, respectively. This stability is independent of $\{c_1,\,c_4,\,c_6\}$. Right: Slices at $c_2=1$ and $2.5$. The parameter points adopted in Figs.~\ref{fig:cosmos:02} and \ref{fig:cosmos:03}, and the sign of $\tau^2-4\delta$ (for $w_\subF=0$), which determines whether the eigenvalues are complex, are marked in the figure. Note that the grey region corresponds to the case of $w_\substd=w_\subF=-1$, although its physical implications are not discussed in this paper.}
  \label{fig:cosmos:01}
\end{figure*}

Without the trigger, the EDE field is required to dilute away during the radiation- and matter-dominated eras. For this analysis, we can substitute $w_\substd=w_\subF={\rm const.}$, $\Lambda=0$, and the constraint equation into Eq.~(\ref{eq:3.5}), yielding a two-dimensional autonomous dynamical system
\begin{widetext}
\begin{subequations}\label{eq:3.6}
\begin{align}
  x_1^\prime &= \frac{1}{\tilde{c}_2}\Big[-3 + 3(1+c_2)x_1 + (2-4c_2+\tilde{c}_{46})x_1^2 - \tilde{c}_{35}x_1x_3 + x_3^2 + \frac{x_1}{2}f(x_1,x_3)\Big], \\
  x_3^\prime &= \frac{x_3}{2\tilde{c}_2}\Big[3(c_2-\tilde{c}_2w_\subF) - 4(2c_2-1)x_1 + f(x_1,x_3)\Big],
\end{align}
\end{subequations}
where $f(x_1,x_3)=(c_2c_6-2c_4)x_1^2 + 2(c_3-c_2c_5)x_1x_3 + (\tilde{c}_2w_\subF-c_2)x_3^2$. Now $\Omega_\subEDE=1-x_3^2/3$. One can verify that $(x_1,x_3)=(0,\sqrt{3})$ is a critical point, which corresponds to $\Omega_\subEDE=0$. We denote this point as $F_1$, with $F$ standing for the finite region. For EDE, the first step is to find the parameter space for which $F_1$ is stable. The source term will be discussed later.

By applying the linear stability theory to Eq.~(\ref{eq:3.6}), we obtain the Jacobian matrix (see Appendix \ref{sec:App.A.1})
\begin{equation}\label{eq:3.7}
  J|_{F_1} =
  \frac{\sqrt{3}}{\tilde{c}_2}
  \begin{bmatrix}
    \displaystyle \frac{\sqrt{3}}{2}(2+c_2+\tilde{c}_2w_\subF)-\tilde{c}_{35}  &  2 \\
    \displaystyle 2-4c_2+\sqrt{3}(c_3-c_2c_5)  &  \sqrt{3}(\tilde{c}_2w_\subF - c_2)
  \end{bmatrix},
\end{equation}
which depends on $\{c_2,\,c_3,\,c_5,\,w_\subF\}$. The two eigenvalues of $J|_{F_1}$ are given by $\lambda_{F_1,\pm}=(\tau\pm\sqrt{\tau^2-4\delta})/2$, where $\delta=\det J|_{F_1}$ and $\tau=\mathrm{trace}\,J|_{F_1}$ \cite{Perko2001.book}. The critical point $F_1$ is stable if the real parts of both eigenvalues are negative~\footnote{Here we do not consider the case of zero eigenvalues.}. Figure \ref{fig:cosmos:01} plots the stable parameter space for $w_\subF=0$ in yellow and for $w_\subF=1/3$ in red. The boundary consists of three planes defined by
\begin{equation}\label{eq:3.8}
  \mathrm{Boundary}|_{F_1\mathrm{\ is\ stable}} = \left\{
  \begin{array}{l}
    c_2=2, \\
    2\tilde{c}_{35} = \sqrt{3}\tilde{c}_2(1+3w_\subF), \\
    \sqrt{3}(c_2+2w_\subF+\tilde{c}_2w_\subF^2) - 2c_3 - 2\tilde{c}_{35}w_\subF = 4/\sqrt{3},
  \end{array}
  \right.
\end{equation}
which coincide with divergent eigenvalues, $\tau=0$, and $\delta=0$, respectively. The expressions of boundaries for $w_\subF=0$ are labeled in Fig.~\ref{fig:cosmos:01} Left. The right panel displays the slices taken at $c_2=1$ and $2.5$. The overlapping region satisfies the requirements for EDE.
\end{widetext}

The EDE dissipation rate is determined by the real part of $\lambda_{F_1,\pm}$. For quantitative analysis, here we derive the general solutions near $F_1$ using the formalism presented in Ref.~\cite{Perko2001.book}. Equation (\ref{eq:3.6}) can be linearized to $\tilde{\mathbf{x}}'=J\tilde{\mathbf{x}}$, where $\tilde{\mathbf{x}}=[x_1,x_3-\sqrt{3}]^{\rm T}$ and the superscript $\mathrm{T}$ denotes the transpose. Corresponding to $\lambda_{F_1,\pm}$, the eigenvectors can be written as $V_{F_1,\pm}=[2\sqrt{3},\,\tilde{c}_2\lambda_{F_1,\pm} + \sqrt{3}\,\tilde{c}_{35} - 3(2+c_2+\tilde{c}_2w_\subF)/2]^{\rm T}$. If $\tau^2-4\delta>0$, both the eigenvalues and eigenvectors are real. We require that $\lambda_-<\lambda_+<0$. We can construct the matrix $P|_{F_1}=[V_{F_1,+}\;V_{F_1,-}]$ such that $P^{-1}JP|_{F_1}=\mathrm{diag}\{\lambda_{F_1,+},\lambda_{F_1,-}\}$. Then the general solution reads
\begin{equation}\label{eq:3.9}
  \tilde{\mathbf{x}}
  =
  P\,
  \mathrm{diag}\{\exp(\tilde{N}\lambda_{F_1,+}),\exp(\tilde{N}\lambda_{F_1,-})\}\,
  P^{-1}\,
  \tilde{\mathbf{x}}_\subini,
\end{equation}
where the subscript $\subini$ denotes the initial time, and $\tilde{N}=N-N_\subini$.  As $\Omega_\subEDE=1-x_3^2/3\propto -\tilde{x}_3$ around $F_1$, and $e^{\lambda N}\propto a^\lambda$, we obtain the dissipation rate $\Omega_\subEDE\propto a^{\lambda_{F_1,+}}$ during a standard-fluid-dominated epoch. If $\tau^2-4\delta<0$, imaginary parts appear, and the matrix $P$ should be defined as $P=[\mathrm{Im}(V_{F_1,+})\;\mathrm{Re}(V_{F_1,+})]$ such that 
\begin{equation}\nonumber
  P^{-1}JP|_{F_1}
  =
  \frac{1}{2}
  \begin{bmatrix}
    \tau & -\sqrt{4\delta-\tau^2} \\
    \sqrt{4\delta-\tau^2} & \tau 
  \end{bmatrix}.
\end{equation}
Then the general solution is given by
\begin{equation}\label{eq:3.10}
  \tilde{\mathbf{x}}
  =
  e^{\tau\tilde{N}/2}\cdot
  P\,
  R(\sqrt{\delta-\tau^2/4}\,\tilde{N})\,
  P^{-1}\,
  \tilde{\mathbf{x}}_\subini,
\end{equation}
where the rotation matrix $R(\theta)=\big[\begin{smallmatrix}
                                       \cos\theta & -\sin\theta \\
                                       \sin\theta & \cos\theta
                                     \end{smallmatrix}\big]$. 
This implies that $\Omega_\subEDE$ exhibits an oscillatory decay, in which the period is equal to $4\pi/\sqrt{4\delta-\tau^2}$ and the amplitude is proportional to $a^{\tau/2}$. The oscillation may trigger parametric resonance in cosmological perturbations \cite{Smith2020.PRD.101.063523}, imprinting characteristic observational signals that merit  future research.  Together with the source term (see the next subsubsection), we can determine the EDE dissipation rate for a given parameter setting. Observationally, one might require that the EDE decays more rapidly than radiation after matter-radiation equality \cite{Poulin2019.PRL.122.221301}.

\begin{figure*}[!t]
  \centering
  \includegraphics[width=0.96\textwidth]{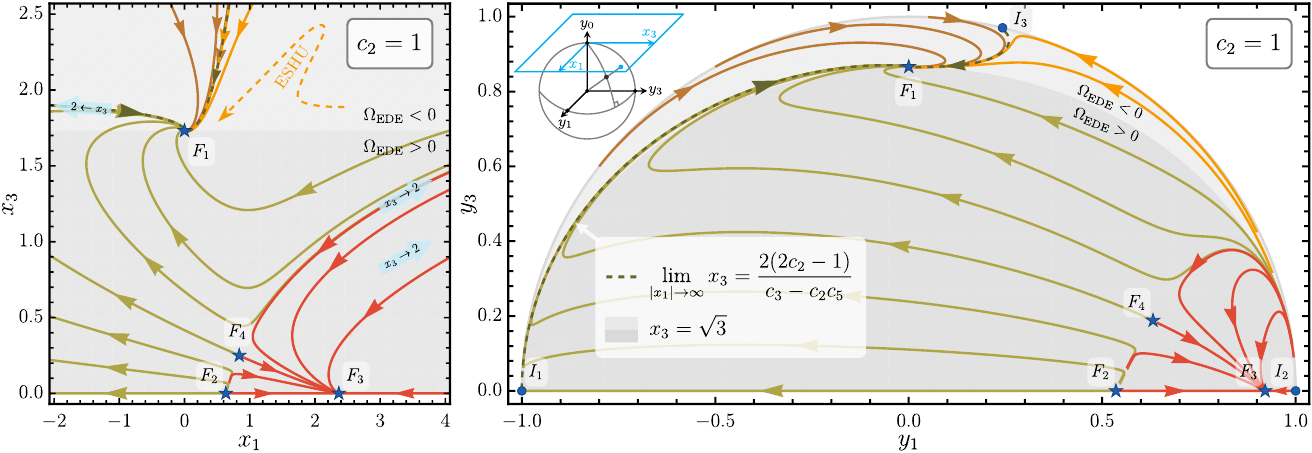}
  \caption{Phase portrait of Eq.~(\ref{eq:3.6}) and its projection onto the $(y_1,y_3)$ axes. The inset illustrates the mapping. The parameters are fixed to $c_2=1$, $c_3=-2$, $c_5=-3$, and $c_4=c_6=w_\subF=0$, with only $c_2$ labeled in the figure for brevity. Trajectories in the two panels are in one-to-one correspondence, where colours and relative positions enable cross-identification. Blue stars mark finite-region critical points ($F_i$), dots mark critical points at infinity ($I_i$), and their coordinates are collected in Table~\ref{tab:02}. The dark-green dashed curves denote attractor trajectories. Equation~(\ref{eq:3.12}) describes the asymptotic behaviour of the $I_1\rightarrow F_1$ attractor and provides an approximate boundary that partitions the phase portrait. The orange dashed curve in the left panel is illustrative of the characteristic behaviour of trajectories in that region and can be used to realise an ESHU, whose key dynamics are captured by the $I_3\rightarrow F_1$ attractor in the right panel (see Sec.~\ref{sec:III.C}).}
  \label{fig:cosmos:02}
\end{figure*}

The above analysis focuses on the local stability of $F_1$. To reveal the global property of the system, we now turn to the phase portrait. We only consider the expanding Universe, which corresponds to the upper half-plane with $x_3\geqslant0$~\footnote{This follows directly from the definition of $x_3$.}. Furthermore, we can project the upper half-plane onto the Poincar\'e sphere \cite{Bahamonde2018.PhysRep.775.1,Perko2001.book}, which enables the analysis of possible critical points at infinity~\footnote{Our initial motivation for introducing the Poincar\'e sphere was to clarify the behavior of trajectories in the upper-right corner of Fig.~\ref{fig:cosmos:02} Left, which ultimately leads the analysis to the critical points at infinity.}. The geometric interpretation and mathematical details of the projection are provided in Appendix~\ref{sec:App.A.3}. In summary, for Eq.~(\ref{eq:3.6}), we have $y_i=x_i/\sqrt{1+x_1^2+x_3^2}$, where $i=1,3$. The upper half-plane corresponds to a half-disk, defined by $y_3\geqslant0$ and $y_1^2+y_3^2\leqslant1$, and infinity is located at $y_1^2+y_3^2=1$. Figure \ref{fig:cosmos:02} depicts a representative phase portrait of Eq.~(\ref{eq:3.6}) in both the $x_i$ and $y_i$ coordinates, with parameters chosen so that $F_1$ is an asymptotically stable spiral.

The global structure of the phase portrait is mainly characterized by the critical points and their stability, the attractor trajectories, and the separatrices that partition the phase space into distinct basins of attraction. We therefore begin by calculating the coordinates of all critical points in a phase portrait. For simplicity, we restrict attention to the case $c_4=c_6=0$, which is sufficient for the EDE analysis. The critical points in the finite region can be obtained by solving $\{x_1^\prime=0,\,x_3^\prime=0\}$. Besides $F_1$, the system admits three further equilibria in the finite region (see Table~\ref{tab:02} for a summary of the coordinates). Two equilibria lie on the $x_1$-axis and describe a caloric-field-dominated state, while the third is off-axis and represents a scaling solution \cite{Copeland1998.PRD.57.4686}. The definition and calculation method for the critical points at infinity can be found in Appendix~\ref{sec:App.A.3}. For Eq.~(\ref{eq:3.6}), there are three critical points at infinity, with two on the $y_1$-axis and one off-axis (denoted $I_3$). The stability of these critical points in Fig.~\ref{fig:cosmos:02} is listed in Table~\ref{tab:02}. A comprehensive analysis of these critical points, covering existence, stability, and cosmological applications, lies beyond the scope of this paper. What we have completed is only the analysis of $F_1$ (EDE in Sec.~\ref{sec:III.A}) and $I_3$ (ESHU in Sec.~\ref{sec:III.C}).

\begin{table*}[!t]
  \caption{Critical points of the dynamical system~(\ref{eq:3.6}) with $c_4=c_6=0$ and $x_3\geqslant0$. For $F_4$, the $x_3$ appearing in the first coordinate is to be evaluated using the value of the second coordinate listed in the same row. We analyse neither the existence nor the stability of the critical points in general and report only the results shown in Figs.~\ref{fig:cosmos:02} and \ref{fig:cosmos:03}. Note that the Stable/Unstable/Saddle designations are based solely on the qualitative behaviour seen in the figures and are not the result of an eigenvalue analysis. The stability of $I_1$ and $I_2$ in Fig.~\ref{fig:cosmos:03} remains unclear and is therefore not tabulated here. The last column lists critical-point-based cosmological applications and highlights the constructive methodology adopted in Sec.~\ref{sec:III}.}
  \label{tab:02}
  \begin{tabular*}{\hsize}{@{\ \ }@{\extracolsep{\fill}}lllll@{\ \ }}
    \hline\hline
    Labels         &  Coordinates \hfill Stability in  &  Fig. \ref{fig:cosmos:02}  &  Fig. \ref{fig:cosmos:03}  &  Applications  \\
    \hline
    Finity         &  $(x_1,x_3)$  \\
    $F_1$          &  $(0,\sqrt{3})$  &  Stable  &  Stable  &  Early dark energy  \\[0.3mm]
    $F_2$          &  $\bigg(\displaystyle\frac{3c_2+3-\sqrt{9c_2^2-30c_2+33}}{4(2c_2-1)}, 0\bigg)$  &  Unstable  &  Stable  &  \\[3mm]
    $F_3$          &  $\bigg(\displaystyle\frac{3c_2+3+\sqrt{9c_2^2-30c_2+33}}{4(2c_2-1)}, 0\bigg)$  &  Stable  &  Unstable  &  \\[3mm]
    $F_4$          &  $\bigg(\displaystyle\frac{(c_2-\tilde{c}_2w_\subF)(x_3^2-3)}{2c_3x_3+4-2c_2(4+c_5x_3)},
      \displaystyle\frac{3\tilde{c}_2w_\subF^2+3c_2+6w_\subF-4}{2(c_3+\tilde{c}_{35}w_\subF)}\bigg)$  &  Saddle  &  Saddle  &  Scaling solution  \\[3mm]
    \hline
    Infinity       &  $(y_1,y_3)$  \\
    $I_1$          &  $(-1,0)$  &  Saddle  &  \\
    $I_2$          &  $(1,0)$  &  Unstable  &  \\[1.5mm]
    $I_3$          &  $\bigg(\displaystyle\frac{\sgn\,\tilde{c}_{35}}{\sqrt{1+\tilde{c}_{35}^2}},\displaystyle\frac{|\tilde{c}_{35}|}{\sqrt{1+\tilde{c}_{35}^2}}\bigg)$  &  Saddle  &  Saddle  &  Early static hot Universe  \\[3mm]
    \hline\hline
  \end{tabular*}
\end{table*}

Figure~\ref{fig:cosmos:02} Left reveals a constant-$x_3$ separatrix ($x_3=2$) in the limit $x_1\rightarrow\pm\infty$~\footnote{Roughly speaking, $x_3=2$ is an asymptote of the red and green trajectories in Fig.~\ref{fig:cosmos:02}. Extending the range of the phase portrait makes the trend clear~\cite{NotShown}.}. Here we outline the derivation of this limiting value. Guided by numerical evidence, we posit that the limit exists. Then, considering $|x_1|\gg x_3$ and $x_3$ is finite in Eq.~(\ref{eq:3.6}), we obtain
\begin{equation}\label{eq:3.11}
  \lim_{|x_1|\rightarrow\infty} x_i^\prime = \frac{x_1x_i}{\tilde{c}_2}\left[2 + c_3x_3 - c_2(4+c_5x_3)\right],
\end{equation}
for $i=1,3$ and $c_4=c_6=0$. The right-hand side is proportional to $x_1$ or $x_1^2$. Setting the coefficient to zero yields a special value
\begin{equation}\label{eq:3.12}
  \lim_{|x_1|\rightarrow\infty} x_3 = \frac{2(2c_2-1)}{c_3-c_2c_5}.
\end{equation}
As $x_3$ varies and the coefficient in Eq.~(\ref{eq:3.11}) changes sign, $x_i^\prime$ switches between $\pm\infty$, thereby generating the separatrix in the phase portrait. The foregoing derivation is heuristic rather than fully rigorous, as it assumes the existence of the limit. Nevertheless, the result is reliable, and one can verify that this is consistent with the asymptotic value shown in Fig.~\ref{fig:cosmos:02}. In the $y_i$ coordinate, Eq.~(\ref{eq:3.12}) approximately describes the behavior of the $I_1\rightarrow F_1$ attractor trajectory near $I_1$. On the right side ($x_1\rightarrow+\infty$), Eq.~(\ref{eq:3.12}) yields an approximate separatrix between the green and orange trajectories, informing the setting of initial conditions for ESHU (see Table~\ref{tab:05}). The $I_3\rightarrow F_1$ attractor trajectory will be discussed in detail in Sec.~\ref{sec:III.C}.

\begin{figure*}[!t]
  \centering
  \includegraphics[width=0.96\textwidth]{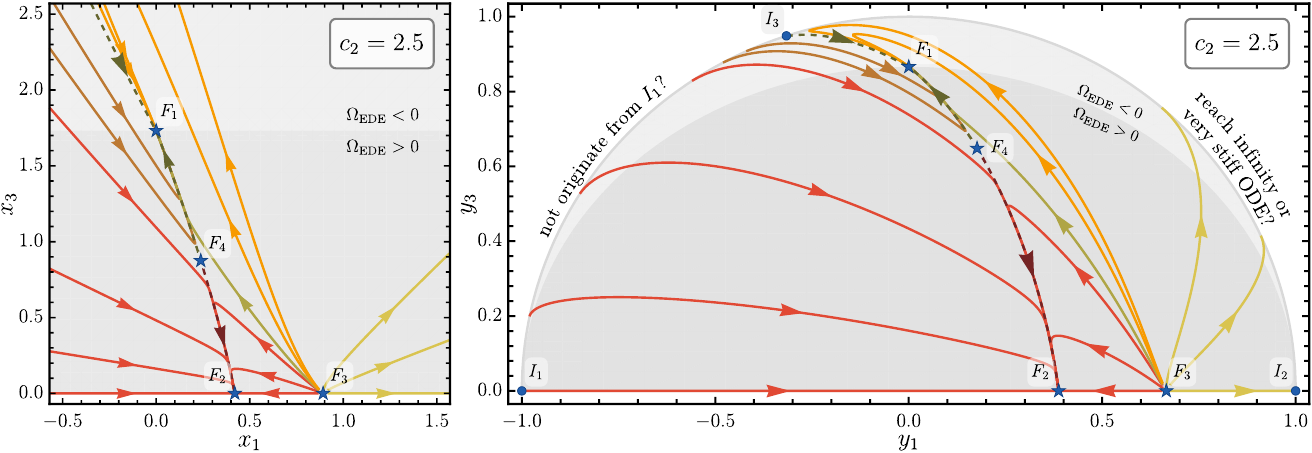}
  \caption{Same as Fig.~\ref{fig:cosmos:02} but with $c_2=2.5$, $c_3=2$, $c_5=2.5$, and $c_4=c_6=w_\subF=0$. Here the behaviour near infinity is more intricate. Whether this reflects numerical instability or additional physics remains unresolved.}
  \label{fig:cosmos:03}
\end{figure*}

To demonstrate generality, Fig.~\ref{fig:cosmos:03} presents the phase portrait for an alternative parameter choice in which the Jacobian at $F_1$ has two real eigenvalues. Overall, Figs.~\ref{fig:cosmos:02} and \ref{fig:cosmos:03} exhibit the same qualitative phase-portrait structure, with some minor differences. The first distinction lies in the ability of $F_1$ to attract nearby trajectories. In Fig.~\ref{fig:cosmos:02}, following a finite perturbation near $F_1$, $|\Omega_\subEDE|$ exhibits an almost monotonic decay. By contrast, after a finite kick in Fig.~\ref{fig:cosmos:03}, $|\Omega_\subEDE|$ may first increase further and then fall back to zero (depending on the kick direction; see the $x_i$ plots). This may enable one to engineer a small perturbation that drives $\Omega_\subEDE$ to the $10\%$ level. However, the analysis for this issue seems to go beyond the local stability of $F_1$, and we therefore defer a comprehensive treatment to future work. The second distinction is that Eq.~(\ref{eq:3.12}) yields a negative asymptotic value for Fig.~\ref{fig:cosmos:03}, which falls outside the plotted domain. Third, the trajectories in Fig.~\ref{fig:cosmos:03} exhibit pronounced singular behavior near infinity. Currently, our calculations cannot determine whether the yellow trajectory in Fig.~\ref{fig:cosmos:03} remains on the arc representing infinity~\footnote{The continuous infinity arc can serve as an attractor or a source in the phase portrait (see Example~1 on p.~20 in Ref.~\cite{Perko2001.book} and its Poincar\'e map for an illustration).} or instead approaches $I_3$ very closely before evolving toward $F_1$ (as in the orange lines). A similar issue exists for the red trajectory on the left arc as well. In the future, one may be able to answer this question in polar coordinates.

\begin{figure}[!t]
  \centering
  \includegraphics[width=0.48\textwidth]{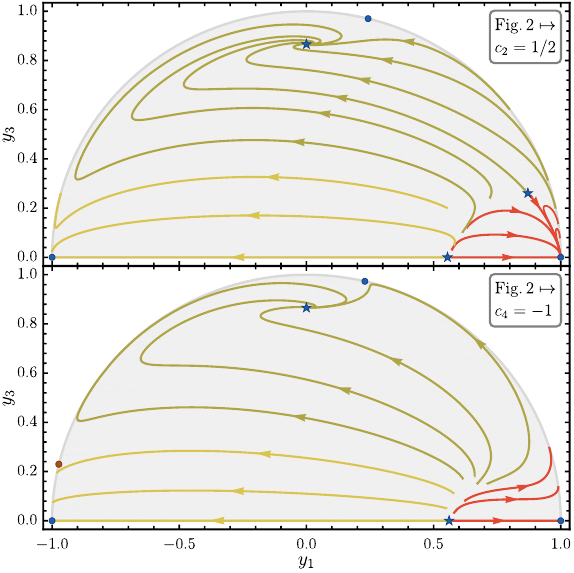}
  \caption{Phase portraits of Eq.~(\ref{eq:3.6}) in the $(y_1,y_3)$ axes. Each panel differs from Fig.~\ref{fig:cosmos:02} by only one parameter and all others remain unchanged. The top panel shows the special case $c_2=1/2$, while the bottom illustrates the effect of a nonzero $c_4$.}
  \label{fig:cosmos:04}
\end{figure}

Figure \ref{fig:cosmos:04} shows the phase portraits for two additional parameter sets. The top panel adopts $c_2=1/2$, for which $F_3$ is absent, $F_2$ remains well-defined in the limiting view, and the asymptotic value given by Eq.~(\ref{eq:3.12}) vanishes. The bottom one adopts a nonzero $c_4$, eliminating two equilibria from the finite region and introducing a new one at infinity (see Sec.~\ref{sec:III.C.3} for the coordinates). If one does not attempt a full stability analysis, reducing the number of equilibria may offer a computationally inexpensive route to making $F_1$ the unique global attractor. This consideration motivates the inclusion of the figure, although our attempt is unsuccessful as shown in the phase portraits.

\subsubsection{Radiation-matter transition}\label{sec:III.A.3}
The previous subsubsection analyzed the system with constant $w_\subF$ and delineated the region of parameter space compatible with an EDE phase (see the overlap region in Fig.~\ref{fig:cosmos:01}). To quantitatively assess the dependence of EDE on the $c_1$ source term, we now numerically integrate the equations with a time-dependent $w_\subF$ that captures the radiation-matter transition. During the transition, the Universe is dominated by photons, relativistic neutrinos, baryons, and dark matter. Modeling these components as perfect fluids, the total EoS is taken as \cite{Tian2021.PRD.103.043518}
\begin{equation}\label{eq:3.13}
  w_\subF(N) = \frac{1/3}{1+e^{N-N_\subeq}},
\end{equation}
where $N_\subeq=-8.13$ \cite{Aghanim2020.AaA.641.A6} corresponds to matter-radiation equality. Note that EDE may slightly shift the best-fit value but does not alter the qualitative conclusions. After the equality, the source term $c_1(w_\subF^\prime)^2/3$ entering $\Omega_\subEDE$ scales as $e^{-2N}\propto a^{-2}$ and therefore decays faster than the radiation fraction. This sets a bound on how fast $\Omega_\subEDE$ can dilute after equality. The physical decay rate depends on the source term and the eigenvalues $\lambda_{F_1,\pm}$. For a realistic early Universe, the cosmological constant $\Lambda$ is negligible and we can set $x_4=0$ and $w_\substd=w_\subF$. Here, the calculation adopts the $\{x_1,x_2,x_3\}$ system in Eq.~(\ref{eq:3.5}), and consistency is checked against the constraint equation.

\begin{figure*}[!t]
  \centering
  \includegraphics[width=0.96\textwidth]{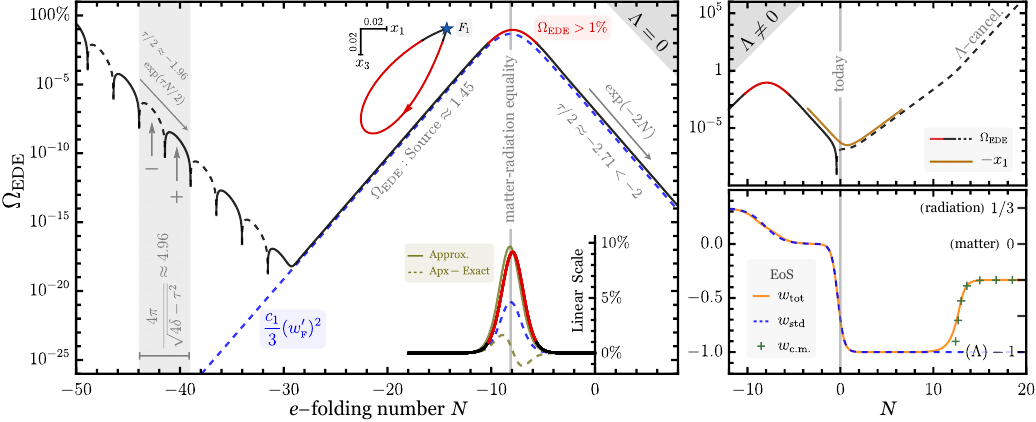}
  \caption{Evolution of $\Omega_\subEDE$ as computed by numerically integrating Eq.~(\ref{eq:3.5}) with $w_\subF$ given by Eq.~(\ref{eq:3.13}), modelling the cosmic radiation-matter transition. In both panels we set $c_1=20$, take the remaining $c_i$ as in Fig.~\ref{fig:cosmos:02}, and choose initial conditions $\{x_1=0,x_3=1.5\}$ at $N_\subini=-50$, with the initial $x_2$ fixed by the constraint equation. The left panel assumes $\Lambda=0$, whereas the right panel takes $\Lambda\neq0$, for which $w_\substd$ and the initial $x_4$ are chosen to match current dark-energy observations (see the main text). Two key epochs, \textit{matter-radiation equality} and \textit{today}, are labeled in the figure. The black-red curve shows $\Omega_\subEDE$, with black dashed segments corresponding to the absolute value for $\Omega_\subEDE<0$ and red segments indicating $\Omega_\subEDE>1\%$. The blue dashed curve shows the source term. In the left panel, the upper inset shows the corresponding trajectory in the $(x_1,x_3)$ phase plane near the equality. On linear axes, the lower inset also plots the approximate solution (\ref{eq:3.15}) (solid olive) and the residual between the approximate and exact solutions (dashed olive). To quantify the $\Lambda$-cancelling solution, the bottom-right subpanel displays the EoS $w_\subtot$, $w_\substd$, and $w_\subcm$. Here, the centre-manifold EoS $w_\subcm$ is obtained by substituting the approximate solution (\ref{eq:3.24}) into Eq.~(\ref{eq:3.21}), with $z_0(N)$ reconstructed from the numerical solution.}
  \label{fig:cosmos:05}
\end{figure*}

Figure~\ref{fig:cosmos:05} Left presents a typical evolution of $\Omega_\subEDE$ across the radiation-matter transition together with, for reference, the source contribution $c_1(w_\subF^\prime)^2/3$. Model parameters and initial conditions are specified in the caption. Deep in the radiation era ($N\lesssim-30$ for this setup), the source term is negligible and the dynamics are governed by the local stability of $F_1$. With $w_\subF\approx1/3$, a direct calculation gives $\tau^2-4\delta<0$ for $J|_{F_1}$. Consequently, $\Omega_\subEDE$ exhibits oscillatory decay, with the amplitude scaling as $\exp(\tau N/2)$, where $\tau/2=-1.96$, and the period given by $4\pi/\sqrt{4\delta-\tau^2}=4.96$. The numerical solution agrees with the previous eigenvalue analysis. Near $N\approx-30$, the source term becomes dominant in the vector-field dynamics, elevating its energy density to a comparable level and pushing the phase-space trajectory away from $F_1$. During the rising phase, the ratio of the total $\Omega_\subEDE$ to the source term is $1.45$, and the trajectory in the phase portrait is also indicated in Fig.~\ref{fig:cosmos:05}. The source term peaks exactly at matter-radiation equality, whereas the total $\Omega_\subEDE$ attains its maximum very near that epoch. The minor offset can be interpreted as a finite response time due to the dynamical inertia of the field. This result is sufficient to resolve the EDE coincidence problem. After the peak, with $w_\subF\approx0$, the source-free dynamics continues to exhibit damped oscillations with $\tau/2=-2.71<-2$. This numerical inequality implies that the post-peak decay of $\Omega_\subEDE$ is controlled by the source term and scales as $e^{-2N}$ without oscillation.

To enable an intuitive evaluation of $\Omega_\subEDE$ near matter-radiation equality, here we present an approximate solution. With $c_1\neq0$ and $x_4=0$, we start from the $\{x_1,x_3\}$ system given by Eq.~(\ref{eq:3.5}) together with the $x_2$-solution given below Eq.~(\ref{eq:3.3}). Our analysis proceeds under two assumptions: $c_1(w_\subF^\prime)^2/3\ll1$ and $|w_\subF^{\prime\prime}|\ll|w_\subF^\prime|$. The first one confines the EDE trajectories to a neighborhood of $F_1$, thereby justifying a linearization of the system. For notational clarity, we introduce an auxiliary infinitesimal $\varepsilon$ and set $\varepsilon=1$ after the series expansion. Mathematically, we set $\mathbf{x}\equiv[x_1,x_3]^\mathrm{T}=[0,\sqrt{3}]^\mathrm{T}+\varepsilon\tilde{\mathbf{x}}$ and replace $c_1$ by $\varepsilon c_1$ in the equations. Linearizing in $\varepsilon$ then yields
\begin{equation}\label{eq:3.14}
  \tilde{\mathbf{x}}^\prime = J|_{F_1} \tilde{\mathbf{x}} + \mathbf{s},
\end{equation}
where the source vector $\mathbf{s}=c_1(w_\subF^\prime)^2/\tilde{c}_2\cdot[1,-\sqrt{3}]^\mathrm{T}$. The general solution of the above equation is too complex to yield an intuitive evaluation. For further simplicity, we adopt the second approximation ($|w_\subF^{\prime\prime}|\ll|w_\subF^\prime|$), although it holds only near matter-radiation equality \cite{NotShown}. Considering the form of $\mathbf{s}$, a series solution of Eq.~(\ref{eq:3.14}) can be written as $\tilde{\mathbf{x}}=\mathbf{v}_1(w_\subF^\prime)^2+\mathbf{v}_2w_\subF^\prime w_\subF^{\prime\prime}+\cdots$, where $\mathbf{v}_i$ are time-independent vectors. The leading term is a solution of $J|_{F_1} \tilde{\mathbf{x}} + \mathbf{s}=0$, which further gives
\begin{equation}\label{eq:3.15}
  \Omega_\subEDE \approx \frac{2c_1(2\sqrt{3}c_5+3w_\subF+5)(w_\subF^\prime)^2} {9\tilde{c}_2(1-w_\subF^2)+6\sqrt{3}(c_3+\tilde{c}_{35}w_\subF)-6(1+3w_\subF)}.
\end{equation}
For consistency with the first assumption, only the $\mathcal{O}(\varepsilon)$ terms are retained here. On linear axes, the lower inset in Fig.~\ref{fig:cosmos:05} Left displays the exact $\Omega_\subEDE$, the source term, the approximate solution derived above, and the residual (approximate minus exact). For our parameter setting, $\Omega_\subEDE$ can reach $10\%$ while the absolute residual remains below $2\%$, establishing Eq.~(\ref{eq:3.15}) as a suitable approximation for our purpose. 

\subsection{$S_\mu$ confronts $\Lambda$}\label{sec:III.B}
Observations of late-time cosmic acceleration point to the existence of dark energy in the Universe \cite{Riess1998.AstronJ.116.1009,Perlmutter1999.ApJ.517.565}. For our theory, a natural question is how the caloric field behaves in the dark-energy-dominated epoch. This subsection is devoted to analysing this issue. For simplicity, we work with the simplest model of dark energy, namely a cosmological constant $\Lambda$ \cite{Einstein1917.NotEngJ.1917.142}. We first present an illustrative calculation and then proceed to a detailed analysis of its properties.

A brief remark is in order here. In Fig.~\ref{fig:cosmos:01} Right, the grey region shows the parameter space in which $F_1$ is stable for $w_\subF=-1$. This corresponds to $\Lambda=0$ and dark energy being described by a special \textit{perfect fluid} with $\mathrm{EoS}=-1$. In this subsection, and in the default convention adopted in GCT, dark energy is instead described by a geometric $\Lambda$, while the \textit{fluid} sector includes only radiation and pressureless matter. The essential difference is whether dark energy is included in $w_\subF$ and $\rho_\subF$.

Here we summarize the parameter dependence relevant for this subsection. In the $\Lambda$-dominated regime, the $\{c_1,c_3,c_5\}$-dependent terms are negligible because they are tied to the fluid sector. Accordingly, the dynamical system (\ref{eq:3.5}) can be reduced to a two-dimensional system for $(x_1,x_4)$, with $\{c_2,c_4,c_6\}$ as the remaining parameters. The subsequent complete stability analysis is performed on a simpler but still physically relevant parameter slice $c_4=c_6=0$, with $c_2$ left arbitrary.

\subsubsection{Growth evolution of $S_\mu$}\label{sec:III.B.1}
To investigate the impact of a purely geometric $\Lambda$ on the $S_\mu$ dynamics, the full Eq.~(\ref{eq:3.5}) is employed, while the constraint equation is used to check the numerical results. In this scenario, $w_\subF$ is still given by Eq.~(\ref{eq:3.13}). A nonzero $\Lambda$ leads to $w_\substd\neq w_\subF$. Using energy conservation for each component, one obtains
\begin{equation}
  w_\substd = \frac{1/3-(\Omega_{\Lambda,0}/\Omega_{{\rm m},0})e^{4N-N_\subeq}}{1+e^{N-N_\subeq}+(\Omega_{\Lambda,0}/\Omega_{{\rm m},0})e^{4N-N_\subeq}},
\end{equation}
where $\Omega_{\Lambda,0}$ and $\Omega_{{\rm m},0}$ denote the present values of the relative energy densities of dark energy and matter, respectively. Note that $\Omega_{{\rm m},0}/\Omega_{{\rm r},0}=e^{-N_\subeq}$, where the subscript $\mathrm{r}$ denotes radiation. The initial conditions are chosen such that $\Omega_\Lambda\approx0.7$ and $\Omega_{\rm m}\approx0.3$ at $N=0$ (today) \cite{Aghanim2020.AaA.641.A6}. To this end, we consider the ratio $\Omega_\Lambda/\Omega_{\rm r} = (\Omega_{\Lambda,0}/\Omega_{{\rm m},0}) e^{4N-N_\subeq}$ in the expanding Universe. Evaluating this ratio at $N=N_\subini$ and imposing $\Omega_{\rm r}\approx\Omega_\subF=x_3^2/3$ in the deep radiation era yields
\begin{equation}\label{eq:3.17}
  x_{4,\subini} = \Omega_{\Lambda,\subini}
  \approx \frac{\Omega_{\Lambda,0}x_{3,\subini}^2}{3\Omega_{{\rm m},0}} \cdot e^{4N_\subini-N_\subeq}.
\end{equation}
The above results do not depend on whether EDE exists. For a given $\{x_{1,\subini}, x_{3,\subini}\}$ and cosmological parameters, $x_{4,\subini}$ is first computed from Eq.~(\ref{eq:3.17}), and the constraint equation is then used to determine $x_{2,\subini}$. The approximation used in Eq.~(\ref{eq:3.17}) causes the value of $\Omega_\Lambda$ at $N=0$ obtained from the numerical integration to differ slightly from the input $\Omega_{\Lambda,0}$, while the constraint equation remains exactly satisfied. Note that the late-time evolution is insensitive to $\{x_{1,\subini}, x_{3,\subini}\}$, since trajectories are attracted to $F_1$, which effectively erases any memory of these initial values.

Within the above framework, Fig.~\ref{fig:cosmos:05} Right provides an illustrative calculation. The model parameters and part of the initial conditions are the same as in the left panel (see the caption). When $N\lesssim-2$, the results in the left and right panels are almost identical, since $\Lambda$ is negligible at early epochs. Once $\Lambda$ becomes dynamically significant, $\Omega_\subEDE$ flips from positive to negative and then continues to evolve toward negative infinity (see the black dashed curve). A negative caloric energy can cancel the $\Lambda$, effectively realizing the self-tuning mechanism \cite{Dolgov1985.JETPLett.41.345,Ford1987.PRD.35.2339,Charmousis2012.PRL.108.051101}. To quantify this dynamics, the bottom subpanel displays the total EoS via $w_\subtot=-(2h+3)/3$ \cite{Amendola2010.book}, which characterizes the behavior of the net effective energy density. Around the present epoch, $w_\subtot$ remains close to $w_\substd$, indicating that the finite-time evolution of $S_\mu$ has little impact on the observed low-redshift expansion history. In the far future, $w_\subtot$ approaches $-1/3$, pointing to novel asymptotic dynamical behavior in GCT.

During the matter-$\Lambda$ transition ($N\approx0$), integrating Eq.~(\ref{eq:3.5}) requires very high accuracy to avoid numerical instabilities. We have not yet clarified the mathematical origin of this instability. Instead, in Fig.~\ref{fig:cosmos:05}, we plot $x_1$ to hint at the presence of this issue.

\subsubsection{$\Lambda$-cancelling solution}
Figure~\ref{fig:cosmos:05} Right shows that, in the far future, the contribution of $\Lambda$ to the Hubble expansion rate is almost exactly canceled by the energy density of $S_\mu$. We refer to this as the $\Lambda$-cancelling solution. This subsubsection is devoted to a detailed analysis of its properties.

The physical setup allows the dynamical equations to take a simple form. By \textit{far future} we mean the regime in which the energy density of $\Lambda$ is much larger than that of the fluid components. Therefore, all fluid-related quantities in Eq.~(\ref{eq:3.5}) can be set to zero, i.e., $x_3=0$ and $c_1=c_3=c_5=0$, so that $w_\substd=-1$. The remaining terms are independent of $w_\subF$. Further substituting the constraint equation into Eq.~(\ref{eq:3.5}) to eliminate $x_2$ yields
\begin{subequations}\label{eq:3.18}
\begin{align}
  x_1^\prime &= \frac{1}{\tilde{c}_2} \Big[-3 + 3(x_1+c_2x_1+x_4) + (2-4c_2+\tilde{c}_{46})x_1^2 \nonumber\\
    &\qquad - 3x_1x_4 + (\frac{c_2c_6}{2}-c_4)x_1^3\Big], \\
  x_4^\prime &= \frac{x_4}{\tilde{c}_2} \Big[6 + (4-8c_2)x_1 - 6x_4 + (c_2c_6-2c_4)x_1^2\Big].
\end{align}
\end{subequations}
For a positive cosmological constant, the physically relevant region is $x_4\geqslant0$~\footnote{Based on the definition of $x_4$, the case where the equality holds can be understood from a limiting perspective.}.

\begin{figure*}[!t]
  \centering
  \includegraphics[width=0.96\textwidth]{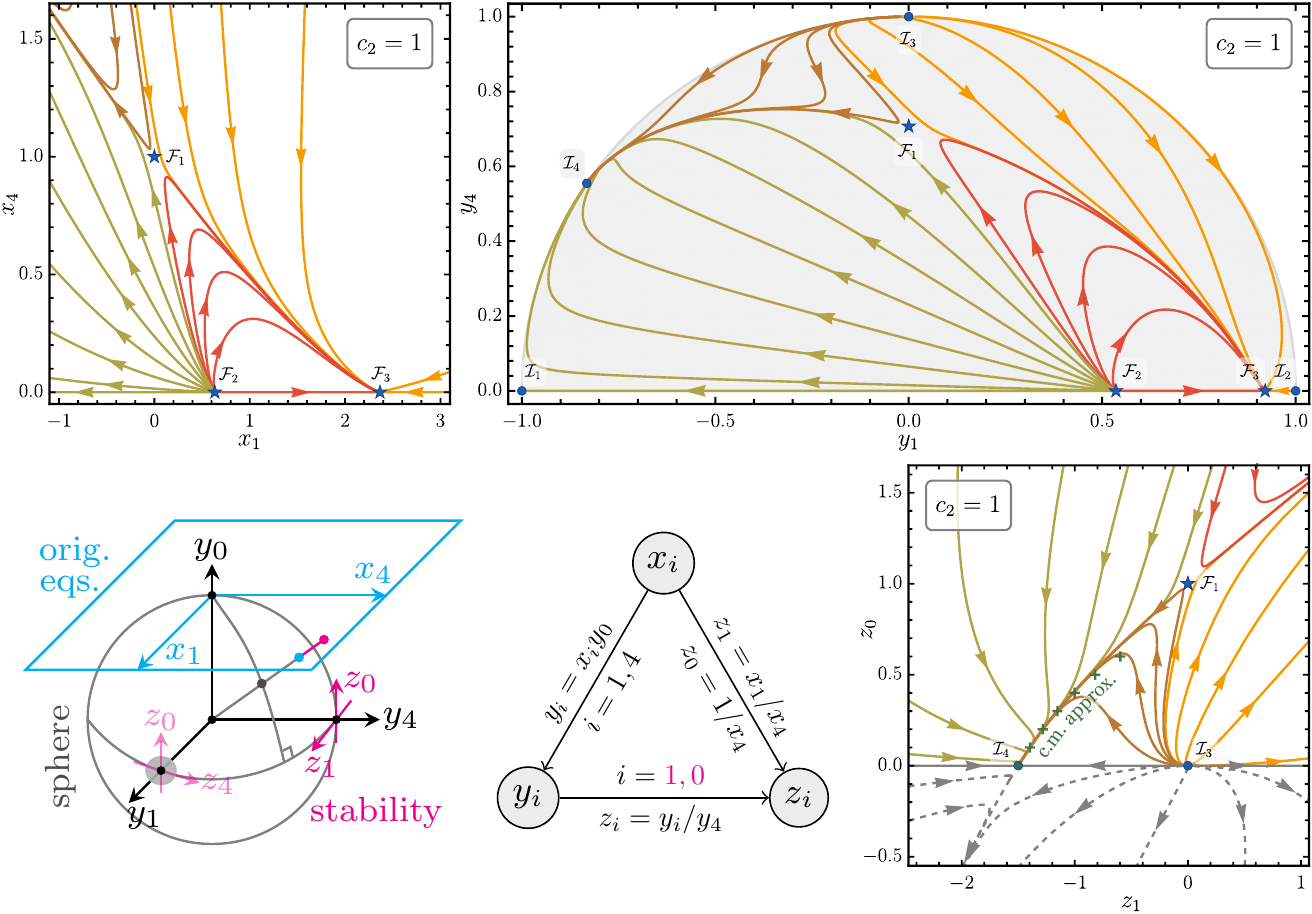}
  \caption{Phase portrait of Eq.~(\ref{eq:3.18}) and its projections onto the $(y_1,y_4)$ and $(z_1,z_0)$ axes. The parameters are fixed to $c_2=1$ and $c_4=c_6=0$. Trajectories in the three panels are in one-to-one correspondence, except for the grey lines in the $(z_1,z_0)$ axes, which represent a purely mathematical continuation without physical reality. Part of the brown and red curves lies outside the plotting region. In the $z_i$ panel, the centre manifold given by Eqs.~(\ref{eq:3.20}) and (\ref{eq:3.24}) is plotted as plus symbols. The critical points listed in Table~\ref{tab:03} are marked in the panels, illustrating the consistency. The two bottom insets summarise the coordinate transformations $x_i\rightarrow y_i\rightarrow z_i$, with the left showing a schematic diagram and the right giving the explicit formulae. The mapping is constructed along the straight line that passes through the grey ($y_i$), blue ($x_i$), and red ($z_i$) dots. Note that the $x_i$ and $z_i$ coordinates are defined on planar charts, while the $y_i$ coordinates lie on the sphere $y_1^2+y_4^2+y_0^2=1$.}
  \label{fig:cosmos:06}
\end{figure*}

Applying the dynamical analysis to the system (\ref{eq:3.18}) may provide a quantitative characterization of the $\Lambda$-cancelling solution. To accommodate the parameter setting in Fig.~\ref{fig:cosmos:05} and to simplify the discussion, here we fix $c_4=c_6=0$ and perform a complete analysis for arbitrary $c_2$. Solving $\{x_1^\prime=0,x_4^\prime=0\}$ yields the critical points in the finite region, and Table~\ref{tab:03} lists those that satisfy $x_4\geqslant0$. Among these critical points, $\mathcal{F}_1$ represents a $\Lambda$-dominated solution, while $\mathcal{F}_2$ and $\mathcal{F}_3$ are $S_\mu$-dominated solutions. Therefore, none of them corresponds to the $\Lambda$-cancelling solution. For completeness, the stability of these three critical points is analyzed using linear stability theory (see Appendix~\ref{sec:App.A.1}), and the results are summarized in Table~\ref{tab:03}. Figure~\ref{fig:cosmos:06} plots the phase portrait for $c_2=1$, serving as an illustration and a consistency check of the stability classification in Table~\ref{tab:03}. For $c_2=1$, $\mathcal{F}_1$ is a saddle, consistent with Fig.~\ref{fig:cosmos:05}, where the $\Lambda$-dominance drives $\Omega_\subEDE$ away from zero.

\begin{table*}[!t]
  \caption{Critical points and stability analysis of the dynamical system (\ref{eq:3.18}) with $c_4=c_6=0$ and $c_2\in\mathbb{R}$. Finite points $\mathcal{F}_i$ are analysed by linear stability theory, while the points at infinity $\mathcal{I}_i$ are classified using the stability theorem at infinity and centre manifold theory (or the equivalent phase portrait for a specific parameter). At infinity, we further distinguish between the two cases: $y_4,y_0,z_0\in\mathbb{R}$ and $y_4,y_0,z_0\geqslant0$. The former is treated as a purely mathematical analytic continuation (see the inset at the bottom of this table), whereas physically relevant solutions exist only in the latter region (see the solid lines in the inset). The last column highlights the cosmological application of $\mathcal{I}_4$ (see Sec.~\ref{sec:III.B}).}
  \label{tab:03}
  \begin{tabular*}{\hsize}{@{\ \,}@{\extracolsep{\fill}}llcccl@{\ \,}}
    \hline\hline
    Labels           &  Coordinates  &  Stable node or spiral  &  Saddle  &  Unstable node  &  Applications  \\
    \hline
    Finity           &  $(x_1,x_4)$  &  \multicolumn{3}{c}{Linear stability theory}         &  \\
    \cline{3-5}
    \multirow{2}{*}{$\mathcal{F}_1$}  &  \multirow{2}{*}{$(0,1)$}  &  \stablespiral $2<c_2<14/3$  &  \multirow{2}{*}{$c_2<2$}  &  \multirow{2}{*}{$\varnothing$} \\
                                      &                            &  \stablenode $14/3\leqslant c_2$  \\[0.5mm]
    $\mathcal{F}_2$  &  $\bigg(\displaystyle\frac{3c_2+3-\sqrt{9c_2^2-30c_2+33}}{4(2c_2-1)}, 0\bigg)$  &  $\varnothing$  &  $2<c_2$  &  $c_2<2$  \\[3.6mm]
    $\mathcal{F}_3$  &  $\bigg(\displaystyle\frac{3c_2+3+\sqrt{9c_2^2-30c_2+33}}{4(2c_2-1)}, 0\bigg)$  &  \stablenode $c_2<2$ \& $\neq\displaystyle\frac{1}{2}$  &  $\varnothing$  &  $2<c_2$  \\[3mm]
    \hline
    Infinity         &  $(y_1,y_4)$    &  \multicolumn{4}{l}{ \hspace{-2.5cm} Stability theorem at infinity \& Centre manifold theory; Trajectories for $y_4,y_0,z_0\in\mathbb{R}$}  \\
    \cline{3-5}
    $\mathcal{I}_1$  &  $(-1,0)$       &  $\varnothing$  &  $c_2\in\mathbb{R}$  &  $\varnothing$  &  \\
    $\mathcal{I}_2$  &  $(1,0)$        &  $\varnothing$  &  $c_2\in\mathbb{R}$  &  $\varnothing$  &  \\
    $\mathcal{I}_3$  &  $(0,1)$        &  \stablenode $2<c_2$  &  $\varnothing$  &  $c_2<2$       &  \\[1.5mm]
    $\mathcal{I}_4$  &  $\bigg(\displaystyle\frac{3\,\sgn(1-2c_2)}{\sqrt{9+4(1-2c_2)^2}},\displaystyle\frac{2\,|1-2c_2|}{\sqrt{9+4(1-2c_2)^2}}\bigg)$  &  $\varnothing$  &  $c_2\neq1/2$  &  $\varnothing$  &  \\[3.5mm]
    \cline{2-6}
                     &  \multicolumn{1}{c}{A typical $z_i$ phase portrait}  &  \multicolumn{4}{l}{ \hspace{-0.45cm} Same method as above; Trajectories for $y_4,y_0,z_0\geqslant0$}  \\
    \cline{3-5}
    $\mathcal{I}_1$  &  \multicolumn{1}{c}{\multirow{4}{*}{\includegraphics[width=0.113\linewidth]{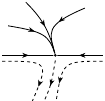}}}  &  \stablenode $c_2=1/2$  &  $c_2\neq1/2$  &  $\varnothing$  &  \\
    $\mathcal{I}_2$  &                 &  \stablenode $c_2=1/2$  &  $c_2\neq1/2$   &  $\varnothing$  &  \\
    $\mathcal{I}_3$  &                 &  \stablenode $2<c_2$    &  $\varnothing$  &  $c_2<2$        &  \\[0.5mm]
    $\mathcal{I}_4$  &                 &  \stablenode $c_2<2$ \& $\neq\displaystyle\frac{1}{2}$  &  $2<c_2$  &  $\varnothing$  &  $\Lambda$-cancelling solution \\[2mm]
    \hline\hline
  \end{tabular*}
\end{table*}

In Fig.~\ref{fig:cosmos:05} Right, $\Omega_\subEDE$ and $\Omega_\Lambda$ evolve toward $\pm\infty$. This behaviour motivates the following analysis of the critical points at infinity of Eq.~(\ref{eq:3.18}). Based on the methods and theorems summarized in Appendix~\ref{sec:App.A.3}, their coordinates are computed in the $y_i$ variables, defined by $y_i=x_i/\sqrt{1+x_1^2+x_4^2}$ for $i=1,4$, and the results (labeled as $\mathcal{I}_i$; compatible with $x_4\geqslant0$) are listed in Table~\ref{tab:03}. Three of these points lie on the axes and correspond to $\Lambda$- or $S_\mu$-dominated solutions. The remaining point has nonzero $y_1$ and $y_4$ and is a natural candidate for the $\Lambda$-cancelling solution. The $y_i$-phase portrait is also shown in Fig.~\ref{fig:cosmos:06}, providing a consistency check of the coordinates of $\mathcal{I}_i$. Before presenting the stability analysis, a brief note on nonzero $c_4$ and $c_6$ is in order. If $c_4\neq0$, $c_6\neq0$ and $c_2c_6-2c_4\neq0$, one can verify that Eq.~(\ref{eq:3.18}) admits only three critical points at infinity, $\{\mathcal{I}_1,\mathcal{I}_2,\mathcal{I}_3\}$. None of these points can realize the $\Lambda$-cancelling solution. This observation justifies the choice $c_4=c_6=0$ for the dynamical analysis in this part. The special case $c_2c_6-2c_4=0$ is not pursued further here.

To simplify the stability analysis, we do not directly employ Theorems~1 and 2 in Sec.~3.10 of Ref.~\cite{Perko2001.book}, but instead use the method summarized in Appendix~\ref{sec:App.A.3} to explicitly map the system into the $z_i$ coordinates, defined on a plane tangent to the Poincar\'e sphere at an equatorial point (see the illustration in Fig.~\ref{fig:cosmos:06}). This eliminates the need for the redundant case-by-case discussion of sign choices in Ref.~\cite{Perko2001.book}. The essential purpose of this mapping is to unfold the trajectories near $\mathcal{I}_i$. We first examine the stability of $\mathcal{I}_3$. To this end, the $z_i$ plane is chosen tangent to the sphere at $\mathcal{I}_3$, i.e., the $(z_1,z_0)$ coordinates shown in Fig.~\ref{fig:cosmos:06}. In this chart, the variables are related by the explicit mapping $z_1=x_1/x_4$ and $z_0=1/x_4$, which allows Eq.~(\ref{eq:3.18}) with $c_4=c_6=0$ to be rewritten as
\begin{subequations}\label{eq:3.19}
\begin{align}
  \frac{\dx z_1}{\dx\sigma} &= \frac{3}{\tilde{c}_2} \big[z_1+z_0 + \frac{4c_2-2}{3}z_1^2 - z_0^2 + (c_2-1)z_1z_0\big], \\
  \frac{\dx z_0}{\dx\sigma} &= \frac{2z_0}{\tilde{c}_2} \big[3 - 3z_0 - (2-4c_2)z_1\big], \label{eq:3.19b}
\end{align}
\end{subequations}
where $\dx\sigma=z_0^{-1}\dx N$ with derivatives taken along the trajectory under consideration. This differential relation implements a \textit{time rescaling}, and serves to eliminate the divergence of $z_i^\prime$ in the limit $z_0\rightarrow0$. Under this transformation, the condition $x_4\geqslant0$ translates into $z_0\geqslant0$. Therefore, for physical solutions, the time rescaling does not change the positive direction of time. This guarantees that the dynamical systems defined by Eq.~(\ref{eq:3.18}) and Eq.~(\ref{eq:3.19}) share the same stability properties at the corresponding critical points when restricted to trajectories with $x_4\geqslant0$. In what follows, trajectories in the region $z_0<0$ and their impact on the stability of the critical points of Eq.~(\ref{eq:3.19}) are also considered separately. This is included purely for mathematical completeness and has no physical application here. Accordingly, the (possible) reversal of the time direction induced by the time rescaling for $z_0<0$ is not taken into account~\footnote{As an illustration, in the $z_i$ panel of Fig.~\ref{fig:cosmos:06}, the grey dashed curves represent trajectories with $z_0<0$, while the arrows still indicate the direction of increasing $\sigma$.}. After these preparations, we can proceed to discuss the stability of $\mathcal{I}_3$, which is equivalent to the stability of the corresponding point $(z_1,z_0)=(0,0)$ in Eq.~(\ref{eq:3.19}). For the latter, linear stability theory yields two nonzero eigenvalues, $6/\tilde{c}_2$ and $3/\tilde{c}_2$. Consequently, $\mathcal{I}_3$ is a stable node for $c_2>2$ and an unstable node for $c_2<2$. Note that this result is independent of whether trajectories in the region $z_0<0$ are taken into account. In the $z_i$ coordinates, Fig.~\ref{fig:cosmos:06} displays the phase portrait of Eq.~(\ref{eq:3.19}), providing a consistency check of the above results and of the subsequent stability analysis of $\mathcal{I}_4$.

For the stability analysis of $\mathcal{I}_1$ and $\mathcal{I}_2$, Eq.~(\ref{eq:3.19}) is not applicable, since both points lie at infinity in this chart. We can project the trajectories in the $y_i$ sphere onto the corresponding tangent planes, e.g., the $(z_4,z_0)$ coordinates in Fig.~\ref{fig:cosmos:06} for $\mathcal{I}_2$. However, this would require introducing two additional coordinate charts, which risks complicating the notation. For clarity, our analysis is carried out directly in the $(y_1,y_4,y_0)$ coordinates. Similar to Eq.~(\ref{eq:A.09}), we can derive $\dx y_i/\dx\sigma$ with $i=1,4,0$ for Eq.~(\ref{eq:3.18}) \cite{NotShown}. The stability depends only on trajectories in an infinitesimal neighbourhood, which allows us to consider the approximate mapping (see Ref.~\cite{Lefschetz1957.book} for an example). For $\mathcal{I}_1$, the stability analysis is equivalent to studying the point $(y_4,y_0)=(0,0)$ in the $\{\dx y_4/\dx\sigma,\dx y_0/\dx\sigma\}$ system with $y_1=-\sqrt{1-y_4^2-y_0^2}\approx-1+y_4^2/2+y_0^2/2$. Here the $(y_4,y_0)$ plane is locally equivalent to the $z_i$ plane in the infinitesimal neighbourhood. Second-order terms are retained here so that the subsequent analysis of the zero-eigenvalue case is self-consistent in view of the series expansion~\footnote{For the zero-eigenvalue case, we simply assume that the stability could be determined by the quadratic terms, without proof from centre manifold theory.}. Linear stability theory yields two eigenvalues $\pm(4c_2-2)/\tilde{c}_2$. If we consider the region with $y_4,y_0\in\mathbb{R}$, then $\mathcal{I}_1$ is a saddle if $c_2\neq1/2$. The special case gives zero eigenvalues, which in principle requires centre manifold theory. For simplicity, the phase portrait is instead plotted \cite{NotShown}, and it clearly shows that $\mathcal{I}_1$ still behaves as a saddle. This result, namely that $\mathcal{I}_1$ is a saddle for $c_2\in\mathbb{R}$ in view of trajectories for $y_4,y_0\in\mathbb{R}$, is summarized in Table~\ref{tab:03}. However, the physically relevant solutions (corresponding to an expanding Universe) occupy only a subset of the coordinates: $x_4\geqslant0$ implies $y_4\geqslant0$, and the $x_i\rightarrow y_i$ transformation imposes $y_0\geqslant0$. Restricting to trajectories in the region $y_4,y_0\geqslant0$ and using the phase portrait together with the eigenvectors \cite{NotShown}, we find that $\mathcal{I}_1$ turns from a saddle into a stable node only at $c_2=1/2$, whereas for all other values of $c_2$ the stability classification remains unchanged. For $\mathcal{I}_2$, the stability analysis is equivalent to studying the point $(y_4,y_0)=(0,0)$ in the $\{\dx y_4/\dx\sigma,\dx y_0/\dx\sigma\}$ system with $y_1\approx1-y_4^2/2-y_0^2/2$. The stability result is identical to that of $\mathcal{I}_1$.

Equation~(\ref{eq:3.19}) is applicable for the stability analysis of $\mathcal{I}_4$. The reason is that such analysis does not require the $z_i$ plane to be exactly tangent to the sphere at the critical point being studied \cite{Perko2001.book}, and mapping to the $(z_1,z_0)$ plane indeed unfolds the surrounding trajectories. Based on the coordinate transformation ($z_1=y_1/y_4$ on the equator), we obtain that $\mathcal{I}_4$ corresponds to $(z_1,z_0)=(3/(2-4c_2),0)$. The eigenvalues at this point are $3/(c_2-2)$ and $0$. A zero eigenvalue implies that centre manifold theory is needed to determine the stability. For simplicity, phase portraits are first plotted for a series $c_2$ \cite{NotShown}. The results show that $\mathcal{I}_4$ is a saddle for $c_2\neq1/2$ when all trajectories with $z_0\in\mathbb{R}$ are taken into account. Note that $\mathcal{I}_4$ does not exist if $c_2=1/2$. Restricting to trajectories with $z_0\geqslant0$, $\mathcal{I}_4$ becomes a stable node for $c_2<2\ \&\neq1/2$, while for $c_2>2$ it remains a saddle. This conclusion is confirmed by the subsequent centre-manifold analysis.

When applying the above stability results to the evolution of the physical Universe, the analysis should be restricted to the sector with $y_4,y_0,z_0\geqslant0$. This is self-consistent as physically relevant trajectories do not cross the boundaries (see the future dynamics in Fig.~\ref{fig:cosmos:05} Right for an indication).

Plotting the solution in Fig.~\ref{fig:cosmos:05} Right on the phase plane shows that the $\Lambda$-cancelling solution converges to $\mathcal{I}_4$. Therefore, its properties can be assessed from the dynamics around $\mathcal{I}_4$, especially the $F_1\rightarrow\mathcal{I}_4$ attractor trajectory shown in Fig.~\ref{fig:cosmos:06}. A key quantitative diagnostic is the asymptotic value of $w_\subtot$, which encodes the behaviour of the net energy density. Considering $w_\subtot=-(2h+3)/3$, Eq.~(\ref{eq:3.4}), and $x_i\rightarrow\infty$, we see that a finite value of $w_\subtot$ must result from cancellations among several divergent contributions. Determining $w_\subtot$ therefore requires information on how the variables $x_i$ approach infinity, not just the coordinates of $\mathcal{I}_4$. Since the stability analysis yields a zero eigenvalue at this point, the calculation should be based on centre manifold theory. The first step is to construct a local centre manifold through $\mathcal{I}_4$, tangent to the eigenvector associated with the zero eigenvalue~\footnote{Our analysis concerns only two-dimensional real systems, so the eigenvector associated with the zero eigenvalue can always be chosen real. For higher-dimensional systems, the term \textit{eigenvector} should be replaced by \textit{eigenspace} for mathematical accuracy.}. We summarize the relevant derivation in Appendix~\ref{sec:App.A.2} and directly quote the result that, for Eq.~(\ref{eq:3.19}), the centre manifold ($\subcm$) can be described by
\begin{equation}\label{eq:3.20}
  z_1|_\subcm = \frac{3-(1+c_2)z_0}{2-4c_2} + \phi(z_0),
\end{equation}
in which both $\phi$ and its first derivative $\dx \phi/\dx z_0$ vanish at $z_0=0$. The boundary conditions, together with Eq.~(\ref{eq:3.19}), determine a unique solution $\phi(z_0)$. Substituting the above equation into $w_\subtot$, we obtain the total EoS along the centre manifold,
\begin{equation}\label{eq:3.21}
  w_\subcm = -\frac{1}{3} + \frac{4-8c_2}{3\tilde{c}_2}\cdot\frac{\phi(z_0)}{z_0}.
\end{equation}
Imposing the boundary conditions on $\phi(z_0)$, this result implies that the asymptotic value of $w_\subtot$ is $-1/3$, in agreement with the numerical result shown in Fig.~\ref{fig:cosmos:05}. This result is independent of the detailed form of $\phi(z_0)$ and of the value of $c_2$. Physically, this $w_\subtot$ corresponds to a linearly expanding Universe, i.e., the $R_{\rm h}=ct$ Universe \cite{Melia2012.MNRAS.419.2579}. Note that, when applied to the real Universe, this asymptotic dynamics belongs to the far future, not to the observed low-redshift history.

For completeness, here we analyse the stability along the centre manifold and derive the solution for $\phi(z_0)$. Substituting Eq.~(\ref{eq:3.20}) into Eq.~(\ref{eq:3.19b}), we obtain
\begin{equation}
  \frac{\dx z_0}{\dx\sigma}\Big|_\subcm = -2z_0^2 + \frac{8c_2-4}{\tilde{c}_2}\cdot z_0\phi(z_0).
\end{equation}
This equation governs the evolution of $z_0$ along the centre manifold. In the physical sector $z_0>0$, the leading term $-2z_0^2$ drives trajectories toward $\mathcal{I}_4$ as $\sigma$ increases. If the analytically continued lower half-plane is drawn with the same $\sigma$-orientation, trajectories with $z_0<0$ instead move away from $\mathcal{I}_4$. This region-dependent centre-manifold behaviour accounts for the two $\mathcal{I}_4$ classifications in Table~\ref{tab:03}: the full analytic continuation gives a saddle, whereas the physical sector may give a stable node in some cases. To find the solution for $\phi(z_0)$, we can differentiate Eq.~(\ref{eq:3.20}) with respect to $\sigma$ and use Eq.~(\ref{eq:3.19}) to eliminate $\dx z_i/\dx\sigma$, thereby obtaining the \textit{orbit} equation
\begin{align}\label{eq:3.23}
  0 &= 2 z_0 [(4c_2-2)\phi(z_0) - \tilde{c}_2z_0] \frac{\dx \phi(z_0)}{\dx z_0} + \frac{\tilde{c}_2^2 z_0^2}{1-2c_2} \nonumber\\
  &\quad + \big[3 + 3(1-c_2)z_0 + (2-4c_2)\phi(z_0)\big] \phi(z_0).
\end{align}
This is a nonlinear ordinary differential equation, which for the initial conditions specified above admits a series expansion solution
\begin{equation}\label{eq:3.24}
  \phi(z_0) = \frac{\tilde{c}_2^2z_0^2}{6c_2-3} \cdot \Big(1 + \frac{5-c_2}{3}z_0 + \frac{7-c_2^2}{3}z_0^2 + \cdots\Big).
\end{equation}
In the bottom-right subpanel of Fig.~\ref{fig:cosmos:05}, the $w_\subcm$ is plotted using the first three terms in Eq.~(\ref{eq:3.24}), with $z_0(N)$ reconstructed from the numerically integrated $x_4(N)$. In the $z_i$ panel of Fig.~\ref{fig:cosmos:06}, the centre manifold is shown as plus symbols, computed with the same truncation of the series in Eq.~(\ref{eq:3.24}). The approximation agrees well with the corresponding exact results in both figures. Centre manifold theory provides a systematic mathematical tool for analysing attractor trajectories.

\subsubsection{Cosmological constant problem}
The preceding subsection treats $\Lambda$ as the standard cosmological constant relevant to the real Universe, and the corresponding $\Lambda$-cancelling dynamics becomes important only in the far future. A separate question is whether the same dynamics can self-tune a huge bare vacuum energy in the early Universe, as required for addressing the old cosmological constant problem \cite{Zeldovich1967.SovPhysJETP.6.316,Weinberg1989.RMP.61.1,Carroll2001.LRR.4.1}. In this hypothetical application, the $\Lambda$-cancelling solution found above can be interpreted as a self-tuning mechanism \cite{Dolgov1985.JETPLett.41.345,Ford1987.PRD.35.2339,Charmousis2012.PRL.108.051101}. This subsubsection briefly reviews the relevant ingredients and discusses how our analysis may contribute to this line of work.

Nearly fifty years ago, many self-tuning constructions were proposed \cite{Dolgov1985.JETPLett.41.345,Ford1987.PRD.35.2339,Peccei1987.PLB.195.183,Barr1988}, but such models have generally proved unsuccessful due to severe fine-tuning or an unacceptably time-varying Newtonian \textit{constant} $G$. Weinberg's no-go theorem shows that, under fairly general assumptions, any attempt to solve the cosmological constant problem within local (static) field theories inevitably requires fine-tuning \cite{Weinberg1989.RMP.61.1}. To evade this theorem, \citet{Charmousis2012.PRL.108.051101,Charmousis2012.PRD.85.104040} relaxed the requirement of Poincar\'e invariance of the field configuration, which is one of Weinberg's key assumptions, and presented a framework for constructing the self-tuning mechanism within Horndeski gravity. At the level of cosmological evolution, this is similar to the strategy of Refs.~\cite{Dolgov1985.JETPLett.41.345,Ford1987.PRD.35.2339}, but admits additional degrees of freedom, which may enable the model to satisfy Solar System constraints (mainly on time-varying $G$) via a screening mechanism. This construction has stimulated many subsequent studies \cite{Bruneton2012,Copeland2012,Linder2013.JCAP.12.032,MartinMoruno2015.PRD.91.084029,Babichev2017,Khan2022.JCAP.10.075}. However, to our knowledge, despite various explicit constructions of self-tuning scalar-tensor theories, a single model that has been demonstrated to realise a fully realistic cosmological history and to remain compatible with Solar System tests via a robust screening mechanism is still lacking. GCT adopts the same strategy to circumvent Weinberg's no-go theorem --- a dynamical field breaks time-translation symmetry. In Sec.~\ref{sec:IV}, the weak-field limit of GCT around the Minkowski background ($S_\mu=0$) is analyzed. Subject to the caveats discussed at the beginning of Sec.~\ref{sec:IV}, this provides a first Solar-System consistency check for the EDE realization, but it does not yet cover the self-tuning scenario ($S_\mu\rightarrow\infty$).

Many self-tuning models suffer from an over-screening problem in which the field cancels not only $\Lambda$ but also the energy density of ordinary matter, preventing the Universe from entering a standard matter-dominated epoch \cite{Linder2013.JCAP.12.032,MartinMoruno2015.PRD.91.084029}. Our model may not exhibit this over-screening problem, but it still fails to realize standard cosmic evolution, since the asymptotic value of $w_\subtot$ is $-1/3$. A satisfactory resolution of this issue is not yet available.

Although our model still exhibits several unresolved issues, the analysis suggests a useful dynamical-system perspective on self-tuning. In suitable variables, a self-tuning branch may be organized by a critical point at infinity. Centre manifold theory can determine not only the stability of this critical point, but also the asymptotic rate at which trajectories diverge. Controlling this rate order by order may help distinguish different self-tuning behaviours and guide model building. Section~\ref{sec:III.C} demonstrates the effectiveness of this idea in engineering the dynamics of ESHU. A more systematic application to self-tuning is left for future work.

\subsection{Early static hot Universe}\label{sec:III.C}
Figure~\ref{fig:cosmos:02} presents the cosmic dynamics when both the caloric field and a perfect fluid are included in GCT~\footnote{Note that Fig.~\ref{fig:cosmos:02} is plotted for $w_\subF=0$ (matter). Taking $w_\subF=1/3$ (radiation) instead does not alter the qualitative structure of the phase portrait \cite{NotShown}.}. Near the critical point $I_3$, one has $\Omega_\subF\propto x_3^2\rightarrow\infty$. On the other hand, since $\Omega_\subF\propto\rho_\subF/H^2$ by definition and $\rho_\subF$ is not increasing in an expanding Universe, the divergence $\Omega_\subF\rightarrow\infty$ implies $H\rightarrow0$. The resulting dynamics are similar to the $\Lambda$-cancelling solution (see Sec.~\ref{sec:III.B}), with $\Lambda$ replaced by a perfect fluid component. Physically, this corresponds to a quasi-static Universe, in the sense that $H^2\ll8\pi G\rho_\subF/3$. Hereafter, the qualifier \textit{quasi} may be omitted when no ambiguity arises. Furthermore, $I_3$ is a saddle in Fig.~\ref{fig:cosmos:02}, thereby providing a natural exit from the static phase. The cosmic evolution can proceed as follows (see the orange dashed curve in Fig.~\ref{fig:cosmos:02} Left for an illustration): starting from suitable finite initial conditions with $\Omega_\subF=\mathcal{O}(1)$, trajectories can first be guided by the attracting direction of $I_3$ toward the quasi-static regime and then exit along its repelling direction toward $F_1$, where the standard fluid-dominated era is recovered~\footnote{If $I_3$ is an unstable node, the following ESHU could still be realized in principle, but only at the expense of highly fine-tuned initial conditions. This scenario is not considered in the present work.}. Applying this scenario to the early Universe, we find that a static phase can exist within the radiation epoch. We refer to this phase as the \textit{early static hot Universe} (ESHU). Note that the hot gas in this scenario preserves its standard thermodynamic properties. A key observation is that ESHU features a large comoving Hubble radius (CHR), potentially allowing a sufficiently large causally connected patch to emerge in the early Universe and thereby resolving the horizon problem~\footnote{This is nontrivial, as the $n=3$ case will be shown below to fail to solve the horizon problem.}. This opens up a new possible alternative to the inflationary paradigm \cite{Guth1981.PRD.23.347,Linde1982.PLB.108.389,Albrecht1982.PRL.48.1220}. This subsection is devoted to the analysis of the horizon problem in ESHU and shows that GCT admits a rich variety of classes of CHR evolution.

A brief model comparison and clarification are in order here. The ESHU scenario bears a clear structural parallel to inflation \cite{Guth1981.PRD.23.347}. First, when viewed backward from the end of the phase, inflation is characterized by nearly constant $H$ and an exponentially decreasing scale factor $a$. By contrast, in ESHU, $a$ remains nearly constant~\footnote{This is a rough description. Section \ref{sec:III.C.3} presents a model, in which $a$ can change by a large factor.} while $H$ drops rapidly. Both scenarios can in principle address the horizon problem because the causal horizon scales as $(aH)^{-1}$ \cite{Baumann2009.arXiv.0907.5424}. Second, concerning the matter temperature and the source of fluctuations, in standard inflation, relativistic particles are rapidly diluted by exponential expansion, rendering the Universe extremely cold. Primordial perturbations are therefore usually attributed to quantum fluctuations rather than thermal fluctuations~\footnote{One exception is warm inflation, in which radiation is produced continuously during inflation, allowing thermal fluctuations to be important \cite{Berera1995.PRL.74.1912,Berera1995.PRL.75.3218}.}. ESHU features a high-temperature gas even when the CHR is large. This provides a suitable platform for studying thermal fluctuations in cosmology. Third, the flatness problem marks a further distinction. The inflationary resolution relies on exponential expansion and reheating \cite{Hu1994.PRD.49.3830}, whereas ESHU would require a different mechanism, which lies beyond the present work and will be discussed in a separate paper.

Similar quasi-static phases also appear in several early-Universe scenarios proposed as alternatives to, or complements of, inflation \cite{Ellis2004.CQG.21.223,Khoury2001.PRD.64.123522,Creminelli2010.JCAP.11.021,Parker1973.PRD.7.2357,Tseytlin1992.NuclPhysB.372.443}. We compare these models with ESHU by focusing on two questions:
\begin{itemize}
  \item whether the quasi-static stage establishes causal contact across the observable Universe and governs the horizon crossing of primordial perturbations;
  \item whether this stage is filled with ordinary hot gas.
\end{itemize}
In the emergent Universe, the Einstein-static phase is a nonsingular precursor to inflation that can resolve the horizon problem \footnote{A caveat should be noted. In the emergent Universe, the Einstein-static phase is imposed as a past-asymptotic initial configuration rather than being obtained as a generic dynamical attractor. It therefore represents a fine-tuned initial state \cite{Ellis2004.CQG.21.223}. In this sense, the horizon problem is not solved dynamically, but shifted to the assumption of a fine-tuned \textit{homogeneous} initial state. By contrast, ESHU corresponds to a saddle-type attractor and can emerge from an ordinary expanding epoch.}, while CMB-scale horizon crossing is governed by the subsequent inflationary evolution \cite{Ellis2004.CQG.21.223,Ellis2004.CQG.21.233}. In the ekpyrotic Universe, inflation is replaced by an ultra-stiff slow-contraction phase with only mild scale-factor evolution, which establishes causal contact and governs the horizon crossing of primordial perturbations \cite{Khoury2001.PRD.64.123522,Lehners2008.PhysRep.465.223}. In the Galilean genesis scenario, the early Universe is described by an asymptotically Minkowski-static phase followed by an initially slow Galileon-driven expansion, a stage that also plays the dual role of providing causal contact and setting the horizon-crossing dynamics \cite{Creminelli2010.JCAP.11.021,Wang2012.JCAP.10.021}. In all three models, ordinary hot gas is not the relevant medium during the \textit{observation-relevant causal-formation} and horizon-crossing stages, while the standard hot-Big-Bang plasma is reached only later through inflationary reheating \cite{Ellis2004.CQG.21.223}, brane collision \cite{Khoury2001.PRD.64.123522}, or Galileon reheating \cite{Creminelli2010.JCAP.11.021}. Accordingly, primordial perturbations in these scenarios generally originate from quantum fluctuations rather than thermal fluctuations \cite{Ellis2004.CQG.21.233,Khoury2001.PRD.64.123522,Creminelli2010.JCAP.11.021,Wang2012.JCAP.10.021}. Bouncing cosmology forms a separate case: the bounce contains an instant at which $H=0$ \cite{Parker1973.PRD.7.2357}, but causal contact and primordial mode evolution are controlled mainly by the pre-bounce contracting epoch rather than by the short bounce itself (a quasi-static interval) \cite{Battefeld2015.PhysRep.571.1,Lilley2015.CRPhys.16.1038}. The closest parallel to ESHU is string gas cosmology, in which the static phase is governed by a (high) Hagedorn-temperature string gas \cite{Brandenberger1989.NuclPhysB.316.391,Tseytlin1992.NuclPhysB.372.443} and is responsible for the relevant horizon dynamics \cite{Nayeri2006.PRL.97.021302}. However, the framework is not yet theoretically complete: its key arguments hinge on T-duality and Hagedorn thermodynamics, while a fully self-consistent gravitational description of the Hagedorn phase remains an open issue \cite{Nayeri2006.PRL.97.021302,Kaloper2006.JCAP.10.006,Brandenberger2011.CQG.28.204005}. Leaving aside theoretical completeness, the \textit{high temperature} of a string gas has a meaning distinct from that of ordinary matter: in the Hagedorn regime, energy is primarily funneled into highly excited states, whose exponentially growing density of states causes the temperature to saturate near the Hagedorn limit, while the pressure approaches zero due to cancellations between momentum and winding modes \cite{Deo1989.PRD.40.2626,Lowe1995.PRD.51.665,Nayeri2006.PRL.97.021302}. An observationally relevant issue is that primordial scale-invariant perturbations can arise from thermal fluctuations of an equilibrium string gas \cite{Nayeri2006.PRL.97.021302,Brandenberger2006.JCAP.11.009}, but controversies exist regarding the phenomenological calculations \cite{Kaloper2006.JCAP.10.006} and the equilibrium assumption \cite{Danos2004.PRD.70.106010}. These uncertainties and ongoing debates limit the applicability of string gas cosmology in modern precision cosmology. Unlike in string gas, \textit{high temperature} in ESHU refers to ordinary relativistic particles with large kinetic energies and efficient scattering, making standard kinetic theory applicable to thermalization and thermal fluctuations \cite{Reif1965.book}.

This subsection is organized as follows. For $c_4,c_6=0$, Sec.~\ref{sec:III.C.1} analyzes the stability of $I_3$ and identifies the region of parameter space in which ESHU lives ($I_3$ is a saddle). The zero-eigenvalue cases are discussed in detail. To characterize the ESHU dynamics, Sec.~\ref{sec:III.C.2} quantitatively computes the CHR evolution and discusses the horizon problem. The $I_3\rightarrow F_1$ attractor shown in Fig.~\ref{fig:cosmos:02} serves as a key mathematical guide for the analysis. Along the attractor, the leading dynamics is captured by $\lim_{x_3\rightarrow\infty}x_3^\prime\propto-x_3^n$, where $n$ is an integer. GCT with $c_4,c_6=0$ could realize models with $n=3,2$ but precludes $n=1$. Motivated by the importance of the CHR evolution corresponding to $n=1$, Sec.~\ref{sec:III.C.3} revisits the above analysis with nonzero $c_4$ and $c_6$. Given the complexity of the derivation and the results, this final step does not aim at a full parameter-space analysis. Instead, it focuses on establishing the existence of parameter choices that realize the $n=1$ dynamics.

\subsubsection{Stability of $I_3$}\label{sec:III.C.1}
For $c_4,c_6=0$, the autonomous dynamical system (\ref{eq:3.6}) admits a unique nontrivial critical point at infinity, $I_3$, whose coordinates are given in Table~\ref{tab:02}. To analyze its stability, following the same procedure as for $\mathcal{I}_4$ in Fig.~\ref{fig:cosmos:06}, the system should be projected onto a plane tangent to the Poincar\'e sphere at a point on the equator. We choose this point to be $(y_1,y_3,y_0)=(0,1,0)$. Intuitively, the relevant geometric sketch can be obtained by taking the inset in Fig.~\ref{fig:cosmos:06} and replacing every subscript $4$ with $3$. The projection plane coincides with the $(z_1,z_0)$ plane shown in Fig.~\ref{fig:cosmos:06}. The variables are related by $z_1=x_1/x_3$ and $z_0=1/x_3$. Then, Eq.~(\ref{eq:3.6}) with $c_4,c_6=0$ can be transformed into
\begin{subequations}\label{eq:3.25}
\begin{align}
  \frac{\dx z_1}{\dx\sigma} &= \frac{z_0}{\tilde{c}_2}\Big[1 - \tilde{c}_{35}z_1 - 3z_0^2 + \frac{3}{2}(2+c_2+\tilde{c}_2w_\subF)z_1z_0\Big], \label{eq:3.25a} \\
  \frac{\dx z_0}{\dx\sigma} &= \frac{z_0}{\tilde{c}_2}\Big[(c_2c_5-c_3)z_1 + \frac{c_2-\tilde{c}_2w_\subF}{2}(1-3z_0^2) \nonumber\\
    &\qquad + (4c_2-2)z_1z_0\Big], \label{eq:3.25b}
\end{align}
\end{subequations}
where $\dx\sigma=z_0^{-2} \dx N$ is given by Eq.~(\ref{eq:A.12}). Physical solutions require $z_0\geqslant0$. All stability requirements below are therefore understood as one-sided conditions in this physical sector. Now we have $(z_1,z_0)|_{I_3} = (1/\tilde{c}_{35},0)$. Moreover, the right-hand side of Eq.~(\ref{eq:3.25}) scales with $z_0$. It follows that $\dx z_i/\dx\sigma=0$ once $z_0=0$, which would complicate the subsequent stability analysis. Note that, at this stage, no additional time rescaling can be used to absorb the factor $z_0$ on the right-hand side. Otherwise, $I_3$ would no longer be a critical point of the system, thereby obstructing the use of standard mathematical tools.

\begin{figure*}[!t]
  \centering
  \includegraphics[width=0.99\textwidth]{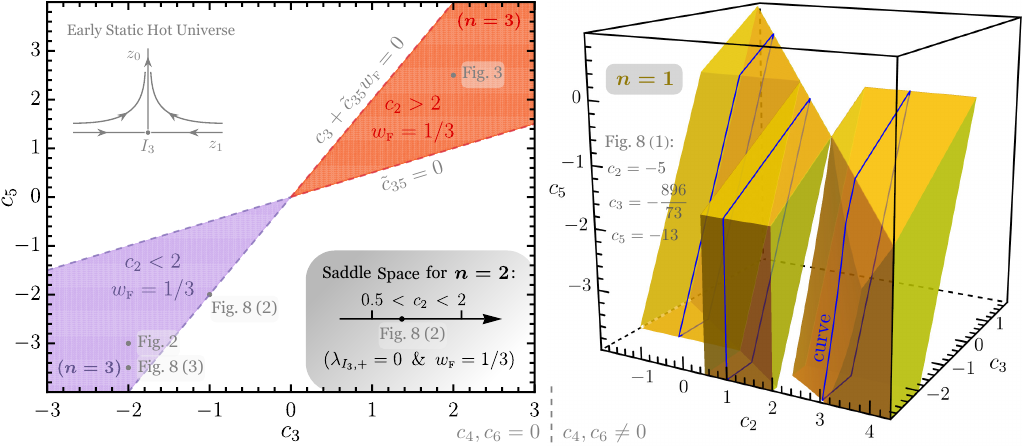}
  \caption{Saddle region of the critical point $I_3$ (see the upper-left inset) for $w_\subF=1/3$, in which ESHU lives. The dynamics can be classified by $n=3,2,1$ (see Sec.~\ref{sec:III.C}). Left with $c_4,c_6=0$: The $n=3$ result is represented by the colored region in the $(c_3,c_5)$ plane, with the upper and lower patches corresponding to $c_2>2$ and $c_2<2$, respectively. The boundary expressions are indicated in the figure. For $n=2$, the result depends explicitly on $c_2$ (see the lower-right inset), with $c_3$ and $c_5$ restricted by the constraint equation $\lambda_{I_3,+}=0$. No parameter point with $c_4,c_6=0$ realises $n=1$. Right with $c_4,c_6\neq0$: Part of the $n=1$ result is shown in the $(c_2,c_3,c_5)$ space, with $c_4$ and $c_6$ determined by the constraint equations given in the main text. The blue lines represent the slices at $c_2=-1,1,3$, respectively. All blue segments are straight lines, except for one curved portion indicated in the figure. The parameter choices used in Figs.~\ref{fig:cosmos:02}, \ref{fig:cosmos:03}, and \ref{fig:cosmos:08}\,$(n)$ are marked in the figure.}
  \label{fig:cosmos:07}
\end{figure*}

One can directly compute the Jacobian matrix of Eq.~(\ref{eq:3.25}) at $I_3$. It has one nonzero eigenvalue $\lambda_{I_3,+}=-(c_3+\tilde{c}_{35}w_\subF)/(2\tilde{c}_{35})$ whose eigenvector is aligned with the $z_0$-axis, and one vanishing eigenvalue whose eigenvector is aligned with the $z_1$-axis. As indicated by Fig.~\ref{fig:cosmos:02} and Fig.~\ref{fig:cosmos:07} Inset, the flow in the vicinity of $I_3$ is expected to be repelled along the $z_0$-axis (which corresponds to the natural exit from ESHU) and attracted along the $z_1$-axis~\footnote{The $z_1$-axis corresponds to the circular arc in Fig.~\ref{fig:cosmos:02}. This expectation hinges on the eigenvectors being precisely aligned with the coordinate axes. If such an alignment is absent, it may be sufficient to require only that $I_3$ be a saddle to realize ESHU.}. Therefore, we require $\lambda_{I_3,+}>0$. The special case $\lambda_{I_3,+}=0$ is deferred to a later discussion. To determine the stability along the $z_1$-direction, it may be appropriate to try centre manifold theory. Building on the calculation in Appendix~\ref{sec:App.A.2}, the centre manifold for Eq.~(\ref{eq:3.25}) can be written in the form
\begin{equation}\label{eq:3.26}
  z_0 = \phi\big(z_1-\frac{1}{\tilde{c}_{35}}\big)
  = \sum_{i=2}^{\infty} \alpha_i \cdot \big(z_1-\frac{1}{\tilde{c}_{35}}\big)^i,
\end{equation}
where the series expansion encodes the boundary conditions and $\alpha_i$ are constant coefficients. Differentiating the above ansatz with respect to $\sigma$ and using Eq.~(\ref{eq:3.25}) to eliminate $\dx z_i/\dx\sigma$ yields an ordinary differential equation for $\phi$. This equation can be solved by a series expansion method. However, the solution is $\alpha_i=0$ for all $i$. Considering Eq.~(\ref{eq:3.25a}), this implies that no nonzero leading term exists along the centre manifold, so the centre manifold theory cannot be used to determine the stability along the $z_1$-axis. This may be viewed as a consequence of the fact that $\dx z_i/\dx\sigma=0$ once $z_0=0$.

The eigenvectors identified above are aligned with the coordinate axes, which permits the stability along the $z_1$-axis to be analyzed independently. In this analysis, the limit $z_0\rightarrow0^{+}$ can be adopted. Under this approximation, Eq.~(\ref{eq:3.25a}) effectively reduces to
\begin{equation}
  \lim_{z_0\rightarrow0^+}\frac{1}{z_0}\frac{\dx z_1}{\dx\sigma} = \frac{1}{\tilde{c}_2}(1 - \tilde{c}_{35}z_1),
\end{equation}
in which the left-hand side corresponds to a further time rescaling $\dx\sigma\rightarrow z_0\dx\sigma$, whereas the right-hand side can be handled using linear stability theory. The equilibrium $z_1=1/\tilde{c}_{35}$ in the above 1D dynamical system corresponds to $I_3$ in Eq.~(\ref{eq:3.25}). This equilibrium is an attractor only if the eigenvalue $-\tilde{c}_{35}/\tilde{c}_2<0$. The trajectories near $I_3$ along the arc in Fig.~\ref{fig:cosmos:02} provide additional support for the reasonableness of the above discussion.

In summary, aside from boundary cases, the conditions for $I_3$ to be a saddle with repulsion in the $z_0$ direction are $\lambda_{I_3,+}>0$ and $-\tilde{c}_{35}/\tilde{c}_2<0$. It will be shown that this corresponds to the ESHU dynamics with $n=3$. For fixed $w_\subF$, the corresponding parameter region can be plotted in the $(c_3,c_5)$ plane and classified according to $c_2\gtrless2$, as illustrated by the $n=3$ \& $w_\subF=1/3$ result in Fig.~\ref{fig:cosmos:07}. The boundaries admit two simple expressions $\tilde{c}_{35}=0$ and $c_3+\tilde{c}_{35}w_\subF=0$. Note that, as discussed at the beginning of this subsection, a successful exit from ESHU also requires $F_1$ to be stable. To clearly present the parameter dependence of the $I_3$ stability, Fig.~\ref{fig:cosmos:01} is not superimposed on Fig.~\ref{fig:cosmos:07}. Instead, the parameter points used in Fig.~\ref{fig:cosmos:08} are chosen to satisfy the constraints from both $I_3$ and $F_1$.

In the above analysis, $\lambda_{I_3,+}$ controls the speed at which the system evolves from infinity toward the finite region. This is a key indicator for describing the ESHU dynamics. The limiting case $\lambda_{I_3,+}=0$ may have more fundamental consequences and merits a detailed investigation. The stability of $I_3$ in this case is analyzed below. Solving $\lambda_{I_3,+}=0$ gives $c_3=2c_5w_\subF/(1+w_\subF)$. Under this parameter constraint, Eq.~(\ref{eq:3.25}) can be transformed into $\{\dx z_i/\dx\tilde{\sigma}\}$ with a further time rescaling $\dx\sigma\rightarrow\dx\tilde{\sigma}=z_0\dx\sigma$. Crucially, $I_3$ remains a critical point of the rescaled system, so standard mathematical tools can be applied. Applying linear stability theory to the rescaled system $\{\dx z_i/\dx\tilde{\sigma}\}$ yields the Jacobian matrix
\begin{equation}
  \tilde{J}|_{I_3} =
  \frac{1+w_\subF}{\tilde{c}_2 c_5 }
  \begin{bmatrix}
    \displaystyle\frac{2 c_5^2}{(1+w_\subF)^2}  &  -\displaystyle\frac{3}{4} (2+c_2+\tilde{c}_2 w_\subF)  \\[3mm]
    \displaystyle\frac{(c_2-\tilde{c}_2w_\subF) c_5^2}{(1 + w_\subF)^2}  &  1-2c_2
  \end{bmatrix}.
\end{equation}
In general, $\tilde{J}|_{I_3}$ admits two nonzero eigenvalues, denoted by $\tilde{\lambda}_{I_3,\pm}$. Since the associated eigenvectors are no longer aligned with the coordinate axes, the stability requirement is reduced to demanding that $I_3$ be a saddle. For a 2D system, this is equivalent to $\tilde{\lambda}_{I_3,+}\cdot\tilde{\lambda}_{I_3,-}=\det\tilde{J}|_{I_3}<0$. A direct calculation shows that this inequality does not depend on $c_5$. For the range of $w_\subF$ of interest, this yields $0<\tilde{c}_2<2(1/3+w_\subF)/(1-w_\subF^2)$ if $-1/3<w_\subF<1$. This case falls into the $n=2$ ESHU class (see Sec.~\ref{sec:III.C.2}). The corresponding parameter region for $w_\subF=1/3$ is plotted in Fig.~\ref{fig:cosmos:07}. Hereafter, $\tilde{\lambda}_{I_3,+}$ denotes the positive one.

If one further imposes $\tilde{\lambda}_{I_3,+}=0$ while maintaining $I_3$ as a saddle, the system can evolve more slowly from $I_3$ to $F_1$, which may enable an $n=1$ realization of the ESHU dynamics. The first step is still to assess whether any parameter choices satisfy these requirements. Solving the equation gives $\tilde{c}_2 = 2(1/3+w_\subF)/(1-w_\subF^2)$, which is the boundary of the previous parameter space. One can verify that $\tilde{\lambda}_{I_3,-}\neq0$ generically in this case. Substituting this $c_2$ solution into the previous $\{\dx z_i/\dx\tilde{\sigma}\}$, we obtain
\begin{subequations}\label{eq:3.29}
\begin{align}
  \frac{\dx z_1}{\dx\tilde{\sigma}} &= \frac{3(1-w_\subF)}{2-6w_\subF} \cdot \frac{\dx z_0}{\dx\tilde{\sigma}}, \\
  \frac{\dx z_0}{\dx\tilde{\sigma}} &= \frac{1-3w_\subF}{1+3w_\subF} \cdot \big[ (1+w_\subF)(1-3z_0^2) \nonumber\\
    &\qquad + 2c_5z_1 + (5+3w_\subF)z_1z_0\big].
\end{align}
\end{subequations}
As set up above, $I_3$ remains a critical point of Eq.~(\ref{eq:3.29}), and its associated eigenvalues consist of one zero eigenvalue and one nonzero eigenvalue. Therefore, it is natural to proceed using centre manifold theory. However, the analysis shows that the flow along the centre manifold vanishes identically~\cite{FootnoteEq329}. As a result, the system cannot evolve from $I_3$ toward $F_1$ \footnote{This conclusion applies to the case of constant $w_\subF$. A time-varying $w_\subF$ could lead to different dynamical behaviour and deserves separate study.}, and the $n=1$ dynamics cannot be realized. The phase portrait corroborates this conclusion \cite{NotShown}. Nonzero $c_4$ and $c_6$ can avoid this issue, and will be discussed in Sec.~\ref{sec:III.C.3}.

\subsubsection{Inverse power-law CHR}\label{sec:III.C.2}
As pointed out at the beginning of Sec.~\ref{sec:III.C}, the static limit $H\rightarrow0$ implies an extremely large CHR. Building on the stability analysis, this subsubsection aims to quantify the evolution of the CHR during ESHU and to assess whether the resulting causally connected region is sufficient to address the horizon problem. For each case of $n$, both analytical and numerical results are presented. A simple analytic expression can make the parameter dependence of the physical conclusions explicit. Quantitative characterization of the CHR is also useful for future analyses of perturbations, in particular their evolution around horizon crossing.

The following conventions are adopted for the variables and reference time (\textit{e}-folding number) points. The CHR is defined as $r_\subH\equiv c/(aH)$, and the comoving size of the causally connected region that can form between $N_1$ and $N_2$ is $r(N_1,N_2)=\int_{N_1}^{N_2}r_\subH\,\dx N$ \cite{Dodelson2020.book}. Physically, ESHU is embedded in the radiation epoch. For a numerical evolution, the initial conditions are set at $N_\subini$, where $\Omega_\subF=\mathcal{O}(1)$. Once the system enters the ESHU regime, $\Omega_\subF$ grows toward infinity ($I_3$), and subsequently decreases back to unity ($F_1$). The start and end of the ESHU phase are defined as the times when $\Omega_\subF=3$ (i.e., $x_3=3$)~\footnote{This value is essentially arbitrary and does not affect the subsequent calculations or the physical conclusions.}, denoted by $N_\subSsta$ and $N_\subSend$, respectively. The time at which $\Omega_\subF$ reaches its maximum ($x_{3,\submax}^2/3$) is denoted by $N_\subSmax$. To facilitate the description of the $I_3\rightarrow F_1$ attractor, it is necessary to introduce $N_\subSinf$, at which $\Omega_\subF=\infty$ in the approximate solution. The times at which $r_\subH$ reaches its local maximum and, subsequently, a local minimum are denoted by $N_\subRmax$ and $N_\subRmin$, respectively. Intuitively, $N_\subSmax\approx N_\subSinf\approx N_\subRmax$, whereas $N_\subSend$ can be much larger than $N_\subRmin$, since the CHR depends on $N$ not only via $x_3$ but also through additional time-dependent factors.

In order to solve the horizon problem, one should require $r(N_\subSsta,N_{\rm dec})\gg r(N_{\rm dec},0)$ \cite{Bassett2006.RMP.78.537}, where $N_{\rm dec}$ and $0$ correspond to the time of CMB decoupling and today, respectively. During inflation, the integral can be approximately carried out as the functional form of $r_\subH(N)$ is well-established. The late-time Universe admits an analogous treatment, which yields $r(N_{\rm dec},0)=\mathcal{O}(1)\times c/(a_0H_0)$. For ESHU, the evolution over $[N_\subSsta,N_\subSmax]$ depends strongly on the initial conditions, and may render the solution complicated. By contrast, the evolution over $[N_\subSmax,N_\subSend]$ is well captured by the $I_3\rightarrow F_1$ attractor (see Fig.~\ref{fig:cosmos:02}), which can suppress the sensitivity to initial conditions and may therefore allow a concise result. The contribution from the standard radiation-dominated era, i.e., $r(N_\subSend,N_{\rm dec})$, is negligible. Accordingly, the analysis below focuses first on the $I_3\rightarrow F_1$ attractor, while the contribution from $[N_\subSsta,N_\subSmax]$ is just illustrated in a later numerical calculation.

Equation~(\ref{eq:3.25}) serves as the starting point for computing $r_\subH$. Since $H$ does not enter explicitly, the relevant physical quantities should be reconstructed from the solution $x_i(N)$. The scale factor follows directly from the definition of \textit{e}-folding number, yielding $a/a_\subini=\exp(N-N_\subini)$. For the Hubble parameter, the definition of $x_3$ gives $H = \varrho_\subF/x_3$. We consider only the radiation fluid ($w_\subF=1/3$), so $\varrho_\subF\propto\exp(-2N)$. Evaluating these relations at the initial time and forming the ratio $H/H_\subini$ then leads to
\begin{equation}\label{eq:3.30}
  H = H_\subini \cdot \frac{x_{3,\subini}}{x_3} \cdot \exp[-2(N-N_\subini)].
\end{equation}
Here, $H_\subini$ sets the overall energy scale of the system and its value is independent of the solution $x_i(N)$. Given a solution for $x_3(N)$, $H$ can be reconstructed from Eq.~(\ref{eq:3.30}), after which $r_\subH$ follows straightforwardly. This reconstruction applies to both approximate and exact solutions.

\begin{figure}[!t]
  \centering
  \includegraphics[width=0.47\textwidth]{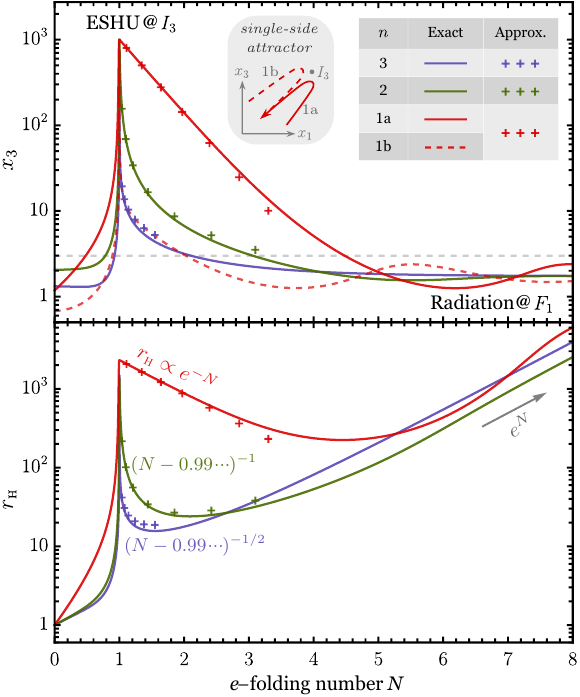}
  \caption{ESHU dynamics obtained by integrating Eq.~(\ref{eq:3.6}) for $w_\subF=1/3$, shown as $x_3(N)$ and $r_\subH(N)$. The critical points $I_3$ and $F_1$ play an important role in shaping the evolution. The index $n$ provides a suitable classification of the post-peak evolution, and the key features of $r_\subH$ near the peak are labeled in the figure. Model parameters are given in the main text, and initial conditions are summarized in Table~\ref{tab:05}. Without loss of generality, we set $c=1$, $a_\subini=1$, $H_\subini=1$ and $N_\subini=0$ in this figure. Solid and dashed curves denote the full numerical (exact) solutions, while plus symbols denote the approximate solutions given by Eqs.~(\ref{eq:3.32}), (\ref{eq:3.36}), (\ref{eq:3.46}) and the corresponding $r_\subH$ reconstructions. In most cases, the approximate solution matches the full numerical evolution well, underscoring the utility of the $I_3\rightarrow F_1$ attractor. However, case 1b is an exception. The inset illustrative phase portrait offers a plausible explanation for this mismatch: the attraction on the 1b-side is comparatively weak.}
  \label{fig:cosmos:08}
\end{figure}

In the first case analyzed in Sec.~\ref{sec:III.C.1}, $\lambda_{I_3,+}>0$ and the associated eigenvector is aligned with the $z_0$-axis, so the leading-order dynamics of the $I_3\rightarrow F_1$ attractor (near $I_3$) is governed by $\dx z_0/\dx\sigma = \lambda_{I_3,+}\cdot z_0$. Using the relevant variable transformations, it follows that
\begin{equation}
  x_3^\prime = - \lambda_{I_3,+} \cdot x_3^3.
\end{equation}
The power-law exponent shows that this case corresponds to $n=3$. A general solution for $x_3>0$ is  
\begin{equation}\label{eq:3.32}
  x_3 = \frac{1}{\sqrt{2\lambda_{I_3,+}\cdot(N-N_\subSinf)}},
\end{equation}
where the integration constant $N_\subSinf$ denotes the \textit{e}-folding number at which $x_3$ diverges. In practice, given the full numerical solution, $N_\subSinf$ can be \textit{fitted} by matching at the maximum, i.e., by substituting $(N_\subSmax,x_{3,\submax})$ into Eq.~(\ref{eq:3.32}). As will be seen in Fig.~\ref{fig:cosmos:08}, this is a sufficiently good approximation. This fitting procedure implies that $N_\subSinf$ is slightly smaller than $N_\subSmax$. Combining the above results, we obtain
\begin{equation}\label{eq:3.33}
  r_\subH = \frac{c}{\sqrt{2\lambda_{I_3,+}}(aH x_3)|_\subini} \cdot \frac{e^{N-N_\subini}}{\sqrt{N-N_\subSinf}}.
\end{equation}
Near $N_\subSinf$, $r_\subH$ is approximately governed by an inverse power law (see the denominator of Eq.~(\ref{eq:3.33})) and can diverge. This is the origin of the title of this subsubsection. Using the relevant definitions, a direct calculation yields $N_\subSend=N_\subSinf+1/(18\,\lambda_{I_3,+})$ and $N_\subRmin=N_\subSinf+1/2$. To obtain a larger causally connected region, one should require $N_\subSend>N_\subRmin$, i.e., a smaller $\lambda_{I_3,+}$. However, roughly speaking, $\lambda_{I_3,+}$ cannot be much smaller than $10^{-3}$, because the real Universe needs to enter the standard hot Big Bang phase before primordial nucleosynthesis \cite{Pitrou2018.PhysRep.754.1,Tian2022.PRD.106.043524}. We can write $r(N_\subSmax,N_\subSend) = r(N_\subSmax,N_\subRmin) + r(N_\subRmin,N_\subSend)$. In the second term, the integrand is dominated by the $e^N$ factor in the numerator of Eq.~(\ref{eq:3.33}). As a result, the integral differs from the standard radiation-dominated estimate only by an $\mathcal{O}(1)$ factor. This contribution is therefore negligible for the horizon problem. In the first term, the denominator $\sqrt{N-N_\subSinf}$ in Eq.~(\ref{eq:3.33}) is crucial. This integral can be evaluated approximately, yielding
\begin{equation}\label{eq:3.34}
  r(N_\subSmax,N_\subRmin) \approx \frac{c\,e^{N_\subSmax-N_\subini}}{(a H x_3)|_{\subini}} \cdot \frac{1.19}{\sqrt{\lambda_{I_3,+}}}.
\end{equation}
This result remains finite even in the limit $N_\subSmax\rightarrow N_\subSinf$ ($r_{\subH,\submax}\rightarrow\infty$). The finiteness follows from the convergence of the integral $\int_{0}^{1}x^{-1/2}\,\dx x$. For the parameters of interest, $r(N_\subSmax,N_\subRmin)\ll r(N_{\rm dec},0)$. In summary, these results suggest that the $n=3$ case is insufficient to resolve the horizon problem~\footnote{As shown in Fig.~\ref{fig:cosmos:08}, the contribution from $[N_\subSsta,N_\subSmax]$ is smaller than that from $[N_\subSmax,N_\subSend]$, and thus is also negligible.}. The key features of this case are summarized in Table~\ref{tab:04}.

\begin{table*}[!t]
  \caption{Key features of the ESHU dynamics along the $I_3\rightarrow F_1$ attractor (see Fig.~\ref{fig:cosmos:02} for an illustration). The first column classifies the leading-order dynamics by the power-law exponent $n$. The second and fifth columns list the main requirements for realizing ESHU and for solving the horizon problem, respectively. The third and fourth columns list the late-ESHU duration and the CHR-decreasing duration, respectively. The fourth column also specifies whether the CHR decreases as a power law or exponentially. The last column highlights the main challenges associated with the ESHU realization within GCT.}
  \label{tab:04}
  \begin{tabular*}{\hsize}{@{\ \ }@{\extracolsep{\fill}}cccccl@{\ \ }}
    \hline\hline
    Exponent $n$  &  ESHU realization  &  Late-ESHU duration  &  CHR {\fontsize{8pt}{0pt}\selectfont$\searrow$} duration  &  Horizon problem  &  \multicolumn{1}{c}{\multirow{2}{*}{Challenges in GCT}}  \\[-0.3mm]
    $x_3^\prime=-\lambda\,x_3^n$  &  requirements  &  $N_\subSend-N_\subSmax$  &  $N_\subRmin-N_\subRmax$  &  requirements  \\[0.2mm]
    \hline
    \multirow{2}{*}{$3$}  &  \multirow{2}{*}{$\lambda_{I_3,+}>0$}  &  \multirow{2}{*}{$(18\,\lambda_{I_3,+})^{-1}$}  &  $1/2$  &  \multicolumn{2}{c}{\multirow{2}{*}{Failure to solve the horizon problem}}  \\[-0.2mm]
      &  &  &  (power law)  &  \\[0.2mm]
    \hline
    \multirow{2}{*}{$2$}  &  \multirow{2}{*}{$\tilde{\lambda}_{I_3,+}>0$}  &  \multirow{2}{*}{$(3\,\tilde{\lambda}_{I_3,+})^{-1}$}  &  $1$  &  \multirow{2}{*}{$x_{3,\submax}\gtrsim e^{e^{70}}$}  &  Too large ini condition  \\[-0.2mm]
      &  &  &  (power law)  &   &  Not robust if $\dot{w}_\subF\neq0$  \\[0.2mm]
    \hline
    \multirow{2}{*}{$1$}  &  \multirow{2}{*}{$\beta_\subcm>0$}  &  \multirow{2}{*}{$\displaystyle\beta_\subcm^{-1}\cdot\log(x_{3,\submax}/3)$}  &  Same as left  &  $x_{3,\submax}\gtrsim e^{70}$  &  \multirow{2}{*}{Not robust if $\dot{w}_\subF\neq0$}  \\[-0.2mm]
      &  &  &  ($\beta_\subcm>1$; $\mathrm{exp.}$)  &  ($\beta_\subcm>1$)  &  \\[0.2mm]
    \hline\hline
  \end{tabular*}
\end{table*}

In the second case analyzed in Sec.~\ref{sec:III.C.1}, $\lambda_{I_3,+}=0$, $\tilde{\lambda}_{I_3,+}>0$ and $\tilde{\lambda}_{I_3,-}<0$. In general, the associated eigenvector is not aligned with the coordinate axes. Nevertheless, the leading-order dynamics along the $I_3\rightarrow F_1$ attractor can be written as $\dx\mathbf{v}/\dx\tilde{\sigma} = \tilde{\lambda}_{I_3,+}\mathbf{v}$, where $\mathbf{v}$ is a displacement vector based at $I_3$ and directed along the eigenvector associated with $\tilde{\lambda}_{I_3,+}$. Projecting $\mathbf{v}$ onto the $z_0$-axis gives $\dx z_0/\dx\tilde{\sigma} = \tilde{\lambda}_{I_3,+}\cdot z_0$, in analogy with the result obtained in the first case. Applying the corresponding variable transformations yields
\begin{equation}
  x_3^\prime = - \tilde{\lambda}_{I_3,+} \cdot x_3^2.
\end{equation}
This scaling corresponds to the $n=2$ case. The general solution on the branch $x_3>0$ reads
\begin{equation}\label{eq:3.36}
  x_3 = \frac{1}{\tilde{\lambda}_{I_3,+} \cdot (N-N_\subSinf)},
\end{equation}
which in turn gives
\begin{equation}
  r_\subH = \frac{c}{\tilde{\lambda}_{I_3,+}(aHx_3)|_\subini} \cdot \frac{e^{N-N_\subini}}{N-N_\subSinf}.
\end{equation}
Near $N_\subSinf$, $r_\subH$ diverges as a power law with exponent $-1$. Based on the definitions, the above two equations yield $N_\subSend=N_\subSinf+1/(3\,\tilde{\lambda}_{I_3,+})$ and $N_\subRmin=N_\subSinf+1$. Considering primordial nucleosynthesis, $\tilde{\lambda}_{I_3,+}$ should not be much smaller than $10^{-2}$. It is natural to set $\tilde{\lambda}_{I_3,+}=\mathcal{O}(0.1)$, which ensures $N_\subSend\gtrsim N_\subRmin$ and does not entail fine-tuning. As suggested by the discussion above Eq.~(\ref{eq:3.34}), the key to ESHU solving the horizon problem hinges on the evolution from $N_\subSmax$ to $N_\subRmin$. The comoving size of the causally connected region accumulated over this interval is approximately given by
\begin{equation}\label{eq:3.38}
  r(N_\subSmax,N_\subRmin) \approx \frac{c\,e^{N_\subSmax-N_\subini}}{(a H x_3)|_{\subini}} \cdot \frac{\log(x_{3,\mathrm{max}})}{\tilde{\lambda}_{I_3,+}}.
\end{equation}
Unlike Eq.~(\ref{eq:3.34}), increasing $x_{3,\submax}$ in Eq.~(\ref{eq:3.38}) can lead to an arbitrarily large causally connected region. For a conservative estimate of $x_{3,\submax}$, we can assume that ESHU takes place at the Planck scale and $e^{N_\subSmax-N_\subini}/(\tilde{\lambda}_{I_3,+}\, x_{3,\subini}) \approx \mathcal{O}(1)$. Standard hot Big Bang theory gives $a_0H_0/(aH)|_{\subini}\approx e^{-70}$. It then follows that, in this case, resolving the horizon problem requires $\log(x_{3,\mathrm{max}})>e^{70}$. Raising the energy scale of ESHU is permissible, but it can only slightly relax this requirement. The above results are summarized in Table~\ref{tab:04}.

In the $n=2$ case, the extremely large $x_{3,\submax}$ required above comes with two caveats. One concern is that reaching such $x_{3,\submax}$ may demand extreme initial data, typically $x_3\gtrsim\mathcal{O}(1)$ with $x_1\gg1$ \footnote{We have not yet demonstrated this conclusion rigorously in numerical calculations due to numerical instabilities. Preliminary calculations \cite{NotShown} suggest that, for the illustrative trajectories shown as the orange dashed curve in Fig.~\ref{fig:cosmos:02}, an extremely large $x_{3,\submax}$ may be obtained with $x_{3,\subini}\gtrsim\mathcal{O}(1)$ and $x_{1,\subini}\gg1$.}. This does not invalidate the physical scenario, but can make the realization less natural. A second concern is robustness. Since the condition $\lambda_{I_3,+}=0$ depends on $w_\subF$, time variation of $w_\subF$ in a realistic early Universe can shift $\lambda_{I_3,+}$ slightly away from zero (see Eq.~(\ref{eq:3.13}) for an example). An extremely large $x_3$ may then lead to $\lambda_{I_3,+}\,x_3^3\gg \tilde{\lambda}_{I_3,+}\,x_3^2$, thereby obstructing an $n=2$ realization within GCT. Without a suitable mechanism to mitigate these two issues, the $n=2$ realization in GCT may remain only a toy model. Possible answers will be explored in future work.

To validate the analytic approximations and to highlight the characteristic CHR behaviour during ESHU, we numerically integrate the system defined by Eq.~(\ref{eq:3.6}) with $w_\subF=1/3$. For the $n=3$ case, the parameters are chosen as $\{c_2=1,\,c_3=-2,\,c_5=-3.5,\,c_4=0,\,c_6=0\}$, which gives $\lambda_{I_3,+}=1/30$. For the $n=2$ case, $\{c_2=1,\,c_3=-1,\,c_5=-2,\,c_4=0,\,c_6=0\}$ is adopted, which enforces $\lambda_{I_3,+}=0$ and yields $\tilde{\lambda}_{I_3,+}\approx0.1350$. These two parameter points are marked in Fig.~\ref{fig:cosmos:07} as \textit{Fig.~8\,(3)} and \textit{Fig.~8\,(2)}, and both ensure that $I_3$ is a saddle and $F_1$ is stable. Without loss of generality, the numerical integrations start from $N_\subini=0$. The initial conditions are fine-tuned so that the maximum satisfies $x_3(1)=1000$, and the values are summarized in Table~\ref{tab:05}. This adjustment is used solely to facilitate comparisons among different solutions and does not represent a fine-tuning problem. In all cases, $\Omega_{\subF,\subini}=\mathcal{O}(1)$. For each numerical evolution, the value of $N_\subSinf$ is \textit{fitted} using this maximum (see the discussion below Eq.~(\ref{eq:3.32})), while the remaining reference times follow directly from their definitions. The sharp peak renders the system numerically stiff, and high-precision integration is adopted to alleviate this difficulty. This is why we do not provide a numerical result with $x_{3,\submax}\approx\exp(\exp(70))$.

\begin{table*}[!t]
  \caption{Numerical summary of the ESHU evolutions shown in Fig.~\ref{fig:cosmos:08}. The second column lists the initial conditions, which are adjusted only to set $(N_\subSmax,x_{3,\submax})=(1,1000)$ for direct visual comparison. The remaining columns report the key time points. The ``Approx." denotes the analytic approximation summarized in Table~\ref{tab:04}. Cases 1a and 1b both correspond to $n=1$, but the approximation does not match the full numerical solution in case 1b.}
  \label{tab:05}
  \begin{tabular*}{\hsize}{@{\ \ }@{\extracolsep{\fill}}llllllll@{}}
    \hline\hline
    \multirow{2}{*}{$n$}  &  $x_{1,\subini}$  &  \multirow{2}{*}{$N_\subSsta$}  &  \multirow{2}{*}{$N_\subSinf$}  &  \multirow{2}{*}{$N_\subSmax$}  &  $N_\subSend$  &      \multirow{2}{*}{$N_\subRmax$}  &  $N_\subRmin$  \\[-1mm]
      &  $x_{3,\subini}$  &  &  &  &  Approx.  &  &  Approx.  \\
    \hline
    \multirow{2}{*}{3}   &  $16.789727023863257240$  &  \multirow{2}{*}{0.92476}  &  \multirow{2}{*}{$0.99999$}  &  \multirow{2}{*}{$1.00000$}  &  $2.10458$  &  \multirow{2}{*}{$1.00000$}  &  $1.55028$  \\
                         &  $1.3155320024181952471$  &  &                               &                              &  $2.66667$  &                              &  $1.50000$  \\
    \hline
    \multirow{2}{*}{2}   &  $8.7227872750560824514$  &  \multirow{2}{*}{0.78354}  &  \multirow{2}{*}{$0.99259$}  &  \multirow{2}{*}{$1.00000$}  &  $3.05853$  &  \multirow{2}{*}{$1.00000$}  &  $2.08947$  \\
                         &  $2.0496807814811174160$  &  &                               &                              &  $3.46837$  &                              &  $2.00000$  \\
    \hline
    \multirow{2}{*}{1a}  &  $0.3283067146551032224$  &  \multirow{2}{*}{0.47965}  &  \multirow{2}{*}{---}         &  \multirow{2}{*}{$1.00000$}  &  $4.47704$  &  \multirow{2}{*}{$1.00121$}  &  $4.44509$  \\
                         &  $1.1781560878228073809$  &  &                               &                              &  $3.90458$  &                              &  $3.90579$  \\
    \hline
    \multirow{2}{*}{1b}  &  \hspace{-0.78em}$-0.1284457614150932353$  &  \multirow{2}{*}{0.89147}  &  \multirow{2}{*}{---}  &  \multirow{2}{*}{$1.00000$}  &  $2.02516$  &  \multicolumn{2}{l}{\ NOT exponentially}  \\
                         &  $0.6761412921129327446$                   &  &                        &                              &  $3.90458$  &  \multicolumn{2}{l}{\ \ \ decreasing CHR}  \\
    \hline\hline
  \end{tabular*}
\end{table*}
 
Figure~\ref{fig:cosmos:08} shows the evolution of $x_3(N)$ and $r_\subH(N)$, together with the approximate solutions. The physical picture of ESHU is clearly illustrated: the Universe begins in a radiation-like expanding state with $\Omega_\subF\approx\mathcal{O}(1)$, rapidly approaches $I_3$ as $x_3\rightarrow\infty$ and $H\rightarrow0$, and is then drawn toward $F_1$, thereby recovering the standard radiation-dominated evolution. The exponent $n$ serves primarily to characterize the post-peak dynamics. For both the $n=3$ and $n=2$ cases, the analytic approximations in Eqs.~(\ref{eq:3.32}) and (\ref{eq:3.36}) match the full numerical solutions well across the late-ESHU epoch~\footnote{In light of the mismatch between the analytic approximation and the numerical evolution in case 1b discussed later, it is necessary to point out that the $n=3$ and $n=2$ examples displayed here are chosen on the same side in the phase portrait as case 1a, corresponding to the orange dashed curve in Fig.~\ref{fig:cosmos:02}. The opposite side has not been tested for these parameter points.}. The agreement carries over to the reconstructed CHR, which exhibits an inverse power-law evolution after the peak. Table~\ref{tab:05} lists the key reference times for both evolutions. The corresponding analytic estimates for $N_\subSend$ and $N_\subRmin$ are also included to demonstrate consistency. For the pre-peak evolution, the numerical results indicate that the rise of $r_\subH$ is too sharp to accumulate a sizable causally connected region. It is therefore justified to focus on the post-peak dynamics when assessing the horizon scale generated during ESHU.

\subsubsection{Exponentially decreasing CHR}\label{sec:III.C.3}
Section~\ref{sec:III.C.1} shows that, for $c_4,c_6=0$, further imposing $\tilde{\lambda}_{I_3,+}=0$ leads to $\dx u_c/\dx\tilde{\sigma}=0$ \cite{FootnoteEq329}. This freezes the centre-manifold flow and hence obstructs a natural exit from the ESHU, rendering such a realization physically unacceptable. We therefore turn to the case $c_4,c_6\neq0$. A full stability analysis is not pursued, and the goal is instead to present representative examples that establish the existence of a realization beyond the $n=3$ and $n=2$ cases. First of all, what is the key feature of the ESHU dynamics when $\tilde{\lambda}_{I_3,+}=0$ but $\dx u_c/\dx\tilde{\sigma}\neq0$? The linear term vanishes, and the leading centre-manifold evolution is expected to be quadratic, i.e., $\dx u_c/\dx\tilde{\sigma}\propto u_c^2$. Projecting $u_c$ onto the $z_0$-axis and accounting for the variable transformations yields $x_3^\prime\propto-x_3$, which corresponds to $n=1$. It will be shown that, in this case, the CHR can decay exponentially after reaching its peak. This motivates the title of the present subsubsection.

The analysis is based on the dynamical system (\ref{eq:3.6}), with free parameters $\{c_2,c_3,c_4,c_5,c_6\}$ and $w_\subF$. Under the Poincar\'e compactification, the equilibria at infinity lie on the circle $y_1^2+y_3^2=1$, with $y_i$ defined in Sec.~\ref{sec:III.A.2}. Applying the theorems summarized in Appendix~\ref{sec:App.A.3}, the coordinates can be fixed by an additional condition $y_3/y_1 = (\tilde{c}_{35} \pm \sqrt{\tilde{c}_{35}^2-4\tilde{c}_{46}})/2$. Consequently, if $\tilde{c}_{46}\neq0$ and $\tilde{c}_{35}^2-4\tilde{c}_{46}>0$, there are generically two such equilibria in the upper half-plane $y_3>0$. The bottom phase portrait in Fig.~\ref{fig:cosmos:04} corroborates this result. In the limit $\tilde{c}_{46}\rightarrow0$, one of the two equilibria approaches the $y_1$-axis, corresponding to $I_1$ or $I_2$, while the other tends to $I_3$ listed in Table~\ref{tab:02}. The following discussion focuses on the latter equilibrium in the ESHU context. For simplicity, it is still denoted by $I_3$ throughout. Using the same variable transformations as in Eq.~(\ref{eq:3.25}), we obtain $\{\dx z_i/\dx\sigma\}$ for nonzero $c_4$ and $c_6$. Motivated by the above expression for $y_3/y_1$, it is convenient to introduce a new positive parameter $\alpha$, defined by $\tilde{c}_{35}^2-4\tilde{c}_{46}=\alpha^2\tilde{c}_{35}^2$. This parametrization allows $c_6$ to be traded for $\alpha$ via
\begin{equation}\label{eq:3.39}
  c_6 = c_4 + \frac{\alpha^2-1}{4}\tilde{c}_{35}^2.
\end{equation}
This replacement will substantially simplify the following zero-eigenvalue constraints. As a case study demonstrating existence of $n=1$, only the parameter range $\tilde{c}_{35}>0$ is considered. In the $c_4,c_6\rightarrow0$ limit, this corresponds to the lavender region in Fig.~\ref{fig:cosmos:07}. In the $z_i$ variables, $I_3$ is located at
\begin{equation}\label{eq:3.40}
  (z_1,z_0)|_{I_3} = \Big(\frac{2}{(1+\alpha)\tilde{c}_{35}},0\Big). 
\end{equation}
This step relies on $\sqrt{\alpha^2}=\alpha$, which makes $\alpha>0$ a necessary constraint for self-consistency in what follows.

A linear stability analysis of the system $\{\dx z_i/\dx\sigma\}$ at $I_3$ for nonzero $c_4$ and $c_6$ yields one zero eigenvalue and one nonzero eigenvalue, with the latter denoted by $\lambda_{I_3,+}$. The associated eigenvectors are aligned with the $z_1$ and $z_0$ axes, respectively, as in the $c_4,c_6=0$ case. Therefore, $\lambda_{I_3,+}$ controls the evolution of the system from infinity toward the finite region. For the present purpose, we should require $\lambda_{I_3,+}=0$, which yields a unique solution
\begin{equation}\label{eq:3.41}
  c_4 = \frac{1+\alpha}{4} \big[2c_3\tilde{c}_{35} + (1+\alpha)w_\subF \tilde{c}_{35}^2\big].
\end{equation}
Setting $c_4=0$ and $\alpha=1$ recovers the corresponding result obtained previously. Imposing $\lambda_{I_3,+}=0$ renders $I_3$ non-hyperbolic with two zero eigenvalues in the system $\{\dx z_i/\dx\sigma\}$, which motivates a further time rescaling $\dx\sigma\rightarrow\dx\tilde{\sigma}=z_0\dx\sigma$. Importantly, $I_3$ remains an equilibrium of the rescaled system $\{\dx z_i/\dx\tilde{\sigma}\}$, so standard dynamical-system methods remain applicable. Generically, the rescaled system has two nonzero eigenvalues at $I_3$, denoted by $\tilde{\lambda}_{I_3,\pm}$. The previous analysis indicates that $\tilde{\lambda}_{I_3,+}>0>\tilde{\lambda}_{I_3,-}$ corresponds to an $n=2$ realization. For $n=1$, we should require $\tilde{\lambda}_{I_3,+}=0$ while maintaining $\tilde{\lambda}_{I_3,-}<0$. A vanishing eigenvalue implies a vanishing Jacobian determinant, which admits a unique solution
\begin{equation}\label{eq:3.42}
  \alpha = \frac{3 (c_3 + w_\subF \tilde{c}_{35}) (2+c_2+w_\subF\tilde{c}_2)}{\tilde{c}_{35} (4 - 6 w_\subF  - 3 c_2 - 3\tilde{c}_2w_\subF^2)}.
\end{equation}
The above two equations provide constraints on $\{c_4,\alpha\}$, leaving $\{c_2,c_3,c_5\}$ as the remaining free parameters.

Having imposed the above two constraints, we should proceed with the centre manifold theory. Following Appendix~\ref{sec:App.A.2}, a new set of variables
\begin{widetext}
\begin{equation}\label{eq:3.43}
  \begin{pmatrix}
    u_h \\
    u_c 
  \end{pmatrix}
  =
  \begin{bmatrix}
    \displaystyle\frac{3(2+c_2+\tilde{c}_2w_\subF)}{8c_2-4}  &  
    \displaystyle\frac{(4-3c_2-6w_\subF-3\tilde{c}_2w_\subF^2)^2}{2(c_3+\tilde{c}_{35}w_\subF)[8c_3-4c_5+3(c_2c_5-c_3)(1-w_\subF)]}  \\
    1  &  1 
  \end{bmatrix}^{-1}
  \begin{pmatrix}
    z_1 - \displaystyle\frac{2}{(1+\alpha)\tilde{c}_{35}} \\
    z_0 
  \end{pmatrix}
\end{equation}
is introduced to diagonalize the corresponding Jacobian matrix (see Appendix~\ref{sec:App.A.2}). The eigenvectors defining this transformation are normalized so that the second row of the above matrix consists entirely of ones~\footnote{This normalization is arbitrary, and the present choice is made for convenience in the discussion below.}. The centre manifold is then represented as $u_h=\phi(u_c)=\sum_{i=2}^{\infty}\alpha_i u_c^i$, and the coefficients can be fixed by substituting the dynamical equations $\{\dx z_i/\dx\tilde{\sigma}\}$ into this ansatz (see Eq.~(\ref{eq:3.23}) for an example). This expansion encodes the standard boundary conditions that $\phi(u_c)$ and its first derivative vanish at $u_c=0$. Along the centre manifold, the leading dynamics is given by $\dx u_c/\dx\tilde{\sigma}\approx\beta_\subcm\cdot u_c^2$ \footnote{In principle, $\beta_\subcm$ does not depend on the explicit form of $\phi(u_c)$. To see this, start from $\dx u_c/\dx\tilde{\sigma} = g_c(u_h,u_c)$, which follows from Eq.~(\ref{eq:A.05}), and expand $g_c(u_h,u_c)$ near the origin. The first nonzero terms are quadratic, of the form $u_h^2$, $u_h u_c$, and $u_c^2$. Since the centre-manifold expansion implies $u_h\propto u_c^2$ near the origin, the terms involving $u_h$ are higher order. Therefore, $\phi(u_c)$ does not enter the leading term of $g_c(\phi(u_c),u_c)$.}, where, for $w_\subF=1/3$, 
\begin{equation}\label{eq:3.44}
  \beta_\subcm = \frac{(1-2c_2) (3c_3-\tilde{c}_2c_5) [(1-2c_2)^2-3(2c_3-c_5)^2]}{(2c_3-c_5) \{4(1-2c_2)^3 + 3(4+c_2)[6c_3^2+\tilde{c}_2c_5^2-(7-2c_2)c_3c_5]\}}.
\end{equation}
The inverse transformation of Eq.~(\ref{eq:3.43}) can be used to project $u_c$ onto the $z_0$-axis. The second row of the transformation matrix gives $z_0=u_h+u_c$. Near $I_3$, the centre-manifold expansion implies $u_h\propto u_c^2$, hence $z_0\approx u_c$ \footnote{Note that the eigenvector associated with $\tilde{\lambda}_{I_3,+}=0$ in the system $\{\dx z_i/\dx\tilde{\sigma}\}$ is not aligned with the $z_0$-axis. The approximation $z_0\approx u_c$ reflects the relative scaling between the $u_c$ and $z_0$ coordinates, since the transformation matrix in Eq.~(\ref{eq:3.43}) includes a stretching.} along the centre manifold. The reduced evolution then takes the form $\dx z_0/\dx\tilde{\sigma}\approx\beta_\subcm\cdot z_0^2$. The system is expected to evolve from $I_3$ to $F_1$ along the centre manifold, which requires $\beta_\subcm>0$.
\end{widetext}

In summary, the ESHU dynamics with $n=1$ can be realized by imposing $\{\tilde{c}_{35}>0,\,\alpha>0,\,\tilde{\lambda}_{I_3,-}<0,\,\beta_\subcm>0\}$ in conjunction with Eqs.~(\ref{eq:3.39}), (\ref{eq:3.41}) and (\ref{eq:3.42}). The first inequality is a condition for case-by-case classification, rather than a stability requirement. A complete analysis would relax this restriction and treat the remaining case separately, which is beyond the scope of this paper. The second inequality is imposed for self-consistency (see Eq.~(\ref{eq:3.40})), whereas the remaining two inequalities encode the stability requirements. For $w_\subF=1/3$, the parameter region satisfying the above criteria is shown in the right panel of Fig.~\ref{fig:cosmos:07}. The three blue lines delineate the boundaries of the slices at $c_2=-1,1,3$, respectively. Note that the stability requirement for $F_1$ is not imposed in this figure.

To quantify the $n=1$ ESHU dynamics, we begin by analysing the $I_3\rightarrow F_1$ attractor. Combining the centre-manifold flow $\dx z_0/\dx\tilde{\sigma}\approx\beta_\subcm\cdot z_0^2$ derived above with the corresponding variable transformations yields
\begin{equation}
  x_3^\prime = - \beta_\subcm\cdot x_3.
\end{equation}
A general solution and the corresponding CHR read
\begin{align}
  x_3 &= x_{3,\submax} \cdot e^{-\beta_\subcm\cdot(N-N_\subSmax)}, \label{eq:3.46}\\
  r_\subH &= \frac{c\,e^{N_\subSmax-N_\subini}}{(aHx_3)|_\subini} \cdot \frac{x_{3,\submax}}{e^{(\beta_\subcm-1)\cdot(N-N_\subSmax)}}, \label{eq:3.47}
\end{align}
where two integration constants $x_{3,\submax}$ and $N_\subSmax$ are introduced for clarity. When comparing with a full numerical solution, the integration constants can be fixed by fitting to the maximum of $x_3(N)$. To address the horizon problem, $r_\subH$ is required to decrease during the late-ESHU era, which in turn imposes $\beta_\subcm>1$. Moreover, Eq.~(\ref{eq:3.47}) implies an exponentially decreasing CHR, resembling the behaviour in inflation \cite{Guth1981.PRD.23.347}. Using the definition in Sec.~\ref{sec:III.C.2}, we obtain $N_\subSend = N_\subSmax + \beta_\subcm^{-1}\cdot\log(x_{3,\submax}/3)$. Since Eq.~(\ref{eq:3.47}) implies a monotonic decrease, it is natural to set $N_\subRmin=N_\subSend$. The comoving size of the causally connected region that can form during the late-ESHU epoch is well approximated by
\begin{equation}
  r(N_\subSmax,N_\subSend) \approx \frac{c\,e^{N_\subSmax-N_\subini}}{(aHx_3)|_\subini} \cdot \frac{x_{3,\submax}}{\beta_\subcm-1}.
\end{equation}
Note that this approximation assumes $\beta_\subcm>1$. Following the argument in Sec.~\ref{sec:III.C.2}, if $N_\subSmax$ is at the Planck scale and $\beta_\subcm\gtrsim\mathcal{O}(1)$, then $x_{3,\submax}>e^{70}$ is required to solve the horizon problem. Relative to the $n=2$ case, this substantially relaxes the requirement on $x_{3,\submax}$ and may eliminate the need for an extreme initial condition (see the first caveat in the $n=2$ discussion). The robustness problem associated with a time-varying $w_\subF$ remains in the present case. These features are collected in Table~\ref{tab:04}.

To provide a comparison between the approximate and full numerical solutions, we integrate Eq.~(\ref{eq:3.6}) numerically for $w_\subF=1/3$. The model parameters are chosen as $\{c_2=-5,\,c_3=-896/73,\,c_5=-13\}$, which in turn fixes $\{c_4=-186956805/2579236,\,c_6=-76932660/644809\}$ through Eqs.~(\ref{eq:3.39}), (\ref{eq:3.41}) and (\ref{eq:3.42}). Rational numbers are used to ensure that the constraints $\lambda_{I_3,+}=\tilde{\lambda}_{I_3,+}=0$ are satisfied exactly. This parameter setting makes $I_3$ a saddle and $F_1$ stable, and it yields $\beta_\subcm\approx1.999992$, implying an inflation-like CHR scaling, $r_\subH\propto e^{-N}$. The conventions for the initial conditions and other setup details follow those adopted in the $n=3$ and $n=2$ cases (see Sec.~\ref{sec:III.C.2}). Here, two sets of initial conditions are considered, labeled 1a and 1b, with their explicit values listed in Table~\ref{tab:05}.

Figure~\ref{fig:cosmos:08} compares the numerical solutions for cases 1a (red solid) and 1b (red dashed) with the late-ESHU approximation (red plus symbols). Although both cases are expected to be described by the same asymptotic form, the agreement is excellent for 1a but very poor for 1b, indicating that the approximation is valid yet has a restricted domain of applicability. The inset provides a schematic phase portrait, illustrating that the two trajectories approach the $I_3\rightarrow F_1$ attractor from opposite sides. This observation suggests that the 1b trajectory is only weakly captured by the $I_3\rightarrow F_1$ attractor, motivating the label \textit{single-side attractor}. A systematic characterization of this behaviour is left for future work.

If one further imposes $\beta_\subcm=0$, the evolution from infinity toward the finite region can be even slower. At present, there is no compelling motivation to consider such ESHU dynamics, so it is not analysed here.

\section{Perturbations of Minkowski spacetime}\label{sec:IV}
This section is devoted to the analysis of perturbations about Minkowski spacetime. Physically, the Minkowski metric could be an exact solution in GCT when all sources vanish, including the fluid, the caloric field, and $\Lambda$. Moreover, this background can be a good approximation to the real Universe, since during the standard matter era both $H$ and $S_0$ can decay toward zero (see Fig.~\ref{fig:cosmos:05} and relevant definitions). Mathematically, this setup implies that the $S_\mu S_\nu$ and $J_\mu S_\nu$ contributions appear only at higher order in perturbations and can be neglected at linear order. This is equivalent to setting $c_3,c_4,c_5,c_6=0$. For the pressure-related sector, it is reasonable to neglect the $c_1$-term while retaining the pressure in $T_{\mu\nu}$. After these simplifications, the only remaining free parameter is $c_2$, and the field equations reduce to 
\begin{subequations}
\begin{gather}
  G_{\mu\nu} + S_{\mu\nu} = \kappa T_{\mu\nu}, \label{eq:4.1a}\\
  \Box S_\mu + \nabla^\nu\nabla_\mu S_\nu = c_2 \nabla_\mu\nabla_\nu S^\nu,
\end{gather}
\end{subequations}
where $S_{\mu\nu} = \nabla_\mu S_\nu + \nabla_\nu S_\mu - c_2 g_{\mu\nu} \nabla_\alpha S^\alpha$, and $T_{\mu\nu}$ denotes the energy-momentum tensor of a perfect fluid. Unlike the electromagnetic field, $S_\mu$ enters the effective gravitational tensor $S_{\mu\nu}$ at linear order, which may have an essential impact on the perturbation dynamics. Under our conventions, the commutation relation of covariant derivatives yields $(\nabla^\nu\nabla_\mu-\nabla_\mu\nabla^\nu)S_\nu = R_{\mu\nu}S^\nu$, which further reduces the vector-field equation to
\begin{equation}\label{eq:4.2}
  \Box S_\mu + (1-c_2) \nabla_\mu\nabla_\nu S^\nu + R_{\mu\nu}S^\nu = 0.
\end{equation}
This result suggests that $c_2=1$ may be a special case and warrants particular attention in what follows.

The perturbation results can be applied to the Newtonian weak-field regime and to gravitational-wave phenomenology. These applications, however, come with several caveats concerning the applicability of the conclusions. First, working on a Minkowski background removes the cosmological time-dependent degrees of freedom. Consequently, the analysis cannot account for effective coupling drifts induced by cosmic dynamics, such as a time-varying $G$ \cite{Brans1961.PhysRev.124.925,Tian2019.PRD.99.064044}. The second caveat concerns the combined background-and-boundary setup. The subsequent analysis treats local systems as perturbations about Minkowski spacetime and imposes standard asymptotically flat boundary conditions. However, the exact solutions in Sec.~\ref{sec:V} show that asymptotically flat spacetimes do exist in GCT, but only along special separatrix branches in the explored sector. This raises the possibility that the Minkowski/asymptotically flat setup may not be the physically appropriate default for local systems in GCT. Altering the background and the boundary conditions can modify the inferred physical implications. This section is confined to the above Minkowski setup, while a systematic treatment of alternative backgrounds and boundary conditions is left for future work.

\subsection{Decoupling of field equations}
As a covariant theory, GCT is invariant under diffeomorphisms, and thus field configurations related by a coordinate transformation represent the same physical system. Such coordinate transformations are referred to as gauge transformations \cite{Flanagan2005.NewJPhys.7.204}. To extract physical conclusions from the perturbation analysis, it is necessary to construct gauge-invariant variables and accordingly decouple the field equations. Hereafter, we proceed in the passive description \cite{Mukhanov1992.PhysRep.215.203} of gauge transformations.

For convenience, this section adopts $x^\mu=(ct,\mathbf{x})$. Partial derivative is denoted by $\p_\mu\equiv\p/\p x^\mu=(c^{-1}\p/\p t,\p_i)$, and an overdot denotes $\p/\p t$. Greek indices run from $0$ to $3$, Latin indices run from $1$ to $3$, and any repeated index is summed over its range, regardless of position. The Minkowski metric is $\eta_{\mu\nu}=\mathrm{diag}\{-1,+1,+1,+1\}$, and the perturbed metric is written as
\begin{equation}
  g_{\mu\nu} = \eta_{\mu\nu} + h_{\mu\nu},
\end{equation}
where $h_{\mu\nu}$ is a dimensionless first-order perturbation. Tensor decomposition provides the mathematical basis for constructing gauge-invariant variables and decoupling the linearized field equations. Within general relativity, the relevant formulas are standard and widely available in the literature. Here we briefly recapitulate the key results for completeness. Using the scalar-vector-tensor decomposition, $h_{\mu\nu}$ can be written as \cite{Flanagan2005.NewJPhys.7.204}
\begin{subequations}\label{eq:4.4}
\begin{align}
  h_{00} &= -2\Psi/c^2, \\
  h_{0i} &= \beta_i + \p_i\Gamma, \\
  h_{ij} &= h_{ij}^\mathrm{TT} + \frac{1}{2}(\p_i\varepsilon_j + \p_j\varepsilon_i) - \frac{2\Phi}{c^2}\delta_{ij} \nonumber\\
    &\quad + \big(\p_i\p_j-\frac{1}{3}\delta_{ij}\nabla^2\big)\Theta,
\end{align}
\end{subequations}
where $[\Psi]=[\Phi]=c^2$, $[\Theta]=\mathrm{length}^2$, $[\Gamma]=[\varepsilon_i]=\mathrm{length}$, and $\beta_i$ and $h_{ij}^\mathrm{TT}$ are dimensionless. There are four scalar functions $\{\Psi,\Phi,\Gamma,\Theta\}$, two three-dimensional vectors $\beta_i$ and $\varepsilon_i$, and a symmetric $3\times3$ tensor $h_{ij}^\mathrm{TT}$, giving a total of $4+3\times2+6=16$ components. These variables are subject to 6 constraints $\p_i\beta_i = 0$, $\p_i\varepsilon_i = 0$, $\p_i h_{ij}^\mathrm{TT} = 0$, and $\delta^{ij}h_{ij}^\mathrm{TT} = 0$, leaving 10 independent degrees of freedom, as required for the symmetric tensor $h_{\mu\nu}$. Consider an infinitesimal coordinate transformation $x^\mu\mapsto\tilde{x}^\mu=x^\mu+\xi^\mu(x)$, treated as a gauge transformation at linear order. Based on the Helmholtz decomposition theorem, the covector $\xi_\mu=\eta_{\mu\nu}\xi^\nu$ can be written as 
\begin{equation}\label{eq:4.5}
  \xi_\mu=(\mathcal{A},\,\mathcal{B}_i+\p_i\mathcal{C}),
\end{equation}
with $\p_i\mathcal{B}_i=0$. The dimensions are $[\mathcal{A}]=[\mathcal{B}_i]=\mathrm{length}$, and $[\mathcal{C}]=\mathrm{length}^2$. At linear order, the $\xi_\mu$-induced gauge transformation of the metric perturbation reads \cite{Flanagan2005.NewJPhys.7.204}
\begin{align}\label{eq:4.6}
  h_{\mu\nu}(x) \mapsto \tilde{h}_{\mu\nu}(\tilde{x}) = h_{\mu\nu}(x) - \p_\mu\xi_\nu - \p_\nu\xi_\mu.
\end{align}
Note that $\tilde{h}_{\mu\nu}$ is defined via $g_{\mu\nu}(x) \mapsto \tilde{g}_{\mu\nu}(\tilde{x}) \equiv \eta_{\mu\nu}+\tilde{h}_{\mu\nu}$, i.e., we keep using $\eta_{\mu\nu}$ as the Minkowski reference metric in the background-perturbation split, and the gauge freedom is carried by the redefined perturbation. Moreover, distinguishing between $\tilde{h}_{\mu\nu}(\tilde{x})$ and $\tilde{h}_{\mu\nu}(x)$ is unnecessary at the order of interest, since their difference enters only at higher order \footnote{Strictly, Eq.~(\ref{eq:4.6}) should be understood as evaluating $\tilde{h}_{\mu\nu}$ at $\tilde{x}$ and all untilded quantities on the right-hand side at $x$. The reason is that, in the passive view, a gauge transformation compares the same physical point described in two coordinate systems \cite{Mukhanov1992.PhysRep.215.203}.}. Substituting Eqs.~(\ref{eq:4.4}) and (\ref{eq:4.5}) into Eq.~(\ref{eq:4.6}) yields
\begin{subequations}\label{eq:4.7}
\begin{align}
  h_{00} \mapsto \tilde{h}_{00} &= -\frac{2\Psi}{c^2} - \frac{2\dot{\mathcal{A}}}{c}, \\
  h_{0i} \mapsto \tilde{h}_{0i} &= \beta_i - \frac{\dot{\mathcal{B}}_i}{c} + \p_i(\Gamma-\mathcal{A}-\frac{\dot{\mathcal{C}}}{c}), \\
  h_{ij} \mapsto \tilde{h}_{ij} &= h_{ij}^\mathrm{TT} + \frac{1}{2}(\p_i[\varepsilon_j-2\mathcal{B}_j] + \p_j[\varepsilon_i-2\mathcal{B}_i]) \nonumber\\ 
    &\quad - \frac{2\delta_{ij}}{c^2}(\Phi+\frac{c^2}{3}\nabla^2\mathcal{C}) \nonumber\\
    &\quad + \big(\p_i\p_j-\frac{1}{3}\delta_{ij}\nabla^2\big)(\Theta - 2\mathcal{C}).
\end{align}
\end{subequations}
Express $\tilde{h}_{\mu\nu}$ in the form of Eq.~(\ref{eq:4.4}) by placing a tilde on each variable, and a comparison with Eq.~(\ref{eq:4.7}) yields
\begin{subequations}
\begin{align}
  \Psi \mapsto \tilde{\Psi} &= \Psi + c\dot{\mathcal{A}}, \\
  \beta_i \mapsto \tilde{\beta}_i &= \beta_i - \dot{\mathcal{B}}_i/c, \\
  \Gamma \mapsto \tilde{\Gamma} &= \Gamma - \mathcal{A} - \dot{\mathcal{C}}/c, \\
  h_{ij}^\mathrm{TT} \mapsto \tilde{h}_{ij}^\mathrm{TT} &= h_{ij}^\mathrm{TT}, \\
  \varepsilon_i \mapsto \tilde{\varepsilon}_i &= \varepsilon_i-2\mathcal{B}_i, \\
  \Phi \mapsto \tilde{\Phi} &= \Phi+c^2\nabla^2\mathcal{C}/3, \\
  \Theta \mapsto \tilde{\Theta} &= \Theta - 2\mathcal{C},
\end{align}
\end{subequations}
which motivate the following gauge-invariant variables
\begin{subequations}\label{eq:4.9}
\begin{align}
  \Psi_\mathrm{S} &= \Psi + c\dot{\Gamma} - \ddot{\Theta}/2, \\
  \Phi_\mathrm{S} &= \Phi + (c^2/6)\nabla^2\Theta, \\
  \Xi_i &= \beta_i - \dot{\varepsilon}_i/(2c), \\
  &\hspace{-3.5mm} \mathrm{and\ \,} h_{ij}^\mathrm{TT}.
\end{align}
\end{subequations}
Here the subscript $\mathrm{S}$ denotes the scalar mode. With the constraints ($\p_i\Xi_i = 0$, $\p_i h_{ij}^\mathrm{TT} = 0$, and $\delta^{ij}h_{ij}^\mathrm{TT} = 0$) imposed, Eq.~(\ref{eq:4.9}) gives 6 independent gauge-invariant variables. The remaining 4 correspond to the coordinate freedom parametrized by $\xi_\mu$. The field equations should be written in terms of gauge invariants. To this end, a direct calculation yields the Einstein tensor components
\begin{subequations}
\begin{align}
  G_{00} &= \frac{2}{c^2}\nabla^2\Phi_\mathrm{S}, \\
  G_{0i} &= - \frac{1}{2}\nabla^2\Xi_i + \frac{2}{c^3}\p_i\dot{\Phi}_\mathrm{S}, \\
  G_{ij} &= - \frac{1}{2}\Box h_{ij}^\mathrm{TT} - \frac{1}{2c}(\p_i\dot{\Xi}_j+\p_j\dot{\Xi}_i) + \frac{2\delta_{ij}}{c^4}\ddot{\Phi}_\mathrm{S} \nonumber\\
    &\quad + \frac{1}{c^2}(\p_i\p_j-\delta_{ij}\nabla^2)(\Phi_\mathrm{S}-\Psi_\mathrm{S}).
\end{align}
\end{subequations}

We now turn to the energy components in the gravitational field equations. Since the background values of $S_\mu$ and $T_{\mu\nu}$ vanish, both can be treated as first-order quantities. Consequently, all of their components are gauge invariant at linear order \cite{Stewart1974.ProcRSocLondA.341.49}. To facilitate the decoupling of Eq.~(\ref{eq:4.1a}), $T_{\mu\nu}$ is decomposed as \cite{Flanagan2005.NewJPhys.7.204}
\begin{subequations}
\begin{align}
  T_{00} &= \mathcal{U}, \\
  T_{0i} &= \mathcal{V}_i + \p_i\mathcal{V}, \\
  T_{ij} &=  \mathcal{X}_{ij} + \frac{1}{2}(\p_i\mathcal{X}_j+\p_j\mathcal{X}_i) + \mathcal{W}\delta_{ij} \nonumber\\
    &\quad + (\p_i\p_j-\frac{1}{3}\delta_{ij}\nabla^2)\mathcal{X},
\end{align}
\end{subequations}
subject to the constraints $\p_i\mathcal{V}_i=0$, $\p_i\mathcal{X}_i=0$, $\p_i\mathcal{X}_{ij}=0$, and $\delta^{ij}\mathcal{X}_{ij}=0$. The dimensions are $[\kappa\mathcal{U}]=[\kappa\mathcal{V}_i]=[\kappa\mathcal{W}]=[\kappa\mathcal{X}_{ij}]=\mathrm{length}^{-2}$, $[\kappa\mathcal{V}]=[\kappa\mathcal{X}_i]=\mathrm{length}^{-1}$, and $[\kappa\mathcal{X}]=1$. Here $\mathcal{V}$ and $\mathcal{V}_i$ denote distinct variables, and the same distinction applies within the $\mathcal{X}$ family. Energy-momentum conservation for the fluid reduces to
\begin{subequations}
\begin{align}
  \nabla^2\mathcal{V} &= \dot{\mathcal{U}}/c, \\
  \nabla^2\mathcal{X} &= \frac{3}{2}\bigl[\dot{\mathcal{V}}/c - \mathcal{W}\bigr], \\
  \nabla^2\mathcal{X}_i &= 2\dot{\mathcal{V}}_i/c.
\end{align}
\end{subequations}
The caloric field can be decomposed as \cite{GaugeFreedom}
\begin{equation}
  S_\mu = (\varphi,\,\zeta_i+\p_i\vartheta),
\end{equation}
with $\p_i\zeta_i=0$. The dimensions are $[\varphi]=[\zeta_i]=\mathrm{length}^{-1}$ and $[\vartheta]=1$. Under the simplifications adopted at the beginning of this section, a direct computation yields
\begin{subequations}
\begin{align}
  S_{00} &= \frac{2-c_2}{c}\dot{\varphi} + c_2\nabla^2\vartheta, \\
  S_{0i} &= \dot{\zeta}_i/c + \p_i(\varphi+\dot{\vartheta}/c), \\
  S_{ij} &= \p_i\zeta_j + \p_j\zeta_i + 2\p_i\p_j\vartheta + c_2\delta_{ij}(\frac{\dot{\varphi}}{c} - \nabla^2\vartheta).
\end{align}
\end{subequations}

After the above preparations, the decoupling of the gravitational field equations can now be addressed. The final results are summarized as 
\begin{subequations}\label{eq:4.15}
\begin{gather}
  \nabla^2 (2\Phi_\mathrm{S} + c_2c^2\vartheta) + (2-c_2)c\dot{\varphi} = \kappa c^2\mathcal{U}, \label{eq:4.15a}\\
  \frac{2}{c^4}\ddot{\Phi}_\mathrm{S} + (2-c_2)\nabla^2\vartheta +\frac{c_2}{c}\dot{\varphi} = \kappa(\mathcal{W} + \frac{2}{3}\nabla^2\mathcal{X}), \label{eq:4.15b}\\
  2\dot{\Phi}_\mathrm{S} + c^2\dot{\vartheta} + c^3\varphi = \kappa c^3\mathcal{V}, \label{eq:4.15c}\\
  \Phi_\mathrm{S}-\Psi_\mathrm{S} + 2c^2\vartheta = \kappa c^2\mathcal{X}, \label{eq:4.15d}\\
  \nabla^2\Xi_i - 2\dot{\zeta}_i/c = -2\kappa\mathcal{V}_i, \label{eq:4.15e}\\
  \dot{\Xi}_i - 2c\zeta_i = - \kappa c\mathcal{X}_i, \label{eq:4.15f}\\
  \Box h_{ij}^\mathrm{TT} = - 2\kappa\mathcal{X}_{ij}. \label{eq:4.15g}
\end{gather}
\end{subequations}
The derivation details are collected below. The $00$ component of Eq.~(\ref{eq:4.1a}) directly yields Eq.~(\ref{eq:4.15a}). The $0i$ component contains both vector and scalar sectors. Contracting it with $\p_i$ isolates the scalar part and leads to $\nabla^2(2\dot{\Phi}_\mathrm{S} + c^2\dot{\vartheta} + c^3\varphi - \kappa c^3\mathcal{V})=0$. For a local gravitational system, it is natural to impose that all perturbations vanish at spatial infinity~\footnote{In principle, this boundary condition should be viewed as an assumption of the present Minkowski-based perturbative setup, rather than as an established default for local systems in GCT (see the discussion at the beginning of Sec.~\ref{sec:IV}).}. With these boundary conditions and assuming regularity of the perturbations (no $1/r$ singularity at $r=0$), the Laplacian equation implies that the enclosed quantity must vanish, so $\nabla^2$ can be dropped and Eq.~(\ref{eq:4.15c}) follows. This result can then be used to eliminate the scalar part of the original $0i$ component, yielding Eq.~(\ref{eq:4.15e}). Finally, taking the trace and the divergence of the $ij$ component by contracting with $\delta_{ij}$ and $\p_j$ gives a scalar (trace) equation and a vector equation. Applying $\p_i$ to the trace equation and combining the result with the vector equation can eliminate $\p_i\ddot{\Phi}_\mathrm{S}$. The resulting equation can be written in Laplace form, and $\nabla^2$ can be dropped accordingly. For the obtained result, contracting further with $\p_i$ then cleanly separates the scalar and vector sectors and leads to Eqs.~(\ref{eq:4.15d}) and (\ref{eq:4.15f}). Returning to the original trace equation, one can use Eq.~(\ref{eq:4.15d}) to eliminate $\Psi_\mathrm{S}$, yielding Eq.~(\ref{eq:4.15b}). Using these results to eliminate the scalar and vector components from the original $ij$ equation then yields Eq.~(\ref{eq:4.15g}).

At linear order, the vector-field equation (\ref{eq:4.2}) reduces to $\Box S_\mu + (1-c_2) \p_\mu\p_\nu S^\nu=0$. If $c_2=1$, the decoupling is manifest, and the component equations read
\begin{equation}\label{eq:4.16}
  \Box\varphi = \Box\zeta_i = \Box\vartheta = 0.
\end{equation}
If $c_2\neq1$, $\p_\nu S^\nu$ induces additional couplings. A direct computation yields
\begin{subequations}
\begin{gather}
  \Box\varphi + \frac{1-c_2}{c} \bigl(- \frac{\ddot{\varphi}}{c} + \nabla^2\dot{\vartheta}\bigr) = 0, \label{eq:4.17a}\\
  \Box (\zeta_i+\p_i\vartheta) + (1-c_2)\p_i\bigl(- \frac{\dot{\varphi}}{c} + \nabla^2\vartheta\bigr) = 0. \label{eq:4.17b}
\end{gather}
\end{subequations}
Contracting Eq.~(\ref{eq:4.17b}) with $\p_i$ isolates the scalar sector from the vector sector, yielding
\begin{subequations}
\begin{equation}\label{eq:4.18a}
  \Box\vartheta + (1-c_2)\bigl(- \frac{\dot{\varphi}}{c} + \nabla^2\vartheta\bigr) = 0.
\end{equation}
Substituting this back into Eq.~(\ref{eq:4.17b}) gives
\begin{equation}\label{eq:4.18b}
  \Box\zeta_i = 0.
\end{equation}
\end{subequations}
Contracting the linearized vector-field equation with $\p^\mu$ yields $(2-c_2)\cdot\Box(\p_\mu S^\mu)=0$. For $c_2\neq2$, this can be written explicitly as $\Box(\nabla^2\vartheta - \dot{\varphi}/c)=0$. Applying $\Box$ to Eqs.~(\ref{eq:4.17a}) and (\ref{eq:4.18a}) then gives
\begin{equation}\label{eq:4.19}
  \Box^2\varphi = \Box^2\vartheta = 0.
\end{equation}
This result holds for $c_2\neq2$, while the special case $c_2=2$ is not considered in this work. It should be noted that the derivation introduces the $\Box$ operator by hand, thereby promoting a second-order system to fourth order. Solutions of the fourth-order equations must therefore be supplemented by the constraints implied by the original second-order equations. 

\subsection{Ostrogradsky-type instability}\label{sec:IV.B}
The linearized vector-field equation decouples from the gravitational field equation, thus providing a transparent diagnostic of the system's stability. For $c_2=1$, Eq.~(\ref{eq:4.16}) is the standard d'Alembert equation, so no growing mode arises at linear order. For $c_2\neq1$ (and also $2$), Eq.~(\ref{eq:4.19}) is a fourth-order equation and may therefore exhibit an Ostrogradsky-type instability typical of higher-derivative theories. A more discriminating diagnostic is to reconstruct a local effective Lagrangian whose variation reproduces the fourth-order equation. One convenient example is $\mathcal{L}=(\Box\varphi)^2$ for $\varphi$ \cite{deRham2016.JCAP.06.041,ReLagrangian}. This is a non-degenerate higher-derivative Lagrangian, and it generically propagates an Ostrogradsky ghost \cite{Delhom.arXiv.2207.13431}. However, the present conclusion remains tentative. Equation~(\ref{eq:4.19}) is obtained by applying $\Box$ to the original second-order system, which may introduce spurious higher-derivative modes. A reliable stability assessment therefore requires working directly with the original coupled second-order equations (\ref{eq:4.17a}) and (\ref{eq:4.18a}).

In Fourier space, adopting the convention $f(\mathbf{x}) = \int \tilde{f}(\mathbf{k}) e^{i\mathbf{k}\cdot\mathbf{x}} \,\dx^3\mathbf{k}$ for $\varphi$ and $\vartheta$, Eqs.~(\ref{eq:4.17a}) and (\ref{eq:4.18a}) take the form
\begin{equation}\label{eq:4.20}
  \frac{\dx^2}{\dx t^2}
  \begin{pmatrix}
    \tilde{\varphi} \\
    \dot{\tilde{\vartheta}}/c 
  \end{pmatrix}
  +
  \frac{c^2k^2}{\tilde{c}_2}
  \begin{bmatrix}
    1  &  1-c_2  \\
    c_2-1  &  3-2c_2 
  \end{bmatrix}
  \begin{pmatrix}
    \tilde{\varphi} \\
    \dot{\tilde{\vartheta}}/c 
  \end{pmatrix}
  =
  0.
\end{equation}
Here $\p_0$ is applied to Eq.~(\ref{eq:4.18a}). This step does not introduce spurious higher-derivative modes in the present analysis, because no subsequent integration is performed to undo this $\p_0$ operation. In Eq.~(\ref{eq:4.20}), the first term is second order in derivatives, which distinguishes the equation from the first-order dynamical systems used in phase-space analyses. Following Ref.~\cite{Ganz2021.CQG.38.075005}, one can verify that the coefficient matrix is not diagonalizable for $c_2\neq1$, which suggests an instability. To see this explicitly, defining $\tilde{\psi}=\tilde{\varphi}-\dot{\tilde{\vartheta}}/c$ casts the matrix into an upper-triangular form, and Eq.~(\ref{eq:4.20}) becomes
\begin{equation}
  \frac{\dx^2}{\dx t^2}
  \begin{pmatrix}
    \tilde{\varphi} \\
    \tilde{\psi}
  \end{pmatrix}
  +
  c^2k^2
  \begin{bmatrix}
    1  &  (c_2-1)/\tilde{c}_2  \\
    0  &  1
  \end{bmatrix}
  \begin{pmatrix}
    \tilde{\varphi} \\
    \tilde{\psi}
  \end{pmatrix}
  =
  0.
\end{equation}
For $c_2\neq1$, this triangular matrix cannot be further diagonalized, so the two variables cannot be fully decoupled. The general solution reads
\begin{subequations}
\begin{align}
  \tilde{\varphi} &= \bigl(\alpha_1 + \alpha_3\frac{\omega t}{2i}\bigr) e^{i\omega t} 
    + \bigl(\alpha_2  + \alpha_4\frac{\omega t}{2i}\bigr) e^{-i\omega t}, \label{eq:4.22a}\\
  \tilde{\psi} &= \frac{2-c_2}{1-c_2} \bigl(\alpha_3 e^{i\omega t} - \alpha_4 e^{-i\omega t}\bigr),
\end{align}
\end{subequations}
where $\omega=ck$, and $\alpha_i$ are constant (with respect to $t$). Equation~(\ref{eq:4.22a}) contains a mode that grows linearly in $t$, confirming the existence of an instability. A definitive classification of this instability is not established here. For convenience, in view of the reconstructed Lagrangian, we refer to it as an Ostrogradsky-type instability.

The above discussion underscores the distinguished role of $c_2=1$ in ensuring perturbative stability. Nevertheless, it remains worthwhile to explore whether the instability encountered for $c_2\neq1$ can be alleviated. In the classical field approach, this calls for extending the analysis to more general backgrounds. On the other hand, quantization may offer an additional avenue to address this problem \cite{Kaparulin2014.EurPhysJC.74.3072,Raidal2017.NuclPhysB.916.607,Donoghue2021.PRD.104.045010}. These directions are left for future work.

\subsection{Astrophysical applications}
Perturbation analysis serves as a key tool for testing gravitational theories \cite{Will2014.LRR.17.4}. This subsection focuses on applying the decoupled field equations to examine the Newtonian approximation and gravitational waves.

\subsubsection{Newtonian approximation}\label{sec:IV.C.1}
The Newtonian approximation corresponds to a local static configuration and is applicable to Solar-System tests. In this limit, for any $c_2$, each component of the vector-field equation reduces to the standard Laplace equation. Imposing the boundary conditions and requiring regularity of the field then yields $S_\mu=0$ at linear order. As a result, the gravitational field equations reduce exactly to those of general relativity. This property leads to several important physical implications:
\begin{itemize}
  \item The parameter $\kappa$ in Eq.~(\ref{eq:2.1a}) coincides with its value in general relativity, i.e., $8\pi G/c^4$, where $G$ denotes the Newtonian gravitational constant. This is not a trivial result, since modified gravity often rescales the effective gravitational coupling constant in local systems \cite{Brans1961.PhysRev.124.925,Bellini2014.JCAP.07.050,Langlois2017.JCAP.05.033,Tian2019.PRD.99.064044}. This property ensures the purity of the EDE and ESHU models in GCT: the $G$ entering the Friedmann equation~(\ref{eq:3.1}) remains the Newtonian gravitational constant, thereby avoiding an effective scaling-type energy density (or redefinition of mass) in cosmology that would arise from rescaling $G$ \cite{Bellini2014.JCAP.07.050}.
  \item In the post-Newtonian regime relevant to Solar-System tests, the pressure contribution is negligible compared with the energy (mass) density, and hence Eq.~(\ref{eq:4.15d}) yields $\gamma_\mathrm{PPN}=1$. Accordingly, GCT is consistent with the Cassini constraint from the Shapiro time-delay measurement, $\gamma_\mathrm{PPN}=1+(2.1\pm2.3)\times10^{-5}$ \cite{Bertotti2003.Nature.425.374}.
  \item The solution $S_\mu=0$ implies that no caloric charge can be sourced in the weak-field limit. This observation provides part of the motivation for the next section, which is intended to examine compact objects in GCT.
\end{itemize}

Note that, as already emphasized at the beginning of this section, the above conclusions rely on the Minkowski background setup (including the boundary conditions). On the one hand, this setup can be a good approximation. In the real Universe, the present-epoch background value of $S_0$ can be tiny (see Fig.~\ref{fig:cosmos:05} and the associated equations), so any relevant correction may be negligible. On the other hand, a robust physical conclusion requires extending the analysis to more general backgrounds and boundary conditions in future work.

\subsubsection{Gravitational waves}
The detectability and polarization classification of gravitational waves are encoded in the tidal response of freely falling test bodies, as described by the geodesic deviation equation and the associated Riemann tensor \cite{Pirani1956.ActaPhysPol.15.389,Eardley1973.PRD.8.3308}. In terms of gauge-invariant variables, the relevant curvature components can be expressed as \cite{Flanagan2005.NewJPhys.7.204}
\begin{equation}\label{eq:4.23}
  R_{0i0j} = - \frac{\ddot{h}_{ij}^\mathrm{TT}}{2c^2} + \frac{\p_i \dot{\Xi}_j + \p_j \dot{\Xi}_i}{2c} + \frac{\delta_{ij}\ddot{\Phi}_\mathrm{S}}{c^4} + \frac{\p_i\p_j\Psi_\mathrm{S}}{c^2}.
\end{equation}
Without loss of generality, the wave is taken to propagate along the $z$-axis, so that all gauge-invariant variables depend only on $\{t,z\}$. The constraint $\p_j h_{ij}^\mathrm{TT} = \p_z h_{i3}^\mathrm{TT} = 0$ then implies $h_{i3}^\mathrm{TT}=h_{3i}^\mathrm{TT}=0$, and the traceless constraint $\delta^{ij}h_{ij}^\mathrm{TT}=0$ further gives $h_{11}^\mathrm{TT}=-h_{22}^\mathrm{TT}$. Following standard notation, we denote $h_{11}^\mathrm{TT}=h_{+}$ and $h_{12}^\mathrm{TT}=h_\times$. Likewise, the constraint $\p_i\Xi_i=\p_z\Xi_3=0$ reduces to $\Xi_3=0$. With these relations, Eq.~(\ref{eq:4.23}) can be written as
\begin{equation}\label{eq:4.24}
  R_{0i0j} = 
  \frac{1}{2c^2}
  \begin{bmatrix}
    \displaystyle\frac{2\ddot{\Phi}_\mathrm{S}}{c^2} -\ddot{h}_{+}  &  -\ddot{h}_{\times}                          &  c\p_z \dot{\Xi}_1  \\
    -\ddot{h}_{\times}                         &  \displaystyle\frac{2\ddot{\Phi}_\mathrm{S}}{c^2} + \ddot{h}_{+}  &  c\p_z \dot{\Xi}_2  \\
    c\p_z \dot{\Xi}_1                          &  c\p_z \dot{\Xi}_2                           &  \displaystyle\frac{2\ddot{\Phi}_\mathrm{S}}{c^2} + 2\p_z^2\Psi_\mathrm{S}
  \end{bmatrix}.
\end{equation}
Together with the geodesic deviation equation, this decomposition shows that a metric theory admits at most six gravitational-wave polarizations \cite{Eardley1973.PRD.8.3308,Eardley1973.PRL.30.884}. For convenience, the polarization content can be read off by directly comparing Eq.~(\ref{eq:4.24}) with Eq.~(130) in Ref.~\cite{Will2014.LRR.17.4}. For a specific theory of gravity, the field equations determine which components of the perturbations are allowed to propagate.

The gravitational-wave properties in GCT can now be discussed. Given the stability requirement discussed in Sec.~\ref{sec:IV.B}, attention is restricted to the case $c_2=1$. To study wave propagation in the far zone, well outside the fluid source, Eq.~(\ref{eq:4.15}) is evaluated in the fluid-free limit. In addition, $\nabla^2$ is used in place of $\p_z^2$ when no ambiguity arises. In this case, the vector-field equation reduces to Eq.~(\ref{eq:4.16}). For the metric perturbations, Eq.~(\ref{eq:4.15g}) implies that two tensor polarizations are allowed. Combining Eqs.~(\ref{eq:4.15e}) and (\ref{eq:4.15f}) and eliminating $\zeta_i$ yields $\Box\Xi_i=0$, showing that two vector polarizations can also propagate. It can be seen from Eq.~(\ref{eq:4.24}) that $\ddot{\Phi}_\mathrm{S}$ and $\ddot{\Phi}_\mathrm{S} + c^2\nabla^2\Psi_\mathrm{S}$ describe the breathing and longitudinal scalar modes, respectively. Combining Eqs.~(\ref{eq:4.15a}) and (\ref{eq:4.15b}) eliminates the two caloric field components and gives $\Box\Phi_\mathrm{S}=0$, which implies that the breathing mode is permitted. Applying $\Box$ to Eq.~(\ref{eq:4.15d}) and invoking Eq.~(\ref{eq:4.16}) yields $\Box\Psi_\mathrm{S}=0$. As $\Phi_\mathrm{S}$ and $\Psi_\mathrm{S}$ represent two independent gauge-invariant scalar potentials, $\Box\Phi_\mathrm{S}=\Box\Psi_\mathrm{S}=0$ implies the presence of a propagating longitudinal scalar mode. Furthermore, $\ddot{\Phi}_\mathrm{S} = -c^2(\ddot{\vartheta}+c\dot{\varphi})/2$ and $\ddot{\Phi}_\mathrm{S} + c^2\nabla^2\Psi_\mathrm{S} = c^2(\ddot{\vartheta}-c\dot{\varphi})$ can be obtained, which demonstrates the independence of the two scalar wave modes in the caloric-field variables. This differs from Einstein-{\ae}ther theory, where two scalar modes exist but are not independent even at the level of propagation equations \cite{Gong2018.PRD.97.084040,Zhang2020.PRD.101.044002}. Finally, the above wave equations imply luminal propagation for all modes, so the $c_2=1$ branch of GCT is consistent with the GW170817 constraint on the gravitational-wave speed \cite{Abbott2017.PRL.119.161101,Abbott2017.ApJL.848.L13}. In summary, with $c_2=1$, GCT admits six independent gravitational-wave polarizations, all propagating at the speed of light, providing a clean benchmark for future gravitational-wave tests of gravity.

\section{Static spherically symmetric systems}\label{sec:V}
Compact objects provide a sensitive arena for confronting gravitational theories with strong-field physics \cite{Yagi2016.CQG.33.054001,Olmo2020.PhysRep.876.1}. As a first step in this direction for GCT, we restrict attention to static spherically symmetric configurations. This section is intended as a partial analysis of this sector rather than a systematic investigation.

In Sec.~\ref{sec:V.A}, we derive the general structure equations under suitable assumptions on the caloric field profile and single out two tractable vacuum reductions of the system. The remainder of the section explores these vacuum reductions in detail, combining numerical integrations, local series expansions, and several exact solutions. Physically, two features are emphasized:
\begin{itemize}
  \item Over finite radial domains, the caloric field can mimic an effective cosmological constant.
  \item Within the present analysis, the only black-hole branch identified is the Schwarzschild geometry. GCT also naturally accommodates solutions describing naked singularities and wormholes.
\end{itemize}

The notation in this section differs slightly from that of the previous two sections in order to better align with standard conventions in the compact-object literature. Where necessary, symbols are defined at first use to prevent ambiguity.

\subsection{Structure equations}\label{sec:V.A}
This subsection first derives the general structure equations for static spherically symmetric configurations. The resulting complexity discourages a direct analysis and prompts a search for parameter choices and variable transformations that may simplify the equations. It is noted that, in the vacuum (fluid-free) case with $c_i=\delta_{i2}$ (i.e., $c_2=1$ and $c_i=0$ for $i\neq2$), the structure equations can in principle be reduced to a single-variable integro-differential equation. Further setting $\Lambda=0$ and imposing the metric function $\nu=0$ is self-consistent and yields a second-order ordinary differential equation for $\lambda(r)$. Introducing $\lambda\mapsto\mu\equiv1/(r\lambda^\prime)$ then reduces this second-order equation to first order. The resulting equation is an Abel equation of the first kind \cite{Polyanin2018.book} and provides the basis for the global analysis that follows. An analogous transformation is applicable to $\nu(r)$ and can be used to treat the more general case $\nu\neq0$.

\subsubsection{General results}
The most general static spherically symmetric metric can be written as \cite{Tolman1939.PhysRev.55.364,Oppenheimer1939.PhysRev.55.374}
\begin{equation}\label{eq:5.1}
  \dx s^2=-e^\nu c^2\dx t^2+e^\lambda\dx r^2+r^2(\dx\theta^2+\sin^2\theta\dx\varphi^2),
\end{equation}
where the dimensionless functions $\lambda=\lambda(r)$ and $\nu=\nu(r)$. The caloric field is assumed to take the form
\begin{equation}\label{eq:5.2}
  S_\mu=[\alpha(r),\,\beta(r),\,0,\,0],
\end{equation}
where $[\alpha]=\mathrm{time}^{-1}$ and $[\beta]=\mathrm{length}^{-1}$. Note that spherical symmetry requires the effective gravitational tensor $S^\mu_{\ \nu}$ to be rotationally invariant, but it does not directly imply an $r$-only dependence of the components of $S_\mu$. The full generality of Eq.~(\ref{eq:5.2}) for the symmetry requirement is not proved here. Instead, by comparison with the electromagnetic potential $A_\mu$, two key arguments are given to justify its use. The first argument concerns internal gauge freedom. In electromagnetism, one may always choose a gauge in which $A_1=0$ \cite{Pellicer1969.JMathPhysNY.10.1718}. By contrast, no analogous internal freedom has been identified for the caloric field, so both $S_0(r)$ and $S_1(r)$ should be kept. The second concerns magnetic charge. In electromagnetism, a magnetic charge corresponds to $A_3=Q_m\cos\theta$, where $Q_m$ is a constant \cite{Fan2016.PRD.94.124027}. The associated spacetime is described by the Reissner-Nordstr{\"o}m metric \cite{Reissner1916.NotEngJ.355.106,Nordstrom1918.ProcKNAW.20.1238}. For the caloric-field analogue of magnetic charge, our following argument is restricted to the special case $c_i=\delta_{i2}$. Consistency with the gravitational field equations requires the diagonal components of $S^\mu_{\ \nu}$ to depend only on $r$ and the off-diagonal components to vanish. One can verify that assuming $S_3=S_3(r,\theta)$ makes the diagonal components identically zero, and solving the off-diagonal equations yields $S_3=Q_mr^2\sin^2\theta$. This implies that $S_\mu$ appears to admit a magnetic-charge-like structure while giving no contribution to $S^\mu_{\ \nu}$ in this special case. For simplicity, we set $S_3=0$ even in the general case. In addition, $S_2$ is set to zero by default. For the fluid sector, the mass density and pressure are taken to be $\rho_\subF=\rho_\subF(r)$ and $p_\subF=p_\subF(r)$. As the system is static, the four-velocity is $u^\mu=\dx x^\mu/\dx\tau=(e^{-\nu/2},0,0,0)$, where the proper time $\tau$ is defined by $\dx s^2=-c^2\dx\tau^2$. The normalization $u^\mu u_\mu=-c^2$ holds. As in cosmological applications, we introduce $\varrho_\subF(r)=\sqrt{8\pi G\rho_\subF}$ to simplify subsequent expressions. The dimension $[\varrho_\subF]=\mathrm{time}^{-1}$ follows immediately. The general structure equations retain $\Lambda$ for future use.

Substituting the above ansatzes into Eq.~(\ref{eq:2.1a}) yields
\begin{subequations}\label{eq:5.3}
\begin{align}
  \alpha^\prime &= \alpha(\nu^\prime-c_6\beta) + c_5\varrho_\subF\beta e^{\nu/2}, \label{eq:5.3a}\\
  2\beta^\prime &= (\beta-r^{-1})(\lambda^\prime + \nu^\prime) - c_6\beta^2 \nonumber\\
    &\quad + \frac{e^\lambda}{c^2} \Bigl[(1+w_\subF)\varrho_\subF^2 + \frac{2c_5\varrho_\subF\alpha}{e^{\nu/2}} - \frac{c_6\alpha^2}{e^\nu}\Bigr], \label{eq:5.3b}\\
  \tilde{c}_2 \lambda^\prime &= c_2(\nu^\prime - 4\beta) - 2(c_2-1)r\beta\nu^\prime + (2-2e^\lambda)/r \nonumber\\
    &\quad + (c_2c_6-2c_4)r\beta^2 + 2\Lambda r e^\lambda - 2c_1r(w_\subF^\prime)^2 \nonumber\\
    &\quad + \frac{r e^\lambda}{c^2} \Bigl[ (\tilde{c}_2-c_2w_\subF)\varrho_\subF^2 - \frac{2(c_3-\tilde{c}_2c_5)\varrho_\subF\alpha}{e^{\nu/2}} \nonumber\\
    &\quad + \frac{(2c_4-\tilde{c}_2c_6)\alpha^2}{e^\nu} \Bigr], \label{eq:5.3c}\\
  \nu^{\prime\prime} &= 2c_2\beta^\prime - (c_2\beta-r^{-1})(\lambda^\prime-\nu^\prime) + 2c_4\beta^2 - 2\Lambda e^{\lambda} \nonumber\\
    &\quad + 2c_1(w_\subF^\prime)^2 + \frac{(\lambda^\prime-\nu^\prime)\nu^\prime}{2} + \frac{4(c_2-1)\beta}{r} \nonumber\\
    &\quad + \frac{2e^\lambda}{c^2} \Bigl[ w_\subF\varrho_\subF^2 + \frac{c_3\varrho_\subF\alpha}{e^{\nu/2}} - \frac{c_4 \alpha^2}{e^{\nu}} \Bigr], \label{eq:5.3d}
\end{align}
where $\prime\equiv\dx/\dx r$ in this section. Here, Eq.~(\ref{eq:5.3a}) follows from the contravariant-covariant $01$ component, Eqs.~(\ref{eq:5.3b}) and (\ref{eq:5.3c}) are obtained from appropriate combinations of the $00$ and $11$ components, and Eq.~(\ref{eq:5.3d}) comes from the $22$ component. The remaining components impose no additional constraints. The pressure in $T^\mu_{\ \nu}$ is eliminated using $p_\subF=w_\subF\rho_\subF c^2$. The fluid hydrostatic equilibrium equation can be written as
\begin{equation}
  p_\subF^\prime = - (p_\subF + \rho_\subF c^2) \nu^\prime/2.
\end{equation}
\end{subequations}
Together with the equation of state, one obtains a closed system for the six variables $\{\alpha,\beta,\lambda,\nu,\rho_\subF,p_\subF\}$. The expanded form of Eq.~(\ref{eq:2.1b}) is consistent with the above results \cite{NotShown}.

\subsubsection{Vacuum equations for $c_i=\delta_{i2}$}
The complexity of Eq.~(\ref{eq:5.3}) obscures how the caloric field shapes the resulting structure, so it is useful to begin with a simpler setting. We note that imposing the vacuum condition together with the parameter choice $c_i=\delta_{i2}$ enables a substantial decoupling of the variables and a reduction to lower-order differential equations. Here, the parameter slice $c_2=1$ is motivated by two considerations. Physically, Minkowski stability selects this value, and Sec.~\ref{sec:III} further indicates that this slice supports a wide range of cosmological applications. Mathematically, $c_2=1$ simplifies the equations in several respects: the first two terms in Eq.~(\ref{eq:5.3d}) take a form similar to the corresponding terms in Eq.~(\ref{eq:5.3b}), and all contributions proportional to $(c_2-1)$ drop out of Eqs.~(\ref{eq:5.3c}) and (\ref{eq:5.3d}). For the remaining parameters, $c_1=c_3=c_5=0$ is compatible with the vacuum setting, whereas $c_4=c_6=0$ is imposed purely for algebraic simplicity.

Under the above setting, Eq.~(\ref{eq:5.3}) reduces to
\begin{subequations}\label{eq:5.4}
\begin{align}
  \alpha^\prime &= \alpha\nu^\prime, \label{eq:5.4a}\\
  \beta^\prime &= (\beta-r^{-1})(\lambda^\prime + \nu^\prime)/2, \label{eq:5.4b}\\
  \lambda^\prime &= \nu^\prime - 4\beta + (2-2e^\lambda)/r + 2\Lambda r e^\lambda, \label{eq:5.4c}\\
  \nu^{\prime\prime} &= 2\beta^\prime - (\beta-\nu^\prime/2-r^{-1})(\lambda^\prime-\nu^\prime) - 2\Lambda e^{\lambda}. \label{eq:5.4d}
\end{align}
\end{subequations}
Integrating Eq.~(\ref{eq:5.4a}) yields $\alpha\propto e^\nu$. Since $\alpha$ does not enter the remaining equations, it can be omitted. Substituting Eqs.~(\ref{eq:5.4b}) and (\ref{eq:5.4c}) into Eq.~(\ref{eq:5.4d}) eliminates $\beta^\prime$ and $\lambda^\prime$, yielding 
\begin{subequations}\label{eq:5.5}
\begin{equation}\label{eq:5.5a}
  \nu^{\prime\prime} = \Bigl( \Lambda r e^\lambda - \frac{1+e^\lambda}{r} \Bigr)\nu^\prime - 2\Lambda e^{\lambda}.
\end{equation}
Equations (\ref{eq:5.4b}) and (\ref{eq:5.5a}) can be regarded as first-order linear inhomogeneous ordinary differential equations for $\beta$ and $\nu^\prime$, respectively, and can therefore be integrated explicitly. In principle, eliminating $\beta$ and $\nu^\prime$ from Eq.~(\ref{eq:5.4c}) using these solutions reduces the system to an integro-differential equation for $\lambda$ alone. For practical purposes, it is preferable to work with a closed system of ordinary differential equations. Multiplying Eq.~(\ref{eq:5.4c}) by $e^{-(\lambda+\nu)/2}$, differentiating with respect to $r$, and substituting Eq.~(\ref{eq:5.4b}) allows one to eliminate both $\beta$ and $\beta^\prime$ simultaneously. Inserting Eq.~(\ref{eq:5.5a}) into the result then eliminates $\nu^{\prime\prime}$, leading to
\begin{equation}\label{eq:5.5b}
  \lambda^{\prime\prime} = \frac{(\lambda^\prime)^2-(\nu^\prime)^2}{2} + \frac{1+(\Lambda r^2-1)e^\lambda}{r}\lambda^\prime + \frac{2e^\lambda-2}{r^2}.
\end{equation}
\end{subequations}
Equations (\ref{eq:5.5a}) and (\ref{eq:5.5b}) form a closed set of structure equations for GCT with $c_i=\delta_{i2}$. They can be used to study configurations supported by the caloric field, with or without a point mass~\footnote{Note that the vacuum condition does not preclude a point mass, as in the Schwarzschild solution of general relativity.} and $\Lambda$.

With $\Lambda=0$, Eq.~(\ref{eq:5.5a}) admits the particular solution $\nu^\prime=0$~\footnote{Because the constant part of $\nu$ can be absorbed by rescaling the time coordinate in Eq.~(\ref{eq:5.1}), the branch $\nu^\prime=0$ can be fixed to $\nu=0$ without loss of generality.}, under which Eq.~(\ref{eq:5.5b}) reduces to
\begin{equation}\label{eq:5.6}
  \lambda^{\prime\prime} = \frac{(\lambda^\prime)^2}{2} + \frac{1-e^\lambda}{r}\lambda^\prime + \frac{2e^\lambda-2}{r^2}.
\end{equation}
This simplification excludes the Schwarzschild solution and is adopted purely for mathematical convenience, without a physical motivation at present. With the transformation $\lambda\mapsto\mu\equiv1/(r\lambda^\prime)$~\footnote{Solving Eq.~(\ref{eq:5.6}) with \texttt{Maple} (Version 2022) yields this transformation (instead of an explicit solution).}, Eq.~(\ref{eq:5.6}) becomes
\begin{equation}\label{eq:5.7}
  \frac{\dx\mu}{\dx\lambda} = (2-2e^\lambda)\mu^3-(2-e^\lambda)\mu^2-\frac{\mu}{2}.
\end{equation}
This is an Abel equation of the first kind \cite{Polyanin2018.book}. The radial coordinate $r$ is implicit in Eq.~(\ref{eq:5.7}). It can be recovered by viewing the transformation as a differential equation for $r(\lambda)$. Integrating yields $r=r_\subb\times\exp[\int_{\lambda_\subb}^\lambda\mu(\tilde{\lambda})\,\dx\tilde{\lambda}]$, with $\lambda(r_\subb)=\lambda_\subb$ imposed as a boundary condition. In Eq.~(\ref{eq:5.7}), one may treat $e^\lambda$ as a new variable to eliminate the explicit exponentials, but this does not appear to be useful for the subsequent analysis.

The compact form of Eq.~(\ref{eq:5.7}) motivates us to consider whether a similar change of variables can further simplify the more general structure equations. For Eq.~(\ref{eq:5.5}), keeping both $\Lambda$ and $\nu^\prime$ general and further introducing $\eta \equiv 1/(r\nu^\prime)$ leads to
\begin{subequations}\label{eq:5.8}
\begin{align}
  \frac{\dx\mu}{\dx\lambda} &= \big(2 - 2e^\lambda + \frac{1}{2\eta^2}\big)\mu^3 \nonumber\\
    &\quad - \big(2 - e^\lambda + \Lambda r^2e^\lambda\big)\mu^2 - \mu/2, \label{eq:5.8a}\\
  \frac{\dx\eta}{\dx\lambda} &= 2\Lambda r^2e^\lambda\mu\eta^2 + \big(1 - \Lambda r^2\big)e^\lambda\mu\eta. \label{eq:5.8b}
\end{align}
\end{subequations}
Because the right-hand sides depend explicitly on $r$, the equation $\dx r/\dx\lambda=1/\lambda^\prime=r\mu$ is required to close the system. When $\Lambda=0$, this additional equation is unnecessary.

\subsection{Vacuum solutions with $\nu=0$}\label{sec:V.B}
This section is devoted to solving the simplest self-consistent structure equation (\ref{eq:5.6}), or, equivalently, the Abel equation (\ref{eq:5.7}), and to analyzing its physical properties. Attempts to obtain closed-form solutions in full generality with \texttt{Maple 2022} and \texttt{Mathematica 13.2} were unsuccessful \cite{NotShown}, so we proceed with numerical integration and series expansions.

Since the right-hand side of Eq.~(\ref{eq:5.6}) contains $r$ in the denominator, the numerical boundary conditions are imposed at a finite radius rather than at $r=0$ \footnote{Note that, for numerical integration, Eq.~(\ref{eq:5.6}) is preferable to Eq.~(\ref{eq:5.7}) for two reasons. It avoids the need to reconstruct $r$ and remains well behaved at points where $\lambda^\prime$ vanishes, which indeed occur in the solutions discussed below.}. Section~\ref{sec:V.B.1} integrates the solution outward to infinity, revealing a $\Lambda$-like regime at finite radius and showing that infinity can host either a coordinate singularity or a physical singularity, depending on the imposed boundary conditions. Section~\ref{sec:V.B.2} integrates the solution inward toward the center, showing that the interior branch either develops a finite-radius wormhole (a coordinate singularity) or terminates at a physical singularity at $r=0$. Figure~\ref{fig:star:09} illustrates these behaviors with two representative solutions. The caloric de Sitter spacetime connects a wormhole to a coordinate singularity at infinity, while the intermediate region is effectively de Sitter. The caloric anti-de Sitter spacetime connects a physical singularity at the center to a physical singularity at infinity, with an intermediate region that is effectively anti-de Sitter. Section~\ref{sec:V.B.3} provides a global classification of the solution space in the $\lambda$--$\mu$ plane (see Fig.~\ref{fig:star:10}) and identifies two additional classes of solutions. The boundaries between regions correspond to four special solutions, two asymptotically flat at infinity and two approaching (anti-)de Sitter behavior near the center. The singularities associated with these special solutions can be read off from Fig.~\ref{fig:star:10}.

\subsubsection{Effective $\Lambda$ and singularity at infinity}\label{sec:V.B.1}
\begin{figure*}[!t]
  \centering
  \includegraphics[width=0.88\textwidth]{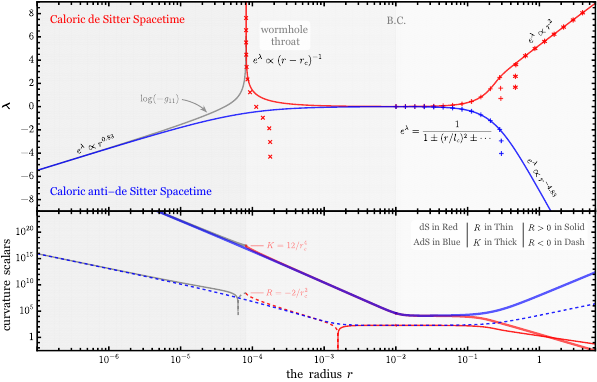}
  \caption{Radial structure of the caloric de Sitter (red) and anti-de Sitter (blue) spacetimes obtained by numerically integrating Eq.~(\ref{eq:5.6}). For the given boundary conditions, the top panel plots $\lambda(r)$, while the bottom panel shows the Ricci scalar $R$ and the Kretschmann scalar $K$. These curvature invariants highlight the approximately constant-curvature regime and help distinguish coordinate singularities from physical singularities. Solid and dashed curves indicate the full numerical solutions, while markers represent the corresponding series expansions. In the figure labels, B.C. marks where the boundary condition is imposed, while dS and AdS denote de Sitter and anti-de Sitter. More details can be found in the main text.}
  \label{fig:star:09}
\end{figure*}

In principle, a dimensionless rescaling of the radial coordinate $r$ is preferable for numerically integrating Eq.~(\ref{eq:5.6}). For simplicity, this rescaling is omitted and both $r$ and the curvature invariants reported below are treated as dimensionless. This choice does not obscure the physical interpretation. The boundary conditions are specified as $\{\lambda=\pm0.01,\,\lambda^\prime=0\}$ at $r=10^{-2}$, and the equation is integrated toward increasing $r$. The resulting profiles are shown in the top light-gray region of Fig.~\ref{fig:star:09}, with red and blue denoting positive and negative boundary values of $\lambda$, respectively. The bottom panel displays the associated Ricci scalar $R$ and Kretschmann scalar $K\equiv R_{\rho\lambda\mu\nu}R^{\rho\lambda\mu\nu}$ using the same color coding. Thin and thick curves represent $R$ and $K$, and solid and dashed styles indicate positive and negative values, respectively.

In the region $r\gtrsim10^{-2}$, both $R$ and $K$ settle to approximately constant values, closely resembling the (anti-)de Sitter geometry \cite{Kim2002}. More precisely, the branches with positive and negative $\lambda$ at the boundary display locally de Sitter and anti-de Sitter behavior, respectively. This indicates that the caloric field in GCT with $c_i=\delta_{i2}$ can mimic an effective $\Lambda$ in certain regions. The difference appears as $r\rightarrow\infty$. For the positive case, $\lambda\rightarrow\infty$ (corresponding to $g_{11}\rightarrow\infty$) as $r\rightarrow\infty$, while $R\rightarrow0$ and $K\rightarrow0$. The vanishing of these curvature invariants suggests that the apparent singularity at infinity may be a coordinate singularity. In this paper, it is simply referred to as a coordinate singularity, without further justification. A detailed analysis of the physical properties of this asymptotic region is deferred to future work. For the negative case, $\lambda\rightarrow-\infty$ (corresponding to $g_{11}\rightarrow0$) as $r\rightarrow\infty$, while both $R$ and $K$ diverge, indicating a genuine physical singularity at infinity. 

Together with the $\Lambda$-cancelling solution in cosmology, these $\Lambda$-related configurations point to a possible deeper connection between $S_\mu$ and $\Lambda$. To emphasize this connection, the two solutions in Fig.~\ref{fig:star:09} are referred to as the caloric de Sitter and caloric anti-de Sitter spacetimes. The emergence of an effective $\Lambda$ in a local gravitational system is non-trivial in GCT: although the field equations are partly inspired by nonlocal RT gravity \cite{Maggiore2014.PRD.89.043008}, the caloric field is not introduced as a cosmological dark-energy sector. By contrast, nonlocal RT gravity was designed to model cosmological dark energy, so $\Lambda$-like behavior is naturally expected in local systems as well \cite{Kehagias2014.JHighEnergyPhys.08.029}. 

To clarify analytically both the $\Lambda$-like behavior observed in Fig.~\ref{fig:star:09} and the asymptotic behavior at infinity, we solve the equation via a series expansion. A naive reading of the present numerics suggests that the $\Lambda$-like behavior around $r\approx10^{-2}$ may persist down to the center, $r=0$ \footnote{Section~\ref{sec:V.B.2} shows that the numerical $\Lambda$-like behavior does not extend smoothly to $r=0$. The expansion about $r=0$ is therefore used only as an analytic probe of the $\Lambda$-like branch, not as a faithful continuation of the plotted profile to the origin.}. This motivates an analytic study of the $\Lambda$-like behavior through an expansion about $r=0$. Choosing an appropriate series ansatz is essential, as it controls both the existence of a consistent local solution and the convergence rate. Near $r=0$, the similarity to (anti-)de Sitter spacetime motivates \footnote{For the regular analytic branch considered here, if odd powers are included in a more general series, the differential equation itself forces their coefficients to vanish.}
\begin{equation}\label{eq:5.9}
  \lambda = - \log \bigg[ \sum_{i=0}^{+\infty} \alpha_i\cdot\Big(\frac{r}{l_c}\Big)^{2i} \bigg],
\end{equation}
where $\alpha_i$ are dimensionless coefficients and $l_c>0$ denotes a characteristic length scale. When $\alpha_1\neq0$, its magnitude can be absorbed into $l_c$, so that one may choose $\alpha_1=\pm1$ without loss of generality. In contrast, $\alpha_0$ cannot be fixed in this manner, since the constant term is independent of $l_c$ and the value of $g_{11}$ carries direct physical meaning, for example in the presence of a topological defect \cite{Barriola1989.PRL.63.341}. Substituting Eq.~(\ref{eq:5.9}) into Eq.~(\ref{eq:5.6}) and expanding about $r=0$ yields the coefficient conditions
\begin{equation}\label{eq:5.10}
\begin{cases}
  \alpha_0 (\alpha_0-1) = 0,                 & \text{for the $r^0$ term,} \\
  \alpha_1 (\alpha_0-1) = 0,                 & \text{for the $r^2$ term,} \\
  \alpha_2 (2\alpha_0+3) - 2\alpha_1^2 = 0,  & \text{for the $r^4$ term,} \\
  \cdots.
\end{cases}
\end{equation}
For the even-power ansatz (\ref{eq:5.9}), the coefficients of all odd powers in the residual equation vanish identically. There are three regular analytic branches of solutions, characterized by $(\alpha_0,\alpha_1)=(1,0)$, $(1,-1)$, and $(1,1)$, which correspond to the Minkowski metric, the caloric de Sitter spacetime, and the caloric anti-de Sitter spacetime, respectively. For the two caloric branches, we compute the expansion coefficients up to $i=128$ in Eq.~(\ref{eq:5.9}). For large $i$, they follow an empirical scaling $\alpha_i\approx\pm e^{-i}/i$, with an apparently irregular sign pattern. Fitting the numerical profiles with the truncated series yields $l_c\approx0.174$ for both cases. In Fig.~\ref{fig:star:09}, the corresponding truncated series solutions are shown as plus markers in the region $r\gtrsim10^{-2}$. These series solutions match the numerical integration well at small $r$, but break down near $r\approx0.29\approx 5l_c/3$. A lower truncation at $i=64$ confirms the location of this divergence \cite{NotShown}, indicating that the series (\ref{eq:5.9}) has a finite radius of convergence.

The expansion at infinity is more complicated. After several trials, we adopt the ansatz
\begin{equation}\label{eq:5.11}
  \lambda = \log \bigg[ \Big(\frac{r}{l_c}\Big)^{n}\cdot\sum_{i=0}^{+\infty} \alpha_i\cdot\Big(\frac{l_c}{r}\Big)^{i} \bigg],
\end{equation}
where the constant $n$ denotes the leading exponent and $\alpha_i$ and $l_c$ are defined as before. Importantly, $n$ is not assumed to be an integer. The logarithm requires $\alpha_0>0$, and when $n\neq0$ we may set $\alpha_0=1$ without loss of generality. Substituting Eq.~(\ref{eq:5.11}) into Eq.~(\ref{eq:5.6}) and expanding at $r=+\infty$ gives
\begin{align}\label{eq:5.12}
  & \Big(\frac{r}{l_c}\Big)^n \cdot \bigg[ \frac{(2-n)\alpha_0}{r^2} + \frac{(3-n)\alpha_1 l_c}{r^3} + \mathcal{O}\big(r^{-4}\big) \bigg] \nonumber\\
  & + \bigg[ \frac{n^2+4n-4}{2r^2} - \frac{(n+3)\alpha_1 l_c}{\alpha_0 r^3} + \mathcal{O}\big(r^{-4}\big) \bigg] = 0. 
\end{align}
Unlike the expansion about $r=0$, matching orders in Eq.~(\ref{eq:5.12}) does not immediately yield a closed recurrence relation for the coefficients because $n$ is unknown. To proceed, we should classify the dominant terms in this equation. If $n>0$, the leading contribution scales as $r^{n-2}$ and its coefficient must vanish, thereby fixing $n=2$. Solving the remaining coefficient conditions yields $\{\alpha_2=-2,\,\alpha_4=-6,\,\alpha_6=-44,\,\cdots\}$, while all odd-index coefficients vanish. We compute the coefficients up to $i=100$ in Eq.~(\ref{eq:5.11}) and, with the fitted value $l_c\approx0.071$, plot the truncated series as sixfold-cross markers in the upper-right corner of Fig.~\ref{fig:star:09}. The result confirms $r^2$ as the correct leading behavior and shows that this series also has a finite convergence boundary. After rescaling the $r$ coordinate, the metric asymptotically takes the form $\dx s^2=-c^2\dx t^2 + \dx r^2 + l\cdot r(\dx\theta^2+\sin^2\theta\dx\varphi^2)$, where $l$ is a characteristic length scale. If $n<0$, the prefactor $(r/l_c)^n$ suppresses the first bracket, and the dominant contribution comes from the $r^{-2}$ term in the second bracket. Setting its coefficient to zero (equivalently solving $n^2+4n-4=0$) gives $n=-2\pm\sqrt{8}$. The branch compatible with $n<0$ is $n=-2-\sqrt{8}\approx-4.83$. However, in this case we do not find a self-consistent higher-order series solution within the form (\ref{eq:5.11}), in the sense that no choice of $\{\alpha_i\}$ makes the coefficients vanish order by order in Eq.~(\ref{eq:5.12}). This invalidates the original series ansatz for the case $n<0$. Nevertheless, the caloric anti-de Sitter numerical solution in Fig.~\ref{fig:star:09} supports $r^{-4.83}$ as the correct leading behavior.

If $n=0$ in Eq.~(\ref{eq:5.12}), then, as in the discussion below Eq.~(\ref{eq:5.9}), the magnitude of $\alpha_1$ can be absorbed into $l_c$, so that one may set $\alpha_1=\pm1$ while treating $\alpha_0$ as a free parameter. In this case, a self-consistent series solution exists, with
\begin{equation}\label{eq:5.13}
  \alpha_0=1, \
  \alpha_1=\pm1, \ 
  \alpha_2=\frac{9}{8}\alpha_1^2, \
  \alpha_3=\frac{21}{16}\alpha_1^3, \
  \cdots.
\end{equation}
Here, the solution (not a convention) $\alpha_0=1$ implies that the metric is asymptotically flat at infinity, whereas $\alpha_1\neq0$ implies the presence of an $r^{-1}$ potential term at large $r$, corresponding to a harmonic solution of the Laplace equation. By contrast, Sec.~\ref{sec:IV} shows that the weak-field regime does not excite the $r^{-1}$ mode in metric perturbations. The nonzero $\alpha_1$ here should therefore originate from strong-field excitation. Neither of the two numerical solutions in Fig.~\ref{fig:star:09} corresponds to Eq.~(\ref{eq:5.13}). The interior structure associated with this series solution, together with its role in the full solution space, is discussed in Sec.~\ref{sec:V.B.3}.

\subsubsection{Wormhole and central singularity}\label{sec:V.B.2}
In Fig.~\ref{fig:star:09}, the series solutions based on Eq.~(\ref{eq:5.9}) match the full numerical solutions well over $r\gtrsim10^{-2}$. The coefficient $\alpha_0=1$ implies $\lim_{r\rightarrow0}\lambda=0$, so that the central value is fixed. However, from the viewpoint of the differential equation, this value should, in principle, be freely specified as part of the boundary conditions. In other words, the expansion contains only one genuine free parameter, $l_c$, whereas Eq.~(\ref{eq:5.6}) is second order and should admit two independent integration constants. This apparent tension motivates a closer examination of whether the series (\ref{eq:5.9}) can truly capture the local behavior in the limit $r\rightarrow0$.

To clarify the above issue, we integrate Eq.~(\ref{eq:5.6}) numerically toward smaller $r$ using the boundary data of the previous subsubsection, and plot the results in Fig.~\ref{fig:star:09}. In both branches, the inward integration eventually departs from the series (\ref{eq:5.9}). For the de Sitter branch, $\lambda$ diverges to $+\infty$ at a finite radius $r\approx8.14875\times10^{-5}$, which we denote by $r_c$. Numerically, the curvature scalars $R$ and $K$ remain finite at $r=r_c$, suggesting that the singular behavior there is a coordinate singularity rather than a physical one. To support this conclusion, we consider a series expansion about $r=r_c$ of the form
\begin{equation}\label{eq:5.14}
  \lambda = \log \bigg[ \Big(\frac{r}{r_c}-1\Big)^{n} \cdot \sum_{i=0}^{+\infty} \alpha_i\cdot\Big(\frac{r}{r_c}-1\Big)^{i} \bigg].
\end{equation}
We impose $n<0$ (with $\alpha_0\neq0$) to capture the divergent behavior as $r\rightarrow r_c^{+}$. Substituting Eq.~(\ref{eq:5.14}) into Eq.~(\ref{eq:5.6}) and expanding about $r=r_c$ yields
\begin{align}
  & \Big(\frac{r}{r_c}-1\Big)^n \cdot \Big[ \frac{n\alpha_0}{1-r/r_c} + \mathcal{O}(1) \Big] \nonumber\\
  & + \bigg[ \frac{n(n+2)}{2(1-r/r_c)^2} + \mathcal{O}(\frac{1}{1-r/r_c}) \bigg] = 0. 
\end{align}
As in the analysis below Eq.~(\ref{eq:5.12}), we should distinguish cases according to the value of $n$. We find that no consistent solution exists for either $-1<n<0$ or $n<-1$. For the unique admissible value $n=-1$, both brackets contribute at leading order, scaling as $(1-r/r_c)^{-2}$, and the resulting coefficients are given by
\begin{equation}\label{eq:5.16}
  \alpha_0=\frac{1}{2}, \
  \alpha_1=-\frac{1}{4}, \
  \alpha_2=-\frac{23}{72}, \
  \cdots.
\end{equation}
We compute coefficients up to $i=48$ in Eq.~(\ref{eq:5.14}) and overlay the corresponding series solution in Fig.~\ref{fig:star:09} using cross markers. An independent check is obtained by plotting $\lambda$ against $\log(r-r_c)$, where the slope verifies the divergent index $n=-1$ and the agreement with the numerical solution near $r=r_c$ is excellent \cite{NotShown}. The convergence radius is less sharply defined, since the truncated series drifts away from the numerical profile smoothly rather than through an abrupt breakdown. The origin of this behavior is not resolved in this paper. Nevertheless, this does not affect the local use of the expansion near $r=r_c$. In particular, the expansion gives $R=-2/r_c^2+\mathcal{O}(r-r_c)$ and $K=12/r_c^4+\mathcal{O}(r-r_c)$, confirming that both remain finite at $r=r_c$. The bottom panel of Fig.~\ref{fig:star:09} marks these constant terms with two light-red segments, illustrating quantitative agreement with the numerical results.

With $\alpha_0=1/2$, the solution can be recast as $e^\lambda=[1+\mathcal{O}(r-r_c)]/(1-r_c^2/r^2)$. Together with $g_{00}=-c^2$, it follows that the leading-order geometry near $r=r_c$ is locally isometric to the Ellis wormhole, which connects two asymptotically Minkowski Universes \cite{Ellis1973.JMathPhysNY.14.104,*Ellis1974}. The same wormhole geometry was later independently rediscovered by Morris and Thorne \cite{Morris1988.AmJPhys.56.395,James2015}. For the caloric de Sitter spacetime, the coordinate transformation $r\mapsto l=\pm\sqrt{r^2-r_c^2}$ \cite{Morris1988.AmJPhys.56.395} makes the wormhole structure manifest. A key difference from the Ellis wormhole is that, in the local region, the throat connects two de Sitter-like regions with approximately constant curvature invariants. The coordinate singularity at infinity may further enrich the global properties of the spacetime and warrants a dedicated analysis in future work.

To obtain a formal continuation of the solution into the region $r<r_c$, we recast Eq.~(\ref{eq:5.6}) as a differential equation for $g_{11}$ \cite{NotShown}. Boundary data are then supplied at a radius slightly below $r_c$ using the local series (\ref{eq:5.14}), and the equation is integrated further inward. In this region, $g_{11}$ becomes negative. We therefore plot $\log(-g_{11})$ in gray in Fig.~\ref{fig:star:09}, together with the corresponding curvature invariants. This part of the solution is not expected to carry a direct physical meaning.

For the anti-de Sitter branch, the inward integration proceeds smoothly from $r=10^{-2}$ to the center at $r=0$, and $\lambda$ diverges to $-\infty$ as $r\rightarrow0$ (see Fig.~\ref{fig:star:09}). Both curvature scalars diverge in this limit, indicating a genuine physical singularity at the center. To characterize the associated asymptotic behavior, we modify the series ansatz (\ref{eq:5.9}) to
\begin{equation}\label{eq:5.17}
  \lambda = \log \bigg[ \Big(\frac{r}{l_c}\Big)^{n}\cdot\sum_{i=0}^{+\infty} \alpha_i\cdot\Big(\frac{r}{l_c}\Big)^{i} \bigg],
\end{equation}
where the leading exponent $n$ is allowed to be non-integer. Substituting Eq.~(\ref{eq:5.17}) into Eq.~(\ref{eq:5.6}) and expanding about $r=0$ may, in principle, allow one to determine $n$ and $\{\alpha_i\}$ order by order. In practice, as in the anti-de Sitter analysis below Eq.~(\ref{eq:5.12}), the dominant term fixes only a special exponent, $n=\sqrt{8}-2\approx0.83$, and no self-consistent higher-order coefficient solution is found within this form. Nevertheless, the numerical profile in Fig.~\ref{fig:star:09} is consistent with $n\approx0.83$. Physically, the resulting metric provides an explicit example of a naked singularity \cite{Penrose1969.NotEngJ.1.252} embedded in an anti-de Sitter-like region.

\subsubsection{Global analysis in the $\lambda$--$\mu$ plane}\label{sec:V.B.3}
\begin{figure*}[!t]
  \centering
  \includegraphics[width=0.8\textwidth]{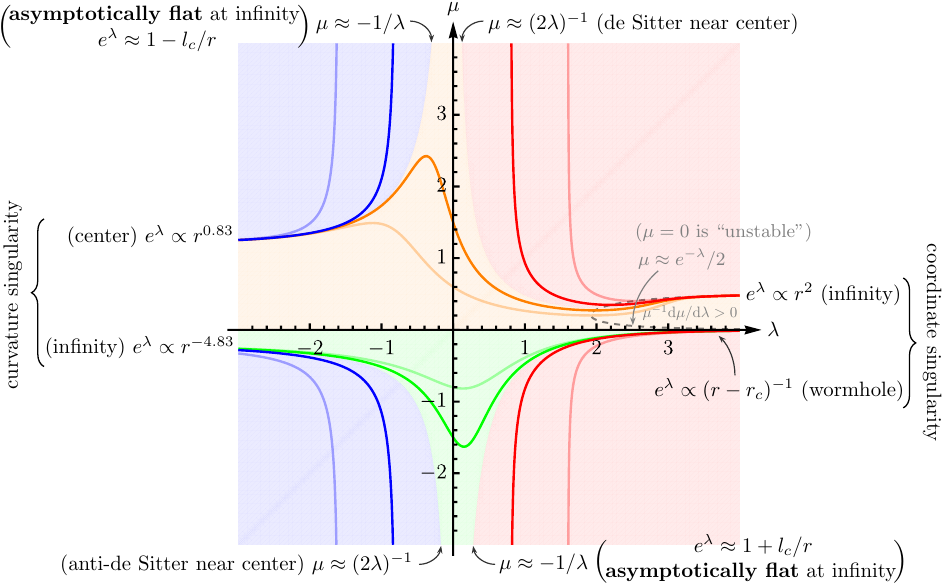}
  \caption{Global classification of the vacuum solutions with $\nu=0$ in the $\lambda$--$\mu$ plane. Each shaded region represents a distinct branch of solutions, and two representative trajectories with different opacity are overlaid for each branch. Boundaries between regions, apart from the $\lambda$-axis, correspond to special separatrix solutions. The associated singularity types and leading asymptotic scalings are indicated in the figure.}
  \label{fig:star:10}
\end{figure*}

Since Eq.~(\ref{eq:5.7}) defines a single-valued slope field for $\dx\mu/\dx\lambda$, integral curves in the $\lambda$--$\mu$ plane do not intersect, making this plane well suited for a global classification of solutions. In Fig.~\ref{fig:star:10}, the red and blue regions indicate the subsets of solution space occupied by the caloric de Sitter and caloric anti-de Sitter 
branches~\footnote{If $\lambda$ is taken too large when $\lambda^\prime=0$, the approximately constant-curvature region may shrink to a very small interval, rendering the $\Lambda$-like behavior barely visible. Even so, the de Sitter-related terminology is retained in the main text.}, respectively. Two representative solutions are overlaid within each region and are distinguished by their opacity. Note that the divergence $\mu\rightarrow\pm\infty$ at finite $\lambda$ corresponds to $\lambda^\prime=0$, and thus the corresponding trajectories should be \textit{linked} between $\mu=\infty$ and $-\infty$.

The asymptotic limits of $\mu$ as $\lambda\rightarrow\pm\infty$ are more physically informative~\footnote{If one assumes that the asymptotic limit of $\mu$ exists and remains finite, then imposing $\dx\mu/\dx\lambda=0$ in Eq.~(\ref{eq:5.7}) yields the candidate limits $\{0,1/2\}$ as $\lambda\rightarrow+\infty$ and $\{0,(1\pm\sqrt{2})/2\}$ as $\lambda\rightarrow-\infty$. Although this argument is not a rigorous proof, it provides a useful consistency check on the numerical behavior.}. In quadrant I, the numerical trajectories approach $\mu\rightarrow1/2$. Accordingly, reconstructing the radial coordinate $r$ shows that $\lambda\rightarrow\infty$ maps to $r\rightarrow\infty$. Moreover, the limiting value $1/2$ implies $e^\lambda\propto r^2$ at large radius, consistent with the series solution discussed below Eq.~(\ref{eq:5.12}). In quadrant IV, the numerical trajectories satisfy $\mu\rightarrow0^{-}$ as $\lambda\rightarrow\infty$. To determine the corresponding $r$, the decay rate of $\mu(\lambda)$ is required. For $\lambda\gg1$ and $|\mu|\ll1$, Eq.~(\ref{eq:5.7}) can be approximated by the Bernoulli equation $\dx\mu/\dx\lambda \approx e^\lambda\mu^2-\mu/2$, whose solution behaves as $\mu\approx1/(\mathrm{const.}\times e^{\lambda/2}-2e^\lambda)\approx -e^{-\lambda}/2$. Integrating $\mu(\lambda)$ yields $r\approx r_c\cdot\exp(e^{-\lambda}/2)$, or equivalently $e^\lambda\approx1/[2\log(r/r_c)]\approx r_c/[2(r-r_c)]$, where $r_c$ is a constant representing the minimal radius reached along the trajectory. This result is consistent with Eq.~(\ref{eq:5.16}). As $\lambda\rightarrow-\infty$, $\mu$ tends to $(1+\sqrt{2})/2$ in quadrant II and to $(1-\sqrt{2})/2$ in quadrant III. These limits corroborate the asymptotic behaviors previously derived near $r\rightarrow0$ and near $r\rightarrow\infty$ for the caloric anti-de Sitter solution.

In Fig.~\ref{fig:star:10}, the orange and green domains correspond to two new solution branches. The orange branch connects a central physical singularity ($e^\lambda\propto r^{0.83}$) to a coordinate singularity at infinity ($e^\lambda\propto r^2$). The green branch connects a finite-radius wormhole to a physical singularity at infinity ($e^\lambda\propto r^{-4.83}$). Their behavior near the $\lambda$-axis, especially for large positive $\lambda$, can be understood directly from Eq.~(\ref{eq:5.7}) \footnote{Such near-$\lambda$-axis behavior may be useful for diagnosing candidate horizons in the more general system (\ref{eq:5.8}), where $\nu$ is allowed to vary.}. The green trajectory follows the Bernoulli asymptotics, yielding $\mu\approx-e^{-\lambda}/2\rightarrow0^{-}$. To analyze the orange trajectory, the region near $\mu=0$ where $\mu^{-1}\dx\mu/\dx\lambda>0$ is first considered, bounded by the gray dashed curve in Fig.~\ref{fig:star:10}. Its lower boundary satisfies $\mu\approx e^{-\lambda}/2$ for $\lambda\gg1$. In the approach to $\mu=0$, the evolution is controlled by the $-\mu/2$ term in the Bernoulli equation, which implies $\mu\propto e^{-\lambda/2}$. This scaling forces the orange trajectory to intersect the lower dashed boundary, after which $\dx\mu/\dx\lambda>0$ and the orbit departs from the $\lambda$-axis instead of approaching it asymptotically. This picture is consistent with the numerical results.

The boundaries separating differently colored regions correspond to special solutions. Their limits as $\lambda\rightarrow\pm\infty$ reproduce the singular behaviors discussed above, while the focus here is the behavior as $\mu\rightarrow\pm\infty$. We conjecture that these boundaries are tied to the series~(\ref{eq:5.9}) with its solutions and to the series~(\ref{eq:5.11}) with solution (\ref{eq:5.13}). The motivation is that these series solutions exist but are not exactly captured by the generic branches discussed above. They should therefore appear as distinguished boundaries in the $\lambda$--$\mu$ plane. A direct calculation gives $\mu\approx1/(2\lambda)$ for series~(\ref{eq:5.9}), independent of whether the branch is de Sitter or anti-de Sitter, and it yields $\mu\approx-1/\lambda$ for series~(\ref{eq:5.11}), independent of the value of $\alpha_1$ in solution (\ref{eq:5.13}). Extending Fig.~\ref{fig:star:10} to a wider range can verify these asymptotic relations \cite{NotShown}. Accordingly, Fig.~\ref{fig:star:10} annotates each boundary solution with its asymptotic behavior in the limit $\mu\rightarrow\pm\infty$ and the associated geometric interpretation inferred from the series expansions (e.g., de Sitter and asymptotically flat). Conversely, this figure helps determine the asymptotic behavior at infinity associated with series~(\ref{eq:5.9}) and the interior structure associated with Eq.~(\ref{eq:5.13}). In particular, the boundary in quadrant II describes a naked singularity with an asymptotically flat structure at infinity.

These separatrix boundaries also provide a useful guide for interpreting the intermediate regime of the generic solutions. For example, if an orange trajectory stays sufficiently close to the boundary while traversing quadrants II and I, it develops an extended intermediate region characterized by $e^\lambda\approx1-l_c/r$ and de Sitter-like behavior.

\subsection{Vacuum solutions with $\nu=\nu(r)$}
The previous section shows that the vacuum system with $\nu=0$ admits a variety of singular configurations, including wormholes and naked singularities. It does not, however, admit black-hole solutions with a regular event horizon, which are of direct astrophysical interest. To investigate black holes within the same $c_i=\delta_{i2}$ vacuum slice, this subsection relaxes the condition $\nu=0$ while retaining $\Lambda=0$ for simplicity. The resulting system is governed by Eq.~(\ref{eq:5.8}). Note that $\Lambda=0$ is assumed in Eq.~(\ref{eq:5.8}) throughout this subsection. Unlike the global classification in the previous subsection, the discussion here is limited to illustrative examples, and no black-hole solutions beyond Schwarzschild are yet identified \cite{NoHairCaveat}.

\subsubsection{Analysis in the $\lambda$--$\mu$--$\eta$ space}\label{sec:V.C.1}
\begin{figure*}[!t]
  \centering
  \includegraphics[width=0.69\textwidth]{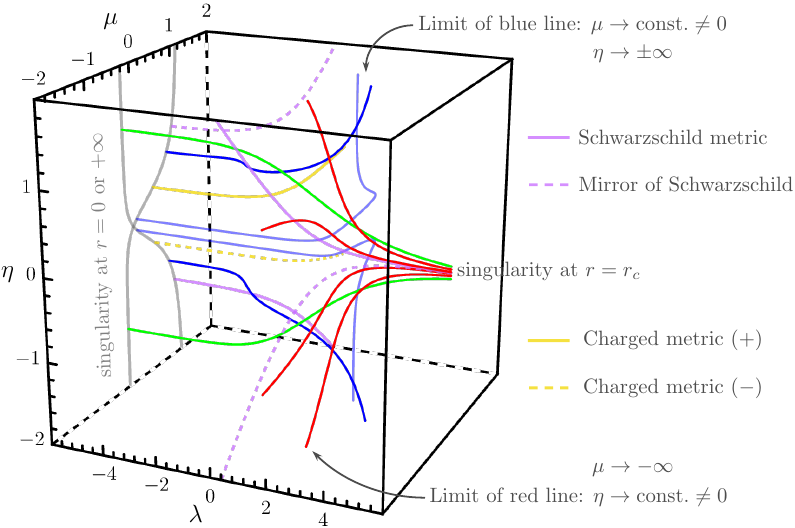}
  \caption{Solutions of Eq.~(\ref{eq:5.8}) for $\Lambda=0$ in the $(\lambda,\mu,\eta)$ space. The red, green, and blue curves are obtained by numerically solving the differential equations, and their limiting behaviours are labeled in the figure. The light-purple solid curve is the Schwarzschild metric ($\lambda<0$ corresponds to negative mass), while the light-purple dashed curve is its mirror with respect to the $\eta=0$ plane and is also an exact solution. The yellow solid and dashed curves show the charged solution in Sec.~\ref{sec:V.C.3}. The trajectories shown are illustrative examples but do not constitute a complete classification.}
  \label{fig:star:11}
\end{figure*}

Equation~(\ref{eq:5.8}) defines a first-order system of ordinary differential equations for $\{\mu,\eta\}$ with $\lambda$ as the evolution variable, so its solutions can be represented as trajectories in the three-dimensional space $(\lambda,\mu,\eta)$. For generic boundary data, Fig.~\ref{fig:star:11} displays three distinct trajectory families in this space, shown in red, green, and blue~\footnote{The boundary data are taken to be $(\lambda,\mu,\eta)=(2,-0.5,\pm0.5)$ and $(2,-0.5,\pm1.5)$ for the red trajectories, $(0,-0.2,\pm1)$ for the green trajectories, and $(0,0.1,\pm0.9)$ and $(2,0.5,\pm2.2)$ for the blue trajectories.}. Despite their differences, they exhibit the same qualitative behavior. For the green and blue trajectories, $\mu$ and $\eta$ both tend to nonzero constants as $\lambda\rightarrow-\infty$. For the red and green trajectories, $\mu\rightarrow0^{-}$ and $\eta\rightarrow0^{\pm}$ as $\lambda\rightarrow+\infty$. In addition, either $\mu$ or $\eta$ may diverge at a finite value of $\lambda$. Each of these features has a direct counterpart in Fig.~\ref{fig:star:10}. It should be emphasized that these three solutions do not cover the full range of solution types.

To obtain the $\lambda\rightarrow-\infty$ asymptotics, we can substitute $\{\lambda=-\infty,\,\mu=\mathrm{const.},\,\eta=\mathrm{const.}\}$ into $\{\dx\mu/\dx\lambda=0,\,\dx\eta/\dx\lambda=0\}$ \footnote{Again, this is not a rigorous proof, since it assumes that $\mu$ and $\eta$ approach finite constants as $\lambda\rightarrow-\infty$.}. The result shows that the $\eta$-equation is satisfied identically, whereas the $\mu$-equation yields
\begin{equation}\label{eq:5.18}
  \lim_{\lambda\rightarrow-\infty} \eta^2 = \frac{\mu^2}{1+4\mu-4\mu^2}.
\end{equation}
This result defines two branches, shown as gray curves in Fig.~\ref{fig:star:11}, and the numerical trajectories are consistent with these asymptotics. Reconstructing the radial coordinate $r$ by integrating $\mu(\lambda)$ indicates that $\lambda\rightarrow-\infty$ corresponds to either $r\rightarrow0$ or $r\rightarrow\infty$. Physically, this asymptotic limit corresponds to a central singularity or a singularity at infinity~\footnote{The present analysis does not establish whether these are physical singularities.}, in direct analogy with Fig.~\ref{fig:star:10}. In the limit $|\eta|\rightarrow\infty$, Eq.~(\ref{eq:5.8}) reduces to Eq.~(\ref{eq:5.7}), and Eq.~(\ref{eq:5.18}) agrees with the limiting values ($\mu=(1\pm\sqrt{2})/2$) reported in Sec.~\ref{sec:V.B.3}.

The blue and red trajectories diverge at nonzero and finite values of $\lambda$. A projection onto the $\mu$--$\eta$ plane \cite{NotShown} shows that the blue trajectory approaches $\eta\rightarrow\pm\infty$ while $\mu\rightarrow\mathrm{const.}\neq0$, whereas the red trajectory approaches $\mu\rightarrow-\infty$ while $\eta\rightarrow\mathrm{const.}\neq0$. From the definitions of $\mu$ and $\eta$, these divergences correspond to the vanishing of either $\lambda^\prime$ or $\nu^\prime$. The manner in which the divergent endpoints are globally connected is left open here.

The asymptotic regime $\lambda\rightarrow+\infty$ is now examined. Motivated by the Schwarzschild solution in Sec.~\ref{sec:V.C.2}, exponential ansatzes are adopted for both $\mu(\lambda)$ and $\eta(\lambda)$ with undetermined coefficients and decay rates. Substituting these forms into Eq.~(\ref{eq:5.8}) and requiring the leading terms to vanish yields two solution branches. The first branch has the asymptotic form $\mu\approx-e^{-\lambda}$ and $\eta\approx\pm e^{-\lambda}$, whereas the second has $\mu\approx-e^{-\lambda}/2$ and $\eta\approx\alpha\cdot e^{-\lambda/2}$, where $\alpha$ is an arbitrary constant. Quantitative calculations \cite{NotShown} confirm that the red and green numerical trajectories in Fig.~\ref{fig:star:11} approach the second branch~\footnote{The Schwarzschild metric corresponds to the first branch and is discussed in Sec.~\ref{sec:V.C.2}.}. For this branch, reconstructing $r$ gives $e^\lambda\approx [2\log(r/r_c)]^{-1} \approx (1-r_c^2/r^2)^{-1}$ and $e^\nu\propto\exp[\sqrt{2\log(r/r_c)}/\alpha]\approx1+\alpha^{-1}\cdot\sqrt{r^2/r_c^2-1}$ near $r=r_c$, where $r_c$ is a positive constant. Hence $\lambda\rightarrow+\infty$ corresponds to $r\rightarrow r_c$, a finite radius at which $g_{11}$ diverges while $g_{00}$ remains finite and nonzero. We also confirm that the curvature invariants $R$ and $K$ remain finite at $r=r_c$, indicating a coordinate singularity. Together with the discussion in the previous subsection and Ref.~\cite{Morris1988.AmJPhys.56.395}, the preceding evidence supports interpreting this solution as a wormhole. This behavior closely parallels that found in quadrant IV of Fig.~\ref{fig:star:10}.

\subsubsection{Mirror of the Schwarzschild metric}\label{sec:V.C.2}
The red, green, and blue trajectories in Fig.~\ref{fig:star:11} exhibit mirror symmetry about the $\eta=0$ plane~\footnote{Equation~(\ref{eq:5.8}) provides a rigorous proof of this symmetry. Note that this symmetry disappears if $\Lambda\neq0$.}. This implies that each solution curve admits a mirror partner under $\mu\mapsto\mu$ and $\eta\mapsto-\eta$. The Schwarzschild metric, which corresponds to $\mu=-\eta=1/(1-e^\lambda)$ and is plotted as the light-purple solid curve, is an exact solution of Eq.~(\ref{eq:5.8}). Its reflection yields the mirror solution $\mu=\eta=1/(1-e^\lambda)$, which is plotted as the light-purple dashed curve in Fig.~\ref{fig:star:11}. Reconstructing $r$ yields the corresponding line element
\begin{equation}\label{eq:5.19}
  \dx s^2 = \frac{- c^2\dx t^2 + \dx r^2}{1-r_c/r} + r^2(\dx\theta^2+\sin^2\theta\dx\varphi^2),
\end{equation}
where $r_c$ is a constant. The $\beta$ component of $S_\mu$ follows from Eq.~(\ref{eq:5.4c}) with $\Lambda=0$, giving $\beta = r_c/[2r(r_c-r)]$. This spacetime is asymptotically flat. The Ricci scalar is $R=r_c^2/[r^3(r_c-r)]$. It follows that, for $r_c>0$, the surface $r=r_c$ is a naked curvature singularity. For $r_c<0$, the curvature divergence instead occurs at the center~\footnote{This is trivial --- we can set the Schwarzschild radius to be negative in the Schwarzschild metric to construct a naked singularity.}.

In the weak-field limit of the metric (\ref{eq:5.19}), the phenomenology differs qualitatively for nonrelativistic massive particles and for photons. For slowly moving particles, the dynamics are governed primarily by $g_{00}$ \cite{Bertschinger2011.PhilTransRSocA.369.4947}, yielding an inverse-square effective force that is either attractive ($r_c<0$) or repulsive ($r_c>0$). For photons, the leading-order deflection vanishes because $\gamma_\mathrm{PPN}=-1$ \cite{Will2014.LRR.17.4,Bertschinger2011.PhilTransRSocA.369.4947}. Observational signatures associated with this property warrant further study.

\subsubsection{An exact charged solution}\label{sec:V.C.3}
Here a \textit{Guess \& Check} strategy is employed to construct new exact solutions. Inspired by the series analysis in Sec.~\ref{sec:V.C.1} and the exact solutions in Sec.~\ref{sec:V.C.2}, consider the ansatz $\mu=[2(\alpha-e^\lambda)]^{-1}$ with an undetermined constant $\alpha$. Substituting this form into Eq.~(\ref{eq:5.8a}) determines $\eta(\lambda)$, after which Eq.~(\ref{eq:5.8b}) provides a self-consistency check and may fix $\alpha$. This step is not guaranteed to succeed, because Eq.~(\ref{eq:5.8b}) may instead rule out the ansatz. For the present choice, the ansatz survives the self-consistency check, fixing $\alpha=1$ and yielding $\mu=[2(1-e^\lambda)]^{-1}$ and $\eta=\pm[8(1-e^\lambda)]^{-1/2}$. This solution is shown in Fig.~\ref{fig:star:11} in yellow, with the solid and dashed curves corresponding to the plus and minus signs, respectively. Note that the form of $\eta$ restricts the solution to the domain $\lambda<0$.

Reconstructing $r$ yields the corresponding line element
\begin{align}\label{eq:5.20}
  \dx s^2 &= - \bigg[\frac{\sqrt{1+r^2/Q}-1}{\sqrt{1+r^2/Q}+1}\bigg]^{\pm\sqrt{2}} \cdot c^2\dx t^2 \nonumber\\
    &\quad + \frac{\dx r^2}{1+Q/r^2} + r^2(\dx\theta^2+\sin^2\theta\dx\varphi^2),
\end{align}
where $Q$ is a positive constant. In the limit $r\rightarrow\infty$, $g_{00}\approx-c^2(1\mp\sqrt{8Q}/r)$. The associated vector-field configuration is $\beta=\pm(2r^2+2r^4/Q)^{-1/2}$. The structure of $g_{11}$ closely resembles that of a purely charged Reissner-Nordstr{\"o}m metric, so it is natural to interpret $Q$ as a charge associated with the caloric field. Evaluating the metric functions shows that the spacetime has no finite-radius black-hole horizon. Instead, it contains a naked singularity at $r=0$, as indicated by $R=-2Q/r^4$.

A nontrivial black-hole solution might be obtained by extending the metric (\ref{eq:5.20}) to include a Schwarzschild-like mass term. We have not yet succeeded in constructing such a solution.

\section{Conclusions and prospects}\label{sec:VI}
This paper proposes the \textit{gravitational caloric theory} (GCT), featuring a new vector field $S_\mu$ that is sourced by fluid thermodynamic quantities and minimally coupled to spacetime. Section~\ref{sec:II} outlines the theory construction, and the resulting field equations are given in Eq.~(\ref{eq:2.1}). The vector field equation is designed so that the standard fluid satisfies the usual energy-momentum conservation law. The key ingredients include a \textit{kinetic} contribution built from $\nabla_\mu S_\nu$ and its symmetric combinations, motivated by the nonlocal RT gravity \cite{Maggiore2014.PRD.89.043008}, a \textit{potential} contribution of the form $S_\mu S_\nu$ suggested by dimensional analysis, and a fluid-field interaction $J_\mu S_\nu$ that controls the stability of the critical point $F_1$ in the cosmological dynamical system. A distinctive feature is that the effective gravitational tensor of the vector field contains terms linear in $S_\mu$, which sets GCT apart from existing generalized vector theories \cite{Tasinato2014.JHighEnergyPhys.04.067,Heisenberg2014.JCAP.05.015}. At present, a Lagrangian corresponding to the field equations has not yet been successfully constructed. The theory contains eight order-unity dimensionless parameters. The roles of six parameters are examined, at different levels, in cosmology, linear perturbations, and static spherically symmetric solutions, while a systematic study of the remaining two parameters is deferred to future work.

The initial motivation for proposing GCT is the EDE coincidence problem: why EDE becomes dynamically important precisely near the epoch of matter-radiation equality \cite{Lin2019.PRD.100.063542,Sakstein2020.PRL.124.161301}. A viable resolution of this issue requires a physical mechanism tied to matter-radiation equality that can trigger EDE, thereby avoiding the need to specify the EDE energy scale by hand or to fine-tune the initial conditions. Following Tian \& Zhu \cite{Tian2021.PRD.103.043518,Tian2023.PRD.107.103507}, the cosmic radiation-matter transition is encoded by the total fluid EoS parameter $w_\subF$, which in turn sources the EDE field $S_\mu$ in GCT. The main advance of the present framework is that the fluid couples to gravity beyond general relativity, while the microphysics of the fluid (including molecular dynamics and thermodynamics) can still be handled within the standard formulation, which facilitates relevant astrophysical applications. To enable a quantitative analysis, Sec.~\ref{sec:III.A} derives the cosmological evolution equations in a flat FLRW framework. In the standard radiation and pressureless-matter eras, $w_\subF=\mathrm{const.}$, and the evolution equations can be recast into a two-dimensional autonomous dynamical system (\ref{eq:3.6}). The EDE scenario requires $\Omega_\subEDE\rightarrow0$ in both eras, which is equivalent to requiring that the critical point $F_1$ be stable. Linearization yields the corresponding viable parameter space (see Fig.~\ref{fig:cosmos:01}) together with the general solution in the vicinity of $F_1$. To capture the global structure, the planar phase portrait is projected onto the Poincar\'e sphere, showing that the critical points at infinity are also crucial for the phase-space flow. A representative phase portrait is shown in Fig.~\ref{fig:cosmos:02}. During the cosmic radiation-matter transition, $w_\subF$ is time-varying, and the two-dimensional system is no longer autonomous. This time dependence can trigger a sizable $\Omega_\subEDE$, and a reliable approximation is obtained in Eq.~(\ref{eq:3.15}). The left panel of Fig.~\ref{fig:cosmos:05} displays a full numerical evolution, directly illustrating the EDE dynamics in GCT and quantitatively validating the approximate solutions in the fluid-dominated eras and in the transition regime. In the future, a cosmological linear-perturbation analysis is required to assess quantitatively whether the caloric EDE model can resolve the Hubble tension. Furthermore, it is worth exploring whether our fluid approach to modified gravity has implications for the $S_8$ tension.

Section~\ref{sec:III.B} includes a cosmological constant, which plays two distinct cosmological roles depending on whether $\Lambda$ is interpreted as the observed dark energy or as a huge theoretical vacuum energy. First, in the role relevant to the main cosmological scenario considered in this paper, $\Lambda$ represents the observed dark energy and becomes dynamically important only in the late-time and future Universe. The right panel of Fig.~\ref{fig:cosmos:05} shows a representative numerical solution. In this case, the relative energy density of $S_\mu$ remains negligible around the present epoch, leaving the observed low-redshift expansion history nearly unchanged. In the far future, however, $S_\mu$ can efficiently counteract $\Lambda$, driving the system toward a linearly (instead of exponentially) expanding asymptotic state with $w_\subtot\rightarrow -1/3$. To characterize this far-future $\Lambda$-cancelling dynamics, we switch off the fluid sector and focus on the $\{\Lambda,S_\mu\}$ subsystem. The cosmological equations then reduce to the two-dimensional autonomous dynamical system (\ref{eq:3.18}). The phase portrait in Fig.~\ref{fig:cosmos:06} shows that the dynamics is organized by a nontrivial critical point at infinity, $\mathcal{I}_4$, together with a unique attractor trajectory from the finite region toward $\mathcal{I}_4$. With an appropriate coordinate transformation, the stability of $\mathcal{I}_4$ and this attractor trajectory can be analyzed using centre manifold theory. For the one-parameter sector with only the $c_2$ term retained, the analysis shows that, wherever the $\Lambda$-cancelling attractor exists, its asymptotic EoS is $-1/3$ and is independent of $c_2$. Second, as a hypothetical application (rather than a description of the real Universe), $\Lambda$ is interpreted as a huge theoretical vacuum energy in the early Universe. The same $\Lambda$-cancelling dynamics can then be viewed as a possible self-tuning mechanism for the old cosmological constant problem \cite{Dolgov1985.JETPLett.41.345,Ford1987.PRD.35.2339,Charmousis2012.PRL.108.051101}. The above result raises a practical challenge: $S_\mu$ can counteract $\Lambda$, but the asymptotic EoS remains $-1/3$, leaving it unclear how the Universe can enter the standard radiation-dominated phase. Thus, the present self-tuning realization is incomplete. Nevertheless, our analysis suggests that critical points at infinity, together with centre manifold theory, can provide an effective framework for analysing self-tuning. In particular, controlling the asymptotic divergence rate of trajectories order by order (as in Sec.~\ref{sec:III.C}) may help probe the finer dynamical structure of self-tuning solutions and guide future model building.

Motivated by the critical point at infinity $I_3$ in Fig.~\ref{fig:cosmos:02}, Sec.~\ref{sec:III.C} proposes the \textit{early static hot Universe} (ESHU) scenario. The defining property is that $S_\mu$ can offset the fluid contribution to the Hubble expansion rate, so that a Universe filled with conventional hot gas can nonetheless exhibit an extremely small expansion rate. This quasi-static dynamics can generate a very large CHR and may therefore offer an alternative to inflation for addressing the horizon problem. Quasi-static cosmic evolution also appears in other inflation alternatives, such as string gas cosmology \cite{Tseytlin1992.NuclPhysB.372.443,Nayeri2006.PRL.97.021302}. The essential difference is that the static stage of ESHU remains compatible with ordinary hot gas, which may leave distinguishable signatures in primordial fluctuations. The present study is restricted to the background evolution in GCT. Since ESHU belongs to the $\{\mathrm{fluid},S_\mu\}$ subsystem, its dynamics is governed by Eq.~(\ref{eq:3.6}). From the perspective of mathematical tractability, the analysis focuses on the $I_3\rightarrow F_1$ attractor (see Fig.~\ref{fig:cosmos:02}), which can capture the late-ESHU dynamics. The dynamical analysis, especially centre manifold theory, shows that this regime can be approximated by $\lim_{x_3\rightarrow\infty}x_3^\prime\propto-x_3^n$, where $n$ is an integer. The corresponding CHR can decay as a power law or exponentially with respect to the \textit{e}-folding number. The parameter regions required to realize $n=3,2,1$ in GCT are shown in Fig.~\ref{fig:cosmos:07}, and the key characteristics, requirements, and challenges of the corresponding ESHU dynamics are summarized in Table~\ref{tab:04}. The natural exit of ESHU, i.e., reentering the hot Big Bang phase, relies on the stability of the critical point $F_1$. Under the stability requirements for both $I_3$ and $F_1$, viable representative parameter points are selected for $n=3,2,1$, respectively. With suitable initial conditions, Fig.~\ref{fig:cosmos:08} presents complete numerical evolutions from the onset of ESHU to its exit, quantitatively demonstrating the validity of the late-ESHU approximations. Although the ESHU scenario appears promising, our model realization still faces several challenges. Quantitative integration indicates that the $n=3$ realization cannot resolve the horizon problem. While the $n=2$ and $n=1$ realizations can do so, they may require extreme initial conditions and are found to be sensitive to the time variation of $w_\subF$, raising robustness concerns. Future work should construct more robust ESHU realizations. It also remains necessary to address the flatness problem and to establish a mechanism for generating a nearly scale-invariant power spectrum.

Self-tuning and ESHU can be regarded as members of a broader family of energy-cancelling solutions: under certain conditions, standard energy components (e.g., radiation, pressureless matter, and dark energy) preferentially excite additional fields (e.g., $S_\mu$ in GCT) rather than generating the corresponding spacetime curvature implied by the Einstein field equations. In parallel with Newton's law of \textit{universal} gravitation \cite{Newton1687.book}, this may suggest the existence of a new \textit{universal} interaction associated with standard energy or mass. Future work should establish more realizations of this phenomenon, particularly beyond cosmology, to strengthen this conclusion. Furthermore, it remains important to explore observable effects of such a possible new interaction.

Section~\ref{sec:IV} formulates the gauge-invariant linear perturbation theory of GCT on a Minkowski background, providing a framework to assess weak-field consistency and gravitational-wave phenomenology. Both the fluid sector and the caloric field are assumed to vanish at the background level, leaving the $c_2$-term as the only relevant contribution at linear order. A full scalar-vector-tensor decomposition of the metric perturbations and a Helmholtz decomposition of $S_\mu$ enable an explicit decoupling of the linearized field equations. The results show that if $c_2\neq1$, the linearized system exhibits an Ostrogradsky-type instability, classified through a reconstructed Lagrangian. Although this does not definitively rule out $c_2\neq1$ because the stability conclusion may depend on the chosen background, it singles out $c_2=1$ as a distinguished case. In the static weak-field regime, for general $c_2$, standard asymptotically flat boundary conditions force the caloric field to vanish at linear order, thereby recovering general relativity and reproducing the Newtonian limit with $\gamma_\mathrm{PPN}=1$. This also identifies the constant $G$ introduced in Eq.~(\ref{eq:2.1}) as Newton's gravitational constant. In the radiative regime, taking $c_2=1$ to avoid the instability, the linear dependence of $S_{\mu\nu}$ on $S_\mu$ implies that propagating caloric perturbations can source metric waves. A detailed analysis shows that GCT admits six independent gravitational-wave polarizations \cite{Eardley1973.PRD.8.3308,Eardley1973.PRL.30.884}, all propagating at the speed of light. To the best of our knowledge, this is the first explicit example of a Riemannian-geometric theory of gravity in which all six polarizations coexist as independent luminal modes. These features provide a clean platform for future gravitational-wave polarization tests.

Section~\ref{sec:V} investigates static spherically symmetric systems in GCT as a first step toward compact-object phenomenology. The generalized Tolman-Oppenheimer-Volkoff equations are straightforward to derive, but their tightly coupled nonlinear structure confines the present analysis to two fluid-free vacuum subsectors. In the simplest slice $\{c_i=\delta_{i2},\,\Lambda=0,\,\nu=0,\,\lambda=\lambda(r)\}$, the structure equations reduce to a single-variable Abel equation (\ref{eq:5.7}) of the first kind. Solving this system and diagnosing the geometry with curvature invariants reveals a rich solution space featuring a central physical singularity, a finite-radius wormhole, an intermediate effective-$\Lambda$ region, and physical and coordinate singularities at infinity. Figure~\ref{fig:star:09} illustrates a representative caloric de~Sitter wormhole and a caloric anti-de Sitter spacetime. Figure~\ref{fig:star:10} provides a global classification of the solution space in the $\lambda$--$\mu$ plane, where distinct domains correspond to different singularity structures and asymptotics. Two separatrix solutions that are asymptotically flat at infinity deserve emphasis: one has a naked singularity in the interior, whereas the other features a wormhole. Relaxing the requirement $\nu=0$ leads to a two-variable first-order system described by Eq.~(\ref{eq:5.8}) with $\Lambda=0$. In contrast to the preceding global classification, the treatment of Eq.~(\ref{eq:5.8}) is exploratory, consisting of several representative numerical trajectories (see Fig.~\ref{fig:star:11}) and two special exact solutions, namely, a mirror solution of the Schwarzschild metric and a purely caloric charged solution. Within the vacuum subsectors explored so far, no black-hole solution beyond Schwarzschild has been identified. This motivates two directions for future work. First, a more complete survey of static vacuum solutions is still required, including both the reduced two-variable system in Eq.~(\ref{eq:5.8}) and the broader full-theory parameter space with $c_4,c_6\neq0$, to determine whether GCT admits a sizeable asymptotically flat branch relevant for astrophysical objects. Second, the stability of the candidate compact-object solutions and their possible formation through dynamical collapse \cite{Penrose1965.PRL.14.57,Hawking1970.ProcRSocLondA.314.529} should be investigated to assess their physical relevance and to clarify whether the present absence of additional black-hole solutions reflects a no-hair-type theorem \cite{Israel1967.PhysRev.164.1776} or instead indicates an obstruction to horizon formation in GCT \cite{NoHairCaveat}.

The exact spherically symmetric solutions in turn shed light on the preceding cosmological and weak-field analyses in two respects. First, the cosmological and local solutions together reveal a recurring $\Lambda$-like role for $S_\mu$: it can offset a bare $\Lambda$ in cosmology and generate approximately constant-curvature regions in static vacuum geometries. This may hint at a deeper relation between $S_\mu$ and $\Lambda$. Second, in the explored exact solution space, asymptotic flatness is confined to special separatrix branches rather than occupying an open set of solutions. This raises the possibility that the Minkowski-based perturbation analysis with standard asymptotically flat boundary conditions may not be the physically appropriate default for local systems in GCT. Further analysis is therefore required to assess the implications of GCT for local gravitational systems.

In closing, we emphasize the present status of GCT. Given the unresolved theoretical issues identified above and the absence of direct fits to observational data, we do not claim that the caloric field $S_\mu$ or the $J_\mu S_\nu$-type couplings actually exist in Nature. Rather, the positive results reported in this paper suggest that GCT can at least serve as a useful \textit{toy model} of gravity, providing a laboratory for studying novel gravitational phenomena, including EDE, ESHU, self-tuning dynamics, gravitational-wave physics and astrophysical structures beyond general relativity. Some of these ingredients may prove to be mathematical artifacts, whereas others may capture effective aspects of a more fundamental theory or of the real Universe.

\section*{Acknowledgements}
This work was supported by the National Natural Science Foundation of China under Grant Nos.~12405050 and 12433001, and the Fundamental Research Funds for the Central Universities. This project was initiated in late 2022, when a preliminary form of the GCT field equations (without the $J_\mu S_\nu$ coupling) was formulated. From then until January 2024, this work was also supported by the China Postdoctoral Science Foundation under Grant No.~2021M700481. S.X.T. is most indebted to his Ph.D. advisor, Prof.~Zong-Hong Zhu, for his continual encouragement and support over the years.

\appendix
\renewcommand{\thesection}{\Alph{section}}
\renewcommand{\theequation}{\thesection.\arabic{equation}}

\section{Mathematics for dynamical systems}
This appendix summarizes the mathematical tools used in Sec.~\ref{sec:III} for the analysis of two-dimensional autonomous dynamical systems. The methods include linear stability theory, with centre manifold theory for nonhyperbolic cases, and coordinate transformations for critical points at infinity. These tools provide a systematic framework for analyzing the critical points as well as \textit{attractor trajectories} in phase portraits \cite{Bahamonde2018.PhysRep.775.1,Perko2001.book,Meiss2007.book}.

The symbols $(x_1,x_2)$ below denote generic coordinates of a two-dimensional phase space and should not be confused with the specific cosmological variable $x_2$ in Eq.~(\ref{eq:3.2}). In Sec.~\ref{sec:III}, they are replaced by the relevant reduced variables, such as $(x_1,x_3)$, $(x_1,x_4)$, or the local chart $(z_1,z_0)$.

\subsection{Linear stability theory}\label{sec:App.A.1}
A two-dimensional autonomous dynamical system can be written as
\begin{equation}\label{eq:A.01}
\begin{aligned}
  x_1^\prime &= P(x_1,x_2), \\
  x_2^\prime &= Q(x_1,x_2),
\end{aligned}
\end{equation}
where $^\prime\equiv\dx/\dx N$ and, in the context of this work, both $P$ and $Q$ are bivariate polynomials. Let $(x_1,x_2)_\ast$ denote a finite-region critical point, defined by $P_\ast=Q_\ast=0$. The Jacobian matrix evaluated at this point is
\begin{equation}\label{eq:A.02}
  J_\ast =
  \begin{bmatrix}
    \p P/\p x_1  &  \p P/\p x_2 \\
    \p Q/\p x_1  &  \p Q/\p x_2 
  \end{bmatrix}.
\end{equation}
Its local stability is determined by the eigenvalues of $J_\ast$ \cite{Perko2001.book}. For a $2\times2$ real matrix, any complex eigenvalues that may arise occur in conjugate pairs. The critical point is classified as a stable node if both eigenvalues are real and negative; a stable spiral if complex with negative real parts; a saddle if the eigenvalues have opposite signs; an unstable node if both are real and positive; and an unstable spiral if complex with positive real parts. 

\subsection{Centre manifold theory}\label{sec:App.A.2}
If one eigenvalue of $J_\ast$ vanishes, the critical point is nonhyperbolic and linear theory alone cannot determine the local dynamics. An appropriate tool for this case is centre manifold theory \cite{Carr1981.book,Bahamonde2018.PhysRep.775.1}, whose construction is summarized below. Denote the nonzero eigenvalue by $\lambda_h$, and the eigenvectors associated with $\lambda_h$ and the zero eigenvalue by the column vectors $\mathbf{v}_h$ and $\mathbf{v}_c$, respectively. The subscripts $h$ and $c$ refer to the \textit{hyperbolic} (nonzero-eigenvalue) and \textit{centre} (zero-eigenvalue) directions. To diagonalize the linear part~\footnote{See Eq.~(2.11) in Ref.~\cite{Bahamonde2018.PhysRep.775.1} for an example.} of Eq.~(\ref{eq:A.01}) around the critical point $(x_1,x_2)_\ast$, we shift the critical point to the origin and introduce the linear change of variables
\begin{equation}\label{eq:A.03}
  \begin{pmatrix}
    u_h \\
    u_c 
  \end{pmatrix}
  =
  S_\ast^{-1}
  \begin{pmatrix}
    x_1 - x_{1,\ast} \\
    x_2 - x_{2,\ast}
  \end{pmatrix},
\end{equation}
where the matrix $S_\ast=[\mathbf{v}_h,\mathbf{v}_c]$. The normalization of eigenvectors is arbitrary, since it only rescales the transformed variables without affecting the stability. Equation~(\ref{eq:A.03}) is an exact change of variables and does not involve any approximation. The linear part is diagonalized as $S_\ast^{-1}J_\ast S_\ast={\rm diag}\{\lambda_h,0\}$. Then Eq.~(\ref{eq:A.01}) gives
\begin{equation}\label{eq:A.04}
\begin{aligned}
  u_h^\prime &= \lambda_h\cdot u_h + g_h(u_h,u_c), \\
  u_c^\prime &= \ \  0\cdot u_c + g_c(u_h,u_c),
\end{aligned}
\end{equation}
where $g_h$ and $g_c$ contain only quadratic and higher-order terms in $u_h$ and $u_c$, and are determined by $P$ and $Q$. So far, the hyperbolic and centre directions are decoupled at linear order. The nonlinear dynamics along the centre direction is characterized by the following theorem:
\begin{theorem}
  For Eq.~(\ref{eq:A.04}), there exists a local invariant curve tangent to the centre direction at the critical point. This curve is called a centre manifold and can be written as $u_h=\phi(u_c)$, with $\phi(0)=0$ and $\dx\phi/\dx u_c|_{u_c=0}=0$. The dynamics restricted to this curve is
  \begin{equation}\label{eq:A.05}
    u_c^\prime = g_c(\phi(u_c),u_c).
  \end{equation}
\end{theorem}
The local form of $\phi(u_c)$ is obtained by imposing the invariance of the centre manifold~\footnote{A centre manifold is not necessarily unique \cite{Meiss2007.book}. This does not affect the local stability analysis because the required series expression is unique and can be fixed order by order. Possible differences between centre manifolds are flat at the critical point, e.g., of order $\exp(-1/u_c^{2})$.}. Differentiating $u_h=\phi(u_c)$ with respect to $N$ and using Eq.~(\ref{eq:A.04}) gives
\begin{equation}\label{eq:A.06}
  \lambda_h\cdot\phi + g_h(\phi,u_c) = g_c(\phi,u_c)\cdot\frac{\dx\phi}{\dx u_c},
\end{equation}
which is a first-order nonlinear ordinary differential equation for $\phi(u_c)$. In practice, it is solved locally by the series ansatz $\phi(u_c) = \sum_{i=2}^{\infty} \alpha_i u_c^i$. The absence of constant and linear terms follows from the conditions stated in the above theorem~\footnote{When transformed back to the original $x_i$ coordinates, the same centre manifold may contain constant, linear, or even fractional powers. Here fractional powers can appear if the curve is parametrized by an inappropriate variable, e.g., rewriting $y\propto x^2$ as $x\propto y^{1/2}$. This is why the initial series ansatz should be imposed in the $u_i$ coordinates, rather than the $x_i$ coordinates.}. Comparing powers of $u_c$ fixes the coefficients order by order, yielding the explicit form of the reduced flow in Eq.~(\ref{eq:A.05}).

In summary, the stability in the hyperbolic direction is determined by the sign of $\lambda_h$, while the stability along the centre direction is determined by the reduced flow in Eq.~(\ref{eq:A.05}). For nonhyperbolic critical points, we adopt the following convention in the main text (e.g., Table~\ref{tab:03}): if nearby phase-portrait trajectories exhibit both attracting and repelling behavior, the point is still referred to as a saddle. A similar convention is adopted for stable and unstable nodes.

In the main text, the above construction is applied after the critical points at infinity have been unfolded in a local $z_i$ chart. For the $\Lambda$-cancelling solution, the relevant point is $\mathcal{I}_4$ in the $z_i$ system (\ref{eq:3.19}), located at $(z_{1},z_{0})|_{\mathcal{I}_4}=(3/(2-4c_2),0)$. Choosing the eigenvectors such that 
$S_\ast=\big[\begin{smallmatrix}
  1 & (1+c_2)/(4c_2-2) \\
  0 & 1
\end{smallmatrix}\big]$ in Eq.~(\ref{eq:A.03}) gives $u_h=z_1-[3-(1+c_2)z_0]/(2-4c_2)$ and $u_c=z_0$, so the ansatz $u_h=\phi(u_c)$ gives Eq.~(\ref{eq:3.20}). For ESHU, the relevant point at infinity is $I_3$ of the autonomous dynamical system (\ref{eq:3.6}). In the simplified case $c_4=c_6=0$, Eq.~(\ref{eq:3.6}) is transformed into the $z_i$ system (\ref{eq:3.25}), where $I_3$ is located at $(z_1,z_0)|_{I_3} = (1/\tilde{c}_{35},0)$. The hyperbolic and centre eigenvectors are aligned with the $z_0$ and $z_1$ axes, respectively. Normalizing the eigenvectors such that $S_\ast=\big[\begin{smallmatrix}
  0 & 1 \\
  1 & 0
\end{smallmatrix}\big]$, Eq.~(\ref{eq:A.03}) gives $u_h=z_0$ and $u_c=z_1-1/\tilde{c}_{35}$. The ansatz $u_h=\phi(u_c)$ then leads to Eq.~(\ref{eq:3.26}). For the more general ESHU case discussed in Sec.~\ref{sec:III.C.3}, the corresponding diagonalizing transformation is given in Eq.~(\ref{eq:3.43}). After the centre-manifold ansatz is translated back into the $z_i$ chart, its coefficients can be fixed directly from the invariance condition using the corresponding $z_i$ dynamical equations, without writing the closed $u_i$ system explicitly.

\subsection{Coordinate transformations for infinity}\label{sec:App.A.3}
Dynamical systems may possess critical points at infinity, which play a key role in shaping the global phase-space structure. To analyze these features, it is common to map infinity onto a finite region using Poincar\'e compactification \cite{Lefschetz1957.book,Perko2001.book}. The mapping of the $(x_1,x_2)$ plane to the Poincar\'e sphere is defined as (see Fig.~\ref{fig:cosmos:06} for an illustration, with the indices $4\to2$ for consistency)
\begin{equation}\label{eq:A.07}
\begin{aligned}
  y_i &= x_iy_0, \mathrm{\ \ for\ } i=1,2, \\
  y_0 &= (1+x_1^2+x_2^2)^{-1/2},
\end{aligned}
\end{equation}
so that $y_1^2+y_2^2+y_0^2=1$. Geometrically, the $x_i$-plane is identified with the tangent plane $y_0=1$ of the unit sphere. The corresponding point $(y_1,y_2,y_0)$ is obtained by drawing the straight line from the center of the sphere to the point on this plane with coordinates $(x_1,x_2)$ and taking its intersection with the upper hemisphere. The derivatives $\{y_1^\prime,y_2^\prime,y_0^\prime\}$ follow directly from this mapping. Infinity in the $x_i$ coordinates corresponds to the equator $y_0\rightarrow0^{+}$. In general, one can verify that $\lim_{y_0\rightarrow 0^{+}}y_i^\prime = \mathcal{O}(y_0^{1-m})$ for $i=1,2$, which represents a divergence in the transformed system. Here, $m$ denotes the maximal degree of the polynomials $P$ and $Q$. This divergence can be removed by a time reparametrization along the trajectories \cite{Meiss2007.book},
\begin{equation}\label{eq:A.08}
  \sigma = \int_{\mathrm{traj.}} y_0^{1-m}\, \dx N,
\end{equation}
with an arbitrary starting point of the integral. Then, Eq.~(\ref{eq:A.01}) can be rewritten as
\begin{equation}\label{eq:A.09}
\begin{aligned}
  \frac{\dx y_1}{\dx\sigma} &= y_0^m \big[(1-y_1^2) P(\frac{y_1}{y_0},\frac{y_2}{y_0}) - y_1y_2 Q(\frac{y_1}{y_0},\frac{y_2}{y_0})\big], \\
  \frac{\dx y_2}{\dx\sigma} &= y_0^m \big[(1-y_2^2) Q(\frac{y_1}{y_0},\frac{y_2}{y_0}) - y_1y_2 P(\frac{y_1}{y_0},\frac{y_2}{y_0})\big], \\
  \frac{\dx y_0}{\dx\sigma} &= - y_0^{1+m} \big[y_1 P(\frac{y_1}{y_0},\frac{y_2}{y_0}) + y_2 Q(\frac{y_1}{y_0},\frac{y_2}{y_0})\big],
\end{aligned}
\end{equation}
which preserves the constraint $y_1^2+y_2^2+y_0^2=1$ and remains regular on the upper hemisphere.

The coordinates of the critical points at infinity can be determined by analyzing the behaviour of Eq.~(\ref{eq:A.09}) on the equator, which is invariant under the flow on the Poincar\'e sphere. In the limit $y_0\rightarrow0^{+}$, imposing $y_1^2+y_2^2=1$, $\lim_{y_0\rightarrow0^{+}}y_0^m P(y_1/y_0,y_2/y_0) = P_m(y_1,y_2)$, and an analogous limit for $Q$, Eq.~(\ref{eq:A.09}) can be reduced to
\begin{equation}
\begin{aligned}
  \lim_{y_0\rightarrow0^{+}}\frac{\dx y_1}{\dx\sigma} &= y_2\big[y_2P_m(y_1,y_2) - y_1Q_m(y_1,y_2)\big], \\
  \lim_{y_0\rightarrow0^{+}}\frac{\dx y_2}{\dx\sigma} &= y_1\big[y_1Q_m(y_1,y_2) - y_2P_m(y_1,y_2)\big],
\end{aligned}
\end{equation}
where $P_m$ and $Q_m$ denote the $m$-th degree terms of $P$ and $Q$ \footnote{One of $P_m$ and $Q_m$ may vanish if the degrees of $P$ and $Q$ differ.}. 
This leads to the following theorem:
\begin{theorem}
  Critical points at infinity are determined by the zeros of the leading angular flow on the equator, namely $y_1^2+y_2^2=1$ and $y_1Q_m(y_1,y_2) - y_2P_m(y_1,y_2) = 0$. If this leading angular flow vanishes identically on the equator, the corresponding next-to-leading-order condition should be imposed instead.
\end{theorem}
The two cases are applied to the critical points listed in Tables~\ref{tab:03} and \ref{tab:02}, respectively.

In principle, Eq.~(\ref{eq:A.09}) can be used to analyze the stability of the critical points at infinity. For convenience, we can instead work with unconstrained planar coordinates by further projecting the Poincar\'e sphere onto a tangent plane. For a critical point in the sector $y_2>0$, a natural choice is the plane $y_2=1$, corresponding to the $z_1$-$z_0$ plane shown in Fig.~\ref{fig:cosmos:06} (with the index $4\rightarrow2$ for consistency). The projection is defined along straight lines passing through the center of the sphere, giving
\begin{equation}
  z_i = y_i/y_2, \mathrm{\ \ for\ }  i=1,0.
\end{equation}
Eliminating the $y_i$ variables gives $z_1=x_1/x_2$, $z_0x_2=1$, and hence $x_1=z_1/z_0$. Infinity in the original $x_i$ variables is mapped to $z_0=0$ in the $z_i$ plane. The expressions for $z_i^\prime$ follow directly from the above relations. In particular, one can verify that, generically, $z_1^\prime=\mathcal{O}(z_0^{1-m})$ and $z_0^\prime=\mathcal{O}(z_0^{2-m})$ as $z_0\rightarrow0^{+}$. To regularize the system near $z_0=0$, we can rescale the time variable as \cite{Meiss2007.book,newsigma}
\begin{equation}\label{eq:A.12}
  \sigma = \int_{\mathrm{traj.}} z_0^{1-m}\, \dx N.
\end{equation}
In terms of the above transformations, Eq.~(\ref{eq:A.01}) becomes
\begin{equation}\label{eq:A.13}
\begin{aligned}
  \frac{\dx z_1}{\dx\sigma} &= z_0^m P(\frac{z_1}{z_0},\frac{1}{z_0}) - z_1z_0^m Q(\frac{z_1}{z_0},\frac{1}{z_0}), \\
  \frac{\dx z_0}{\dx\sigma} &= - z_0^{m+1} Q(\frac{z_1}{z_0},\frac{1}{z_0}).
\end{aligned}
\end{equation}
For the physical trajectories considered in Sec.~\ref{sec:III}, $x_2>0$ implies $z_0>0$, so Eq.~(\ref{eq:A.12}) gives $\dx\sigma/\dx N>0$, preserving the time orientation. Moreover, for Eq.~(\ref{eq:A.13}), since $\dx z_0/\dx\sigma$ is proportional to $z_0$, with a coefficient regular at $z_0=0$, the line $z_0=0$ is an invariant boundary, and trajectories starting from $z_0>0$ remain in this sector. For mathematical classification, one may also consider the analytic continuation to $z_0<0$, but trajectories in this region do not represent physical solutions. Accordingly, the stability classification (e.g., Table~\ref{tab:03}) should specify whether this analytic continuation is included or the analysis is restricted to the physical sector.


\begin{thebibliography}{300}
\bibitem [{Foo()}]{FootnoteEq329}%
  \BibitemOpen
  \href@noop {} {}\bibinfo {note} {Mathematically, adopting the conventions in
  Appendix~\ref{sec:App.A.2}, Eq.~(\ref{eq:3.29}) gives $\dx
  u_c/\dx\tilde{\sigma}$ identically equal to zero, where $u_c\propto
  3(w_\subF-1)z_0 + (2-6w_\subF)(z_1-1/\tilde{c}_{35})$. This is similar to the
  behaviour of Eq.~(\ref{eq:3.25}) along its centre manifold, i.e., the
  $z_1$-axis (see discussions below Eq.~(\ref{eq:3.26})). These two dynamical
  systems impose opposite stability requirements along the centre manifold. One
  requires the centre-manifold direction to be repelling, whereas the other
  requires it to be attracting. This distinction makes the vanishing flow fatal
  for Eq.~(\ref{eq:3.29}), but not for Eq.~(\ref{eq:3.25}).}\BibitemShut
  {Stop}%
\bibitem [{ReL()}]{ReLagrangian}%
  \BibitemOpen
  \href@noop {} {}\bibinfo {note} {This reconstruction is not unique. Here it
  is used only as a diagnostic example, and our final conclusion will rely on
  the original second-order system rather than any particular reconstructed
  Lagrangian.}\BibitemShut {Stop}%
\bibitem [{new()}]{newsigma}%
  \BibitemOpen
  \href@noop {} {}\bibinfo {note} {This $\sigma$ is distinct from the one
  introduced in Eq.~(\ref{eq:A.08}). The same symbol is reused, with its
  meaning fixed by the dynamical system under consideration.}\BibitemShut
  {Stop}%
\end{thebibliography}
%

\end{document}